\documentclass[a4paper,11pt]{article}
\pdfoutput=1 

\usepackage{jcappub} 
                     
\usepackage{aas_macros} 

\usepackage[T1]{fontenc} 
\usepackage[utf8]{inputenc}

\usepackage[english]{babel}
\usepackage{amsmath}
\usepackage{amssymb}
\usepackage{MnSymbol}
\usepackage{nicefrac}
\usepackage{amsfonts}
\usepackage{dsfont}

\usepackage{natbib}
\usepackage{hyperref}

\usepackage[markup=nocolor]{changes}

\usepackage{graphicx}

\usepackage{dcolumn}
\usepackage{bm}

\usepackage{color}
\usepackage{xcolor}
\newcommand{\bea}{\begin{eqnarray}}
\newcommand{\eea}{\end{eqnarray}}
\newcommand{\be}{\begin{equation}}
\newcommand{\ee}{\end{equation}}

\title{From a locality-principle for new physics to image features of regular spinning black holes with disks}

\author[a]{Astrid Eichhorn}
\author[b]{and Aaron Held}

\affiliation[a]{CP3-Origins, University of Southern Denmark,
\\
Campusvej 55, DK-5230 Odense M, Denmark}
\affiliation[b]{Theory Group, Blackett Laboratory,
\\
Imperial College London, SW7 2AZ, London, UK}

\emailAdd{eichhorn@cp3.sdu.dk}
\emailAdd{a.held@imperial.ac.uk}

\abstract{
Current observations present unprecedented opportunities to probe the true nature of black holes, which must harbor new physics beyond General Relativity to provide singularity-free descriptions. To test paradigms for this new physics, it is necessary to bridge the gap all the way from theoretical developments of new-physics models to phenomenological developments such as simulated images of black holes embedded in astrophysical disk environments.
\\
In this paper, we construct several steps along this bridge. We construct a novel family of regular black-hole spacetimes based on a locality principle which ties new physics to local curvature scales. We then characterize these spacetimes in terms of a complete set of curvature invariants and analyze the ergosphere and both the outer event as well as distinct Killing horizon. Our comprehensive study of the shadow shape at various spins and inclinations reveals characteristic image features linked to the locality principle. 
We also explore the photon rings as an additional probe of the new-physics effects. A simple analytical disk model enables us to generate simulated images of the regular spinning black hole and test whether the characteristic image-features are visible in the intensity map.
}

\begin{document}
\maketitle
\flushbottom
\allowdisplaybreaks


\section{Motivation}
An unprecedented multitude of observational opportunities of black-hole spacetimes has recently opened up: Not only can General Relativity (GR) be tested in gravitational waves emitted by binary-black-hole mergers~\cite{Abbott:2016blz}, post-Newtonian physics can also be accessed in stellar orbits around supermassive black holes~\cite{Ghez:1998ph,Ghez:2008ms,Meyer:2012hn,Do:2019txf}, and the predictions of GR can be compared to images of similarly massive black holes obtained with the technique of Radio Very-Long-Baseline-Interferometry (VLBI)
\cite{Falcke:1999pj,Doeleman:2008qh,paper1,paper2,paper3,paper4,paper5,paper6,Wielgus:2020zvj}. 
Although so far GR  has passed all tests, see, e.g., \cite{TheLIGOScientific:2016src,Do:2019txf,LIGOScientific:2019fpa,Cardoso:2019rvt,Abbott:2020jks,Volkel:2020xlc} for recent results and reviews, we already know that it cannot be the full theory: In GR, black holes harbor curvature singularities and the corresponding spacetimes are geodesically incomplete \cite{Penrose:1964wq, Hawking:1965mf}, signaling a breakdown of GR and the need for new physics. Thus, a new theoretical framework is required that will allow us to understand the true nature of black holes.  This has motivated a large body of work on black-hole shadows in settings beyond GR, see, e.g., \cite{
2010PhRvD..81l4045A,2015PhRvL.115u1102C,2014PhRvD..89l4004G,Moffat:2015kva,Amir:2016cen,Abdujabbarov:2016hnw,Cunha:2017eoe,2017PhLB..768..373C,Tsukamoto:2017fxq,Wang:2017hjl,Vetsov:2018mld,Okounkova:2018abo,Mizuno2018,Ayzenberg:2018jip,Wang:2018prk,Stuchlik2019,Held:2019xde,Konoplya:2019goy,Kumar:2019ohr,Allahyari:2019jqz,Shaikh:2019fpu,Dymnikova:2019vuz,Contreras:2019cmf,Kumar:2020owy,Fathi:2020agx,Guo:2020zmf,Brahma:2020eos,Ghosh:2020ece,Konoplya:2020xam,Chen:2020aix,Liu:2020ola,Kumar:2020hgm,Qian:2021qow,Junior:2021atr}.
\\

To explore shadow images in black-hole spacetimes beyond GR, two routes are commonly followed in the literature:
\\
The first could be called the ``parameterized'' approach: Without making specific assumptions about new physics, the proposals in, e.g., \cite{Vigeland:2009pr,Johannsen:2011dh,Vigeland:2011ji,Cardoso:2014rha,Johannsen:2015pca,Lin:2015oan,Konoplya:2016jvv,Papadopoulos:2018nvd} parameterize the deviations of metric coefficients from the form of a Kerr metric. Under the assumption that the dynamics for light and matter is that of standard electrodynamics and General Relativistic Magnetohydrodynamics, simulated images for the Event Horizon Telescope (EHT) can be calculated and deviations from Kerr can, in principle, be constrained.
\\
The second could be called the ``fundamental'' approach: Starting from a specific form of new physics -- either in the form of a classical dynamics beyond GR, see, e.g.,~\cite{DeFelice:2010aj,Capozziello:2011et,Clifton:2011jh,Liebling:2012fv,deRham:2014zqa,Bambi:2015kza, Berti:2015itd, Cardoso:2019rvt} for reviews, or from a proposal for quantum gravity, see, e.g., \cite{1999alfiomartin,Bonanno:2000ep,2006alfiomartin,Nicolini:2008aj,Reuter:2010xb,Falls:2010he,Falls:2012nd,Gambini:2013ooa,Litim:2013gga,Haggard:2014rza,ko14,Kofinas:2015sna,Saueressig:2015xua,Torres:2017ygl,Ashtekar:2018lag,Adeifeoba:2018ydh,Platania:2019kyx} -- a particular metric is derived or motivated.
In this approach, studies typically focus solely on the idealized shape of the shadow boundary \cite{2010PhRvD..81l4045A,2015PhRvL.115u1102C,2014PhRvD..89l4004G,Moffat:2015kva,Amir:2016cen,Abdujabbarov:2016hnw,Cunha:2017eoe,2017PhLB..768..373C,Tsukamoto:2017fxq,Wang:2017hjl,Vetsov:2018mld,Okounkova:2018abo,Ayzenberg:2018jip,Wang:2018prk,Stuchlik2019,Held:2019xde,Konoplya:2019goy,Kumar:2019ohr,Allahyari:2019jqz,Shaikh:2019fpu,Dymnikova:2019vuz,Contreras:2019cmf,Kumar:2020owy,Fathi:2020agx,Guo:2020zmf,Brahma:2020eos,Ghosh:2020ece,Konoplya:2020xam,Chen:2020aix,Liu:2020ola,Kumar:2020hgm,Qian:2021qow,Junior:2021atr} and effects of an accretion disk remain unaccounted, see, however,~\cite{Mizuno2018}.  In many of these studies, spherical symmetry is assumed when black-hole metrics are derived or motivated, and the effects of spin and inclination on black-hole shadows are only partially explored in this approach.
\\
Here and in~\cite{Eichhorn:2021etc}, we  follow a third approach: We formulate a set of new-physics principles that we demand a spinning black hole to satisfy. 
We expect that the latter capture key features from a relatively large class of modifications of GR, both classical and quantum. Most importantly, we assume a locality principle which states that the new physics modifications are tied to local curvature scales. Based on these assumptions, we provide a family of singularity-free, spinning black-hole spacetimes.
\\
We also find that no coordinate transformation of the passive coordinates exists, see App.~\ref{app:BL-form}, that maps our model to the parameterizations in~\cite{Johannsen:2011dh,Cardoso:2014rha,Johannsen:2015pca,Lin:2015oan,Konoplya:2016jvv,Papadopoulos:2018nvd}. In the absence of such a coordinate transformation, the ``parameterized'' approach cannot be used to observationally constrain our spacetimes.

This paper is structured as follows:
In Sec.~\ref{sec:newregblackhole} we present and motivate our locality principle, which is tied to local curvature scales. We then present in Sec.~\ref{subsec:construction}, how Kerr black holes can be upgraded to regular spinning spacetimes following the locality principle. Having proposed a concrete family of line-elements, we ask a number of questions regarding the resulting geometry in Sec.~\ref{subsec:Insideout}: 1) Is the resulting spacetime  regular everywhere? 2) Does it have an outer event horizon? 3) Does the event horizon differ from the Killing horizon? 4) Does the spinning spacetime feature an ergosphere? 5) Is the asymptotic limit in agreement with Newtonian gravity? In addition to answering these questions in the positive, we identify a characteristic feature of the locality principle in the event horizon -- namely a dent in the equatorial plane.

This motivates the question underlying Sec.~\ref{sec:images}, namely, whether black-hole images feature corresponding characteristics.
We thus ask the question, `Are characteristic features which are tied to the locality principle, apparent in black-hole images?' in increasingly realistic settings: First, we investigate the shadow shape at various spins and inclinations, cf.~Sec.~\ref{sec:shapes}. Second, we explore photon rings as a potential probe of the characteristic features in the shadow, cf.~Sec.~\ref{sec:rings}. Third, we add a simple analytically modeled disk as a source of illumination around the black hole and discuss simulated black-hole images, cf.~Sec.~\ref{sec:intensity}. In all of the above, we characterize qualitative deviations from Kerr that can be linked back to the locality principle.

We conclude in Sec.~\ref{sec:conclusions}.
Additionally, we provide an extended appendix \ref{app:curvatureinvnewmetric} that addresses the following questions: 
 What is the form of a complete basis of scalar polynomial curvature invariants (in the following called curvature invariants) for the general family of spacetimes?
 Can the metric be mapped to a Boyer-Lindquist form? Finally in App.~\ref{app:ray-tracing} we provide various technical details of the implementation of numerical ray-tracing.

\section{Regular, spinning black-hole spacetimes based on new physics with a locality principle}\label{sec:newregblackhole}
We build a regular, spinning black-hole spacetime, i.e., a generalization of the Kerr metric, based on the following three physical principles: 
\begin{enumerate}
\item (Newtonian limit) The spacetimes features a correct Newtonian limit.
\item  (Regularity principle) The spacetime is regular everywhere, due to an effective weakening of the gravitational interaction.
\item  (Locality principle) The deviations from Kerr set in beyond a critical value of the local curvature scale.
\end{enumerate}
The physical motivations underlying these assumptions are as follows: The first principle is motivated by numerous tests of gravity in the weak-field regime, see, e.g., \cite{Will:2014kxa,Berti:2015itd,2015ApJ...802...63B}.
\\
The second principle guarantees that all curvature invariants stay finite everywhere which relates to a geodesically complete spacetime.
\\
The third principle is a \emph{locality} assumption in the following sense: the construction is based on the \emph{local} value of the curvature invariants and no non-local information is required. How we implement this general notion in settings with more than one independent curvature invariant is detailed in Sec.~\ref{subsec:construction} directly below.
This is motivated by an effective-field-theory (EFT) point of view which states that modifications of gravity set in at large curvature scales, as they do in quantum and classical modifications of GR: At the quantum level, the imprints of physics beyond GR are captured by higher-order curvature terms which become important beyond a critical curvature scale.  For instance, within perturbative renormalization of GR, counterterms correspond to higher-order curvature operators \cite{tHooft:1974toh,Goroff:1985sz,vandeVen:1991gw}. Similarly, quantum fluctuations in e.g., string theory, see,~e.g.,~\cite{Boulware:1985wk, Zwiebach:1985uq, Bergshoeff:1989de}, or asymptotically safe gravity \cite{Percacci:2017fkn,Eichhorn:2018yfc,Reuter:2019byg} give rise to higher-order curvature terms in an effective action. Additionally, such terms have been explored in view of their renormalizability properties \cite{Stelle:1976gc,Avramidi:1985ki}.
At the classical level, various modifications of GR are based on a similar EFT expansion and postulate the existence of higher-order curvature terms, often motivated by phenomenological considerations in cosmology, see, e.g., \cite{Starobinsky:1980te,DeFelice:2010aj}.
Black holes in EFT settings have been explored in \cite{BjerrumBohr:2002ks,Lu:2015cqa,Cardoso:2018ptl,Bonanno:2019rsq,Konoplya:2020bxa,Borissova:2020knn,deRham:2020ejn,Xie:2021bur}.
\\

We are of course by no means the first to construct deviations from the Kerr spacetime, therefore we highlight two distinguishing points of our construction.
First, our third principle is typically violated by regular spinning black-hole spacetimes that are constructed by applying the Newman-Janis (NJ) algorithm \cite{Newman:1965tw}, see also \cite{Gurses:1975vu,Drake:1998gf}, as in, e.g., \cite{Bambi:2013ufa,Azreg-Ainou:2014pra,Toshmatov:2014nya,Ghosh:2014hea,Abdujabbarov:2016hnw,Torres:2017gix,Lamy:2018zvj,Kumar:2019ohr,Shaikh:2019fpu,Contreras:2019cmf,Liu:2020ola,Junior:2021atr,Mazza:2021rgq}. However, it has major implications for characteristic features of the resulting black-hole spacetime and its image. Conversely, these features are not present in regular black holes obtained through the NJ algorithm. \\
Second, as detailed in App.~\ref{app:BL-form}, we show that our metric cannot be brought into a Boyer-Lindquist form (except in the weak-field regime) through a coordinate transformation of the passive coordinates. Said parameterization is, however, used widely in, see, e.g.,~\cite{Johannsen:2011dh,Cardoso:2014rha,Johannsen:2015pca,Lin:2015oan,Konoplya:2016jvv,Papadopoulos:2018nvd}. Whether a coordinate transformation can be found when these assumptions are loosened is beyond the scope of this paper. For practical purposes, statements about deviations from Kerr obtained in a Boyer-Lindquist form cannot (yet or possibly in principle) apply to our model.
More generally, there are good reasons to work in horizon-penetrating coordinates, including the avoidance of curvature singularities at the event horizon. The model constructed in \cite{Held:2019xde} is an example, where the center of the black hole is regular, but novel curvature singularities occur at the horizon; other examples of the latter are discussed in~\cite{Johannsen:2013rqa,Cardoso:2014rha,Johannsen:2015pca,Lin:2015oan}.
Thus, a generalized form of the line element we explore here, could in the future also serve as testing ground for parameterized deviations from GR.
A more general parameterization in horizon-penetrating coordinates could be important for a comprehensive comparison observational data with new-physics models. 

In the remainder of this section, we show how the three principles listed above can be incorporated into a well-defined black-hole spacetime and investigate the properties of such regular black holes.

\subsection{Construction of a regular, axisymmetric spacetime}\label{subsec:construction}
We now construct our phenomenological model explicitly. To avoid the introduction of spurious curvature singularities at the horizon, see \cite{Held:2021} for further details, we work in horizon-penetrating coordinates.
A family of metrics in ingoing Kerr coordinates $(u, r, \chi, \phi)$ with $\chi = \cos(\theta)$ that manifestly exhibits the Killing vectors related to stationarity and axisymmetry is given by the line element
\bea
\label{eq:regular-spinning-BH-metric-ingoing-Kerr}
ds^2 &=&-\frac{r^2-2M(r, \chi) r +a^2 \chi^2}{r^2+a^2 \chi^2}du^2 +2\,du\, dr - 4\frac{M(r,\chi) a r}{r^2+a^2\chi^2}\left(1-\chi^2 \right) du\, d\phi\nonumber\\
&{}&- 2a\left(1-\chi^2 \right)dr\, d\phi + \frac{r^2+a^2\chi^2}{1-\chi^2}d\chi^2 \nonumber\\
&{}&+ \frac{1-\chi^2}{r^2+a^2\chi^2}\left(\left(a^2+r^2\right)^2 - a^2\left(r^2-2M(r, \chi)r+a^2 \right)\cdot \left(1-\chi^2 \right) \right)d\phi^2.
\eea
By choosing different functions $M(r, \chi)$, rather distinct spacetimes can be obtained. The form in Eq.~\eqref{eq:regular-spinning-BH-metric-ingoing-Kerr}, with appropriate conditions imposed on a nonsingular function $M(r\rightarrow 0,\chi \rightarrow 0)$ results in our parameterization of regular, spinning black-hole spacetimes. In particular, the black-hole-spacetimes in \cite{Modesto:2010rv,Caravelli:2010ff,Bambi:2013ufa,Azreg-Ainou:2014pra,Toshmatov:2014nya,Ghosh:2014hea,Dymnikova:2015hka,Dymnikova:2016nlb,Kumar:2019ohr,Contreras:2019cmf,Liu:2020ola,Junior:2020lya,Junior:2021atr,Mazza:2021rgq} that are obtained by the NJ-algorithm from a non-spinning counterpart, are contained in a special subclass with $M(r,\chi) = M(r)$.
 
We focus on a different subfamily of spacetimes in Eq.~\eqref{eq:regular-spinning-BH-metric-ingoing-Kerr}, namely on those that satisfies all three principles 1)-3).
To do so, we must find an appropriate notion of a local curvature scale to implement principle 3).
In the spherically symmetric case, there is only a single nonvanishing independent curvature invariant, the Kretschmann scalar
\be
\label{eq:Kretsch-spherical}
K = R_{\mu\nu\kappa\lambda}R^{\mu\nu\kappa\lambda}.
\ee
All other curvature invariants are polynomials in $K$. Thus, in the spherically symmetric case, $K^{-1/4}$ sets the relevant length scale, with the power $-1/4$ following from the dimensionality of $K$.
\\
In contrast, for Kerr spacetime, the lower degree of symmetry results in several independent non-zero curvature invariants, see App.~\ref{app:ZM-basis}. In addition, these independent curvature invariants can also change sign, i.e., they are no longer monotonic functions of $r$, unlike in Eq.~\eqref{eq:Kretsch-spherical}.
\\
In accordance with EFT principles, we assume that the new physics does not single out an individual curvature invariant. Further, we treat the sign of a curvature invariant as irrelevant for setting the scale.
Thus, the new-physics scale is set by the maximum of the absolute values of all independent curvature invariants $I_i$ at a given spacetime point, 
 \be
 \overline{K}_{\rm GR} = \underset{i}{\rm max}\,\{|I_i|\}.
 \ee
As we explicitly discuss in App.~\ref{app:KGRfromClassicalKerrInvariants}, for the Kerr spacetime, this quantity can be approximated by
\be
\overline{K}_{\rm GR}\approx
K_{\rm GR} = \left(I_1^2 + I_2^2\right)^{1/2} =\frac{48 M^2}{\left(r^2+a^2\chi^2 \right)^3}.\label{eq:KGR}
\ee
 The two invariants that enter  Eq.~\eqref{eq:KGR} are
\bea
I_1&=&C_{\mu\nu\rho\sigma}C^{\mu\nu\rho\sigma},\\
I_2 &=&\frac{ \sqrt{|g|}}{2}\,\epsilon_{\mu\nu\alpha\beta}C^{\alpha\beta}_{\hphantom{\alpha\beta}\rho\sigma}C^{\mu\nu\rho\sigma},
\eea
with the Weyl tensor $C_{\mu\nu\rho\sigma}$. Here, and for the remainder of the paper, we set the Newton coupling $G=1$.
\\

To incorporate principle 3), we introduce a new-physics length scale $\ell_{\rm NP}$ and construct our model using the dimensionless product $K_{\rm GR} \ell_{\rm NP}^4$.
We leave $\ell_{\rm NP}$ as a free parameter of the model. Often, singularity-resolution is attributed to quantum gravitational effects. However, the theoretical assumption that gravity remains well-described by GR above the quantum-gravity scale, must be tested observationally. It is not a given that classical new physics cannot resolve black hole singularities.
In fact, modifications of gravity in the EFT framework (with and without new fields \cite{Burrage:2017qrf,Brax:2018iyo,Baker:2020apq,Heisenberg:2020xak}) are currently being explored across a wide range of scales, from the quantum-gravity scale to cosmological scales.
As a specific example, the leading-order terms in a gravitational EFT, i.e., curvature-squared terms are explored at inflationary instead of Planckian scales \cite{Starobinsky:1980te}, see \cite{Lu:2015cqa,Lu:2015psa,Holdom:2016nek,Bonanno:2019rsq} for studies of black holes.
Naturally, the quantum-gravity scale remains a candidate for $\ell_{\rm NP}$, see, e.g.,
\cite{tHooft:1984kcu,Ashtekar:1997yu,Gauntlett:1998fz,Horowitz:1998pq,1999alfiomartin,br00,BjerrumBohr:2002ks,Fidkowski:2003nf,Nicolini:2005gy,Nicolini:2008aj,He:2008ku,Gambini:2008dy,Gambini:2013ooa,Rovelli:2014cta}. For black holes, it is actually an open question, what the relevant scale of quantum effects is.
This can be motivated by the black-hole information paradox, some resolutions of which assume an onset of quantum effects long before the Planck scale \cite{Unruh:2017uaw,Marolf:2017jkr,Almheiri:2019psf,Almheiri:2019qdq}. Further, it has been proposed that horizon-scale modifications might arise from quantum gravity and leave observational imprints in the ring-down phase of binary-black hole mergers \cite{Cardoso:2016oxy,Abedi:2016hgu}.
Thus, even if the new physics takes its origin in quantum gravity, it is not settled whether a naive dimensional analysis which would confine departures from GR to transplanckian curvature scales, is actually valid. Following naive dimensional analysis,  i.e., choosing $\ell_{\rm NP} \sim \ell_{\rm Planck}$, none of the effects we explore in the remainder of this paper are, in practice, observable\footnote{Whether dynamical properties of such quantum-gravity inspired spacetimes might have observable imprints is an open question, see, e.g., \cite{Carballo-Rubio:2018jzw}.}. 
\\

Next, we specify that the mass-function depends on the local curvature scale through the above dimensionless product, i.e., $M(r, \chi)=M(K_{\rm GR}\cdot \ell_{\rm NP}^4)$. We constrain $M(K_{\rm GR}\cdot\ell_{\rm NP}^4)$ by requiring the correct Newtonian limit, i.e., principle 1). Thus,
\be
M(K_{\rm GR}) \sim M +\mathcal{O}(K_{\rm GR}\cdot\ell_{\rm NP}^4), \mbox{ for } K_{\rm GR}\cdot \ell_{\rm NP}^4\ll1.
\ee
To accommodate principle 2), we demand that, at points at which the curvature scale is divergent for the Kerr geometry, the mass-function should vanish sufficiently fast, i.e.,
\be
M(K_{\rm GR}) \sim \left( K_{\rm GR}\cdot \ell_{\rm NP}\right)^{-\beta/2}, \mbox { for } K_{\rm GR}\cdot \ell_{\rm NP}^4 \rightarrow \infty,
\ee
with $\beta\geq 1$, see Sec.~\ref{sec:absenceofdiv}.\\
Both requirements can, for instance, be accommodated by setting
\be
M(K_{\rm GR}, \beta)= \frac{M}{1+\left(K_{\rm GR}\ell_{\rm NP}^4 \right)^{\frac{\beta}{2}}}.\label{eq:massfunctiongenbeta}
\ee 
For $\beta=1$, this specific choice reduces to the regular Hayward\cite{Hayward:2005gi} metric in the spherically symmetric limit.
Further members of the family of spinning black-hole spacetimes which satisfy our principles 1)-3) can be constructed by using other functions for $M(K_{\rm GR})$ with the same limits as above. Some of these are presented in our shorter companion paper \cite{Eichhorn:2021etc}.

\subsection{Exploring the black-hole spacetime from the inside out}\label{subsec:Insideout}

Having constructed our regular and spinning black-hole spacetime, we now proceed to analyze its most important features in order to identify how to test the corresponding deviations from GR. We will work our way outwards, starting from the regular core (see Sec.~\ref{sec:absenceofdiv}), via the outer event horizon (see Sec.~\ref{sec:horizon}), via the ergosphere (see Sec.~\ref{sec:ergosphere}), and to the Newtonian limit (see Sec.~\ref{sec:Newtonian-limit}).

\subsubsection{Absence of curvature singularities}\label{sec:absenceofdiv}
Here, we require regularity at the origin of the spacetime, thereby fixing the leading terms of the Taylor expansion of a general mass function $M(r, \chi)$.  We determine the minimum value of $\beta$ in the mass function Eq.~\eqref{eq:massfunctiongenbeta} that ensures the finiteness of the curvature invariants. As an example, we focus on the invariant $I_1=C_{\mu\nu\kappa\lambda}C^{\mu\nu\kappa\lambda}$. A comprehensive discussion of further curvature invariants is presented in App.~\ref{app:curvatureinvnewmetric}, but does not result in any new constraints.

\begin{figure}
\centering
\includegraphics[width=0.48\linewidth]{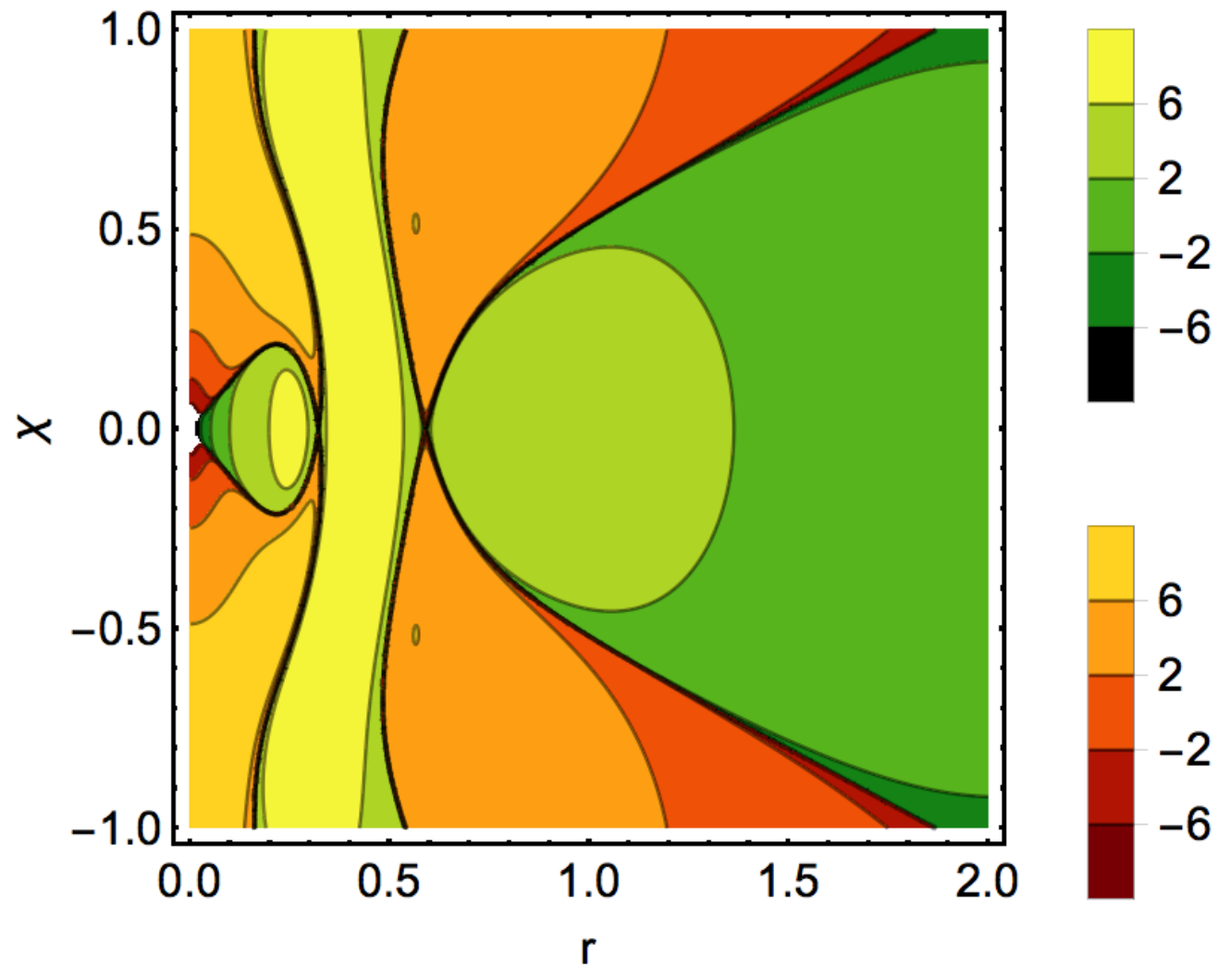}\quad \includegraphics[width=0.48\linewidth]{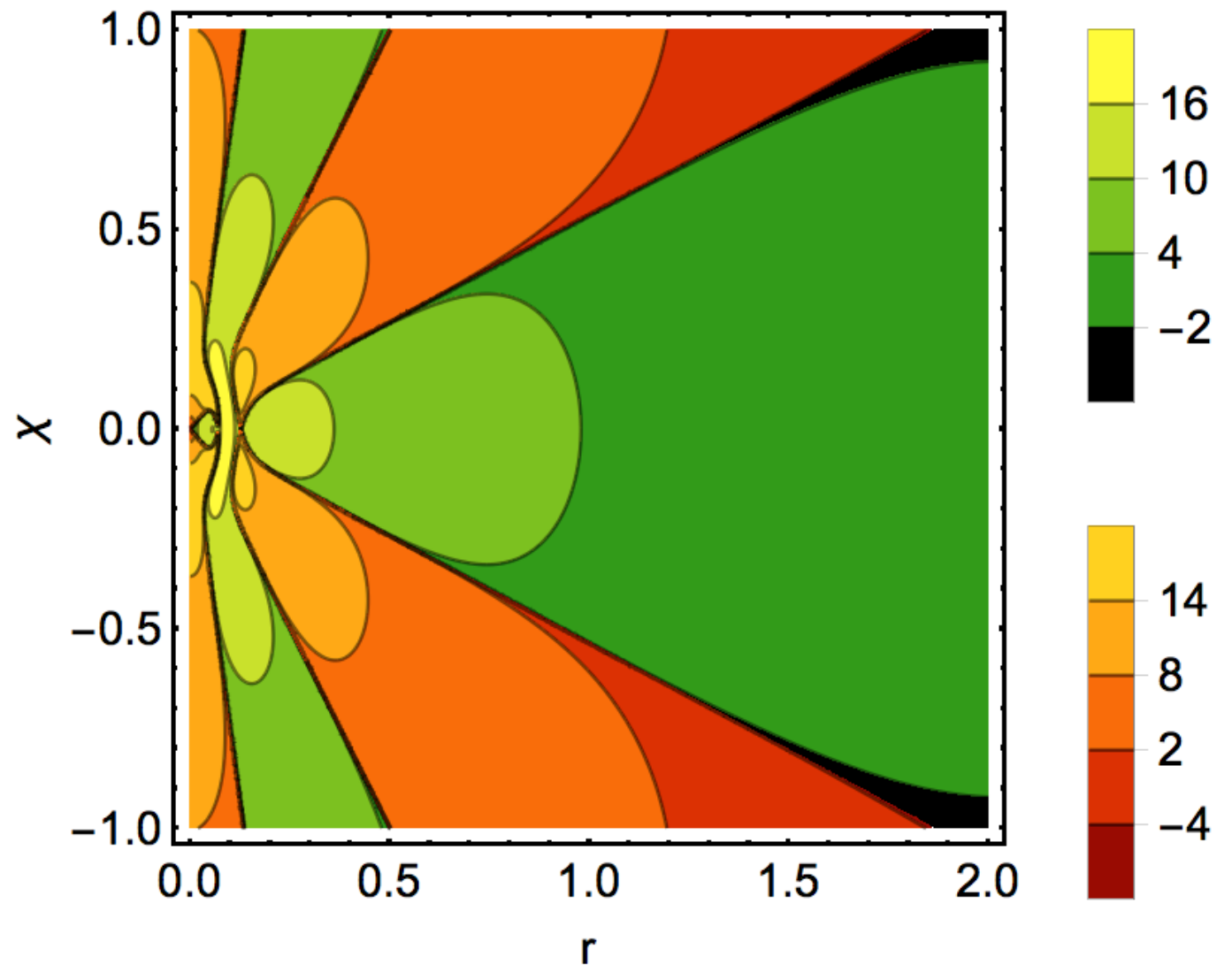}
\caption{\label{fig:I1diva} We show the logarithm of the absolute value of the curvature invariant $I_1$, i.e., the Weyl tensor squared, Eq.~\eqref{eq:Weylsqgenbeta} 
as a function of $r$ and $\chi$ for $M=1, \ell_{\rm NP}=10^{-1}$ (left column) and $\ell_{\rm NP}=10^{-2}$ (right column)
for $a=1/2$. We indicate positive/negative values by green/red tones.
}
\end{figure}

The Kerr spacetime features a ring singularity in the equatorial plane, where, in particular, the Kretschmann scalar and accordingly $I_1$ diverge for $\chi=0$ as $r\rightarrow 0$. We now investigate, which value of $\beta>0$ is required in order to lift this singularity. Sorting the somewhat lengthy expression in terms of powers of $\ell_{\rm NP}$, we obtain
\bea
I_{1}&=&C^{\mu\nu\kappa\lambda}C_{\mu\nu\kappa\lambda}\nonumber\\
&=&\frac{12 M^2}{\left(r^2+a^2\chi^2 \right)^6 \cdot\left( 1+3^{\beta/2}4^{\beta}\left(\frac{\ell_{\rm NP}^2 M}{\left(r^2+a^2\chi^2 \right)^{3/2}} \right)^{\beta}\right)^6} \cdot \Biggl( \frac{\ell_{\rm NP}^{8\beta}\, 2304^{\beta} \,M^{4\beta}}{\left(r^2+a^2\chi^2 \right)^{6\beta}}\cdot \Biggl[ r^6 \left(2-5\beta+3 \beta^2 \right)^2\nonumber\\
&{}&\quad+6r^4a^2\chi^2(\beta-1)(10+\beta(3\beta-13)) + 3r^2 a^4\chi^4(20+\beta(3\beta-20))- 4 a^6 \chi^6\Biggr]\nonumber\\
&{}&+\frac{\ell_{\rm NP}^{6\beta}\,2^{1+6\beta}\,3^{3\beta/2}\, M^{3\beta}}{\left(r^2+a^2\chi^2 \right)^{9\beta/2}}\cdot \Biggl[r^6\left(8+\beta(-30+31\beta-9\beta^3) \right)-12 r^4 a^2\chi^2(10+\beta(7\beta-18))\nonumber\\
&{}&\quad +3r^2a^4\chi^5 (40+3\beta(\beta-10))-8a^6\chi^6 \Biggr]\nonumber\\
&{}&+ \frac{\ell_{\rm NP}^{4\beta} 48^{\beta} M^{2\beta}}{\left(r^2+a^2\chi^2 \right)^{3\beta}}\cdot \Biggl[ r^6(24+\beta(-60+\beta(13+30\beta+9\beta^2)))\nonumber\\
&{}&\quad- 6 r^4 a^2\chi^2 (60+\beta(-72+\beta(5+3\beta)))+9r^2a^4\chi^4(40+\beta(\beta-20))-24 a^6\chi^6\Biggr]\nonumber\\
&{}&+\frac{\ell_{\rm NP}^{2\beta}\, 3^{\beta/2}\, 4^{1+\beta}\,M^{\beta}}{\left(r^2+a^2\chi^2 \right)^{3\beta/2}}\cdot \Biggl[ r^6(4-\beta(5+3\beta))+3r^4a^2\chi^2 (-20+3\beta(4+\beta))\nonumber\\
&{}&\quad-15 r^2a^4\chi^4(\beta-4)-4a^6\chi^6\Biggr]+4\Biggl[r^6-15r^4a^2\chi^2+15r^2a^4\chi^4-a^6\chi^6 \Biggr]\Biggr).\label{eq:Weylsqgenbeta}
\eea
As the Kerr singularity lies in the equatorial plane, we set $\chi =0$ and investigate $r \rightarrow 0$,
\bea
\underset{\chi \rightarrow 0} {\rm lim}\,C^{\mu\nu\kappa\lambda}C_{\mu\nu\kappa\lambda}&=&\frac{12 M^2\, r^{18 \beta}}{r^{12 \beta+6} \left(r^3+3^{\beta/2}\, 4^{\beta}\, \ell_{\rm NP}^{2\beta}M^{\beta} \right)^6}\cdot\Biggl( 
2 r^{6\beta} + 48^{\beta} \ell_{\rm NP}^{4\beta}M^{2\beta}(\beta-1)(3\beta-2)\nonumber\\
&{}&\quad+3^{\beta/2}4^{\beta}\ell_{\rm NP}^{2\beta}r^{3\beta}(4-\beta(5+3\beta))
\Biggr)^2\nonumber\\
&{}& \sim \frac{r^{18 \beta}}{r^{12\beta +6} \left(r^3+3^{\beta/2}4^{\beta}\ell_{\rm NP}^{2\beta}M^{\beta} \right)^6} 12\cdot 48^{\beta} M^{2+2\beta}(\beta-1)(3\beta-2),\label{eq:I1lim}
\eea
for $r \rightarrow 0$.
To yield a finite $r \rightarrow 0$ limit, $\beta$ has to satisfy an inequality arising from the last line in Eq.~\eqref{eq:I1lim}
\bea
18 \beta - 12 \beta -6\geq0 \quad \Leftrightarrow \quad \beta \geq 1.
\eea
In fact, for $\beta \geq 1$, 
\be
\underset{r \rightarrow 0}{\rm lim}\,\underset{\chi \rightarrow 0}{\rm lim}\,C^{\mu\nu\kappa\lambda}C_{\mu\nu\kappa\lambda}=0.\label{eq:I1lim1}
\ee
Requiring singularity-resolution, i.e., $\beta \geq 1$, is actually not the most stringent constraint. In fact, requiring single-valuedness of curvature invariants results in the stronger condition $\beta>1$. This can be seen by considering the opposite order of limits to the one taken in Eq.~\eqref{eq:I1lim}.
In the limit $r \rightarrow 0$, we obtain
\be
\underset{r \rightarrow 0} {\rm lim}\,C^{\mu\nu\kappa\lambda}C_{\mu\nu\kappa\lambda}= - \frac{48 M^2}{a^{12\beta+6}\chi^{12\beta+6}}\frac{a^{18\beta}\chi^{18 \beta}}{\left(1+3^{\beta/2}4^{\beta} \ell_{\rm NP}^{2\beta}M^{\beta} \right)^6} \left((a\chi)^{3\beta} +3^{\beta/2}4^{\beta}\ell_{\rm NP}^{2\beta}M^{\beta} \right)^4.
\ee
For $\beta >1$, the subsequent limit $\chi \rightarrow 0$ results in a vanishing value, i.e., for $\beta>1$, the Weyl-squared invariant is single-valued. For the just-singularity-resolving case $\beta=1$, a peculiarity occurs, which is known from other spinning regular black holes, see, e.g.,~\cite{Bambi:2013ufa}, in that $I_1$ is not single-valued at the origin. In contrast to Eq.~\eqref{eq:I1lim1}, the other order of the limits is non-zero for $\beta=1$:
\be
\underset{\chi \rightarrow 0}{\rm lim}\,\underset{r \rightarrow 0}{\rm lim}\,C^{\mu\nu\kappa\lambda}C_{\mu\nu\kappa\lambda}=
\begin{cases}
0, \quad\quad\,\, \,\,\mbox{for}\, \beta>1,\\
-\frac{1}{\ell_{\rm NP}^4}, \quad \mbox{for}\, \beta=1.
\end{cases}
\ee
Therefore, the minimum value of $\beta$ to provide a completely well-defined geometry, which features neither singular nor multi-valued curvature invariants, is $\beta>1$. In order to work with a simple expression featuring integer powers, we choose $\beta=2$ and
thus work with the following mass function for the remainder of this paper
\be
M(r, \chi) = \frac{M}{1+K_{\rm GR}\ell_{\rm NP}^4},\label{eq:massfunction}
\ee
resulting in $I_1$ as shown in Fig.~\ref{fig:I1diva}.
All other curvature invariants and their behavior at the center of the black hole are discussed in detail in App.~\ref{app:reginvs}, but do not lead to additional constraints.

Away from $(r \rightarrow 0, \chi\rightarrow 0)$, no singularities can exist in any of the curvature invariants as long as $M(K_\text{GR})$ itself is regular. This can be seen by inspecting the expression in Eq.~\eqref{eq:Weylsqgenbeta}, for which the denominator is positive at non-zero $r$ or non-zero $\chi$, see also App.~\ref{app:ZM-basis} and \ref{app:reginvs} for all other independent curvature invariants.

At this point, let us comment on the use of a Boyer-Lindquist-type ansatz in contrast to an ingoing-Kerr-type ansatz: The finiteness of the curvature invariants at the event horizon requires a delicate cancellation of divergent terms in Boyer-Lindquist coordinates for the Kerr black hole. Such a delicate cancellation can be disturbed if the ADM mass $M$ is generalized to a mass function $M(r,\chi)$. Thereby, such constructions can modify coordinate singularities into curvature singularities. In fact, this occurred in \cite{Held:2019xde}, as well as other works, see, e.g., \cite{Johannsen:2013rqa} for multiple examples and \cite{Held:2021} for an in-depth discussion. In contrast, in ingoing Kerr coordinates (or, for that matter, any other choice of horizon-penetrating coordinates), regularity of all curvature invariants at the horizon is automatically ensured, as soon as $M(r, \chi)$ and its first two derivatives with respect to $r$ are non-singular at the horizon. Thus, the metric in \eqref{eq:regular-spinning-BH-metric-ingoing-Kerr} can be generalized to other mass-functions with appropriate limits representing a wider family of regular black-hole spacetimes.

\subsubsection{Horizon of regular, spinning black-hole spacetimes}\label{sec:horizon}

In black hole spacetimes, there exists an event and a Killing horizon. These agree in Kerr spacetime but, as we will see below, not in our regular spinning spacetime.
Surfaces that respect the Killing symmetries (stationarity and axisymmetry) of the spacetime can be parameterized by a function $f(r, \theta)$, where $\theta = {\rm arccos}(\chi)$. 
The event horizon can be described by the condition $f(r, \theta)=0$. 
As the horizon is a null surface, its normal $n^{\mu}= \partial^{\mu}f(r, \theta)$ is null
\bea
0&=&g^{\mu\nu} \left(\partial_{\mu}f(r, \theta)\right)  \left(\partial_{\nu}f(r, \theta)\right)\nonumber\\
&=& g^{rr} \left( \partial_r f(r, \theta)\right)^2 + 2 g^{r \theta} \left(\partial_r f(r, \theta) \right)   
\left(\partial_{\theta} f(r, \theta) \right) + g^{\theta \theta} \left(\partial_{\theta}f(r, \theta) \right)^2.\label{eq:horizoncondition}
\eea
If the location of the event horizon is independent of $\theta$, Eq.~\eqref{eq:horizoncondition} reduces to the condition $g^{rr}=0$, which is the correct condition for the Kerr case. For deviations from Kerr, corrections $\mathcal{O}(\ell_{\rm NP})$ arise and Eq.~\eqref{eq:horizoncondition} must be solved. If we assume that horizons do not cross, then the parameterization of the corresponding null surface can be written in the form
\be
f(r, \theta) = r - H(\theta),
\ee
which provides the location of the horizon, $r_H = H(\theta)$. Since $g^{r \theta}=0$ for our spacetime, the ordinary differential equation that is to be solved reads
\be
g^{rr} (r = H(\theta)) + g^{\theta \theta} (r= H(\theta)) \left (\frac{dH}{d\theta} \right)^2=0, \label{eq:horizonconditionH}
\ee
where we have indicated that the metric components, which depend on $r$, should be evaluated on $r = H(\theta)$.

The solution of this differential equation requires an initial condition. 
Following \cite{Johannsen:2013rqa}, we use that axisymmetry and reflection symmetry about the equatorial plane\footnote{Reflection symmetry about the equatorial plane holds since the family of spacetimes in Eq.~\eqref{eq:regular-spinning-BH-metric-ingoing-Kerr} only depends on $a^2$.} imply that $dH/d\theta=0$ at $\theta=\pi/2$. Thus, the horizon-condition reduces to $g^{rr}=0$ at $\theta=\pi/2$, i.e.,
\be
0=r_H^2+a^2 - 2 M(r, \theta) r_H = r_H^2 +a^2 - 2 M r_H \frac{1}{1+\ell_{\rm NP}^4 \frac{48 M^2}{(r_H^2+a^2 \cos(\theta)^2)^3}}.
\ee
This condition cannot be solved analytically, but can easily be solved numerically to provide the initial conditions for the numerical solution of Eq.~\eqref{eq:horizonconditionH}.
For the solution, it holds that
\be
H(0) = H(\theta),
\ee
due to the symmetries of the spacetime. Before discussing the resulting shape of the event horizon, we investigate the Killing horizon.
\\

A Killing horizon is defined as a null hypersurface $\Sigma$, to which a Killing vector field $\xi$ is orthogonal, i.e.,
\be
g_{\mu\nu}\xi^{\mu}\xi^{\nu}=0.\label{eq:Killingeqgeneral}
\ee
In the Kerr case, for which the event horizon is a Killing horizon, the Killing vector can be written as $\xi^{\mu} = u^{\mu} + \Omega \phi^{\mu}$, where $\Omega$ is an angular velocity. Outside of the Killing horizon, $\xi^{\mu}$ is the four-velocity of a stationary observer, i.e., an observer moving with uniform angular velocity, see, e.g., \cite{Poisson:2009pwt}. In order to be a viable four-velocity, $\xi^{\mu}$ must be timelike, which limits $\Omega$ to be within the range $\Omega_1 < \Omega < \Omega_2$, where 
\be
\Omega_{1,2} = - \frac{g_{u\phi}}{g_{\phi\phi}} \pm \sqrt{\left(\frac{g_{u\phi}}{g_{\phi\phi}} \right)^2 - \frac{g_{uu}}{g_{\phi\phi}}},
\ee
are the solutions to
\be
0 = g_{uu} + 2 \Omega g_{u\phi} + \Omega^2 g_{\phi\phi}.\label{eq:Killingeq}
\ee
At the Killing horizon, the two solutions $\Omega_{1,2}$ are equal. Thus, the location of the Killing horizon is determined by
\be
g_{u\phi}^2 - g_{uu}g_{\phi\phi}=0.\label{eq:Killinghorizon}
\ee
For the case of GR, where the Hawking rigidity theorem \cite{Hawking:1973uf} holds, event and Killing horizon coincide. Thus, in GR a stationary observer must be in a state of corotation with the event horizon of a Kerr black hole, as $\Omega_1=\Omega_2$ is the angular velocity of the event horizon and at the same time the only possible angular velocity for a stationary observer.

\begin{figure}[!t]
\centering
\includegraphics[width=0.48\linewidth]{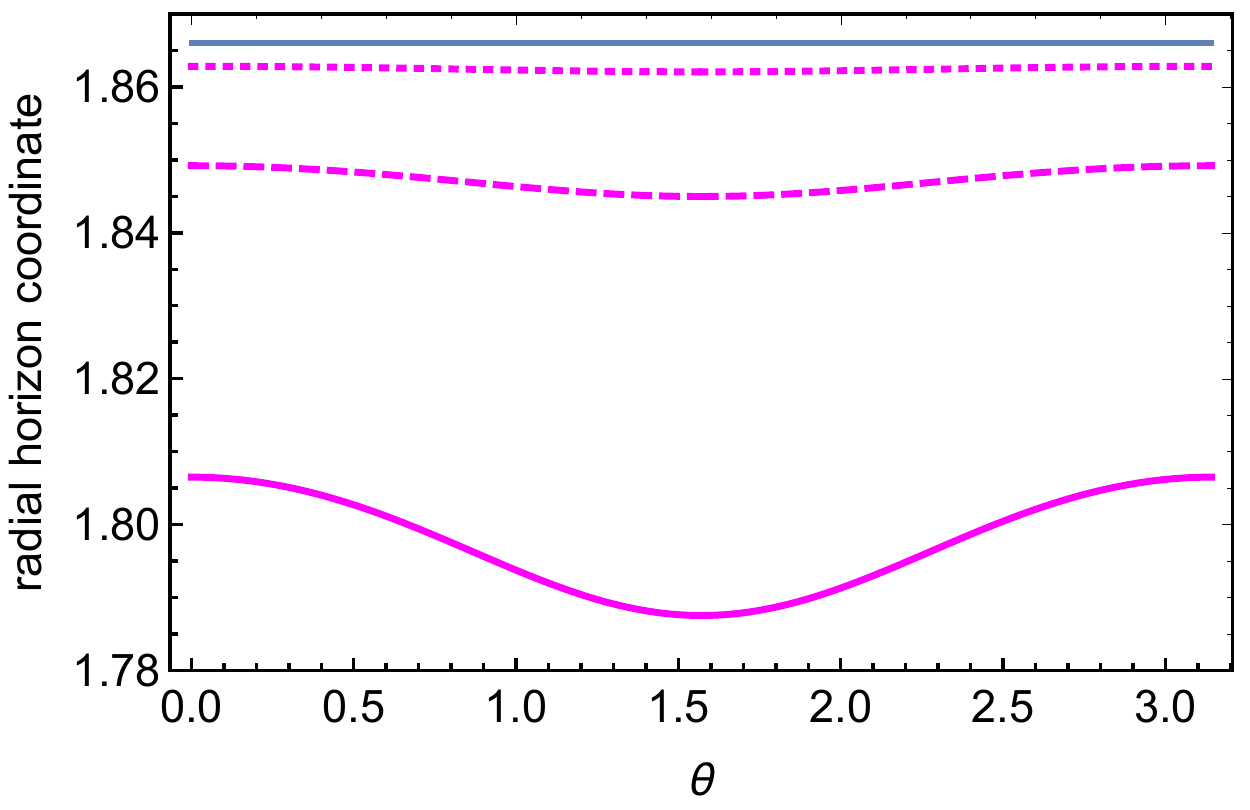}\hfill\includegraphics[width=0.48\linewidth]{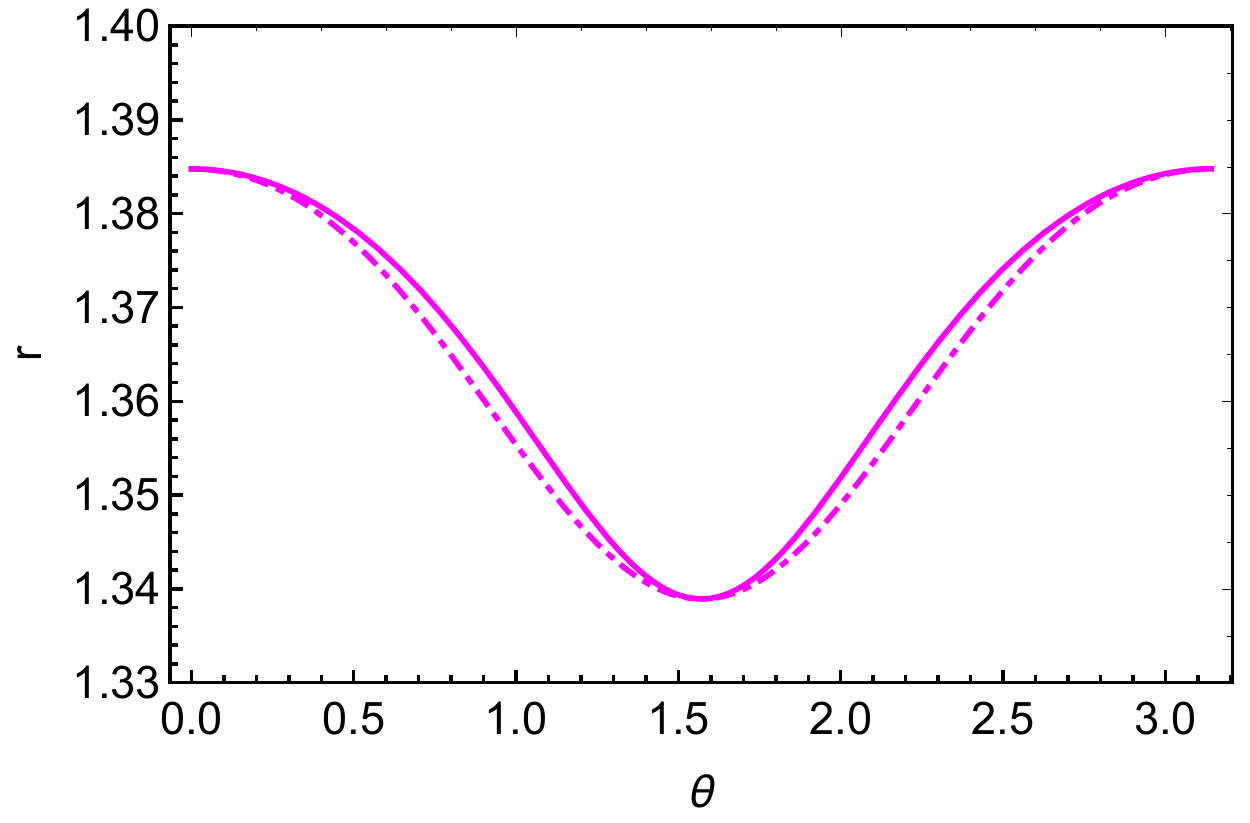}
\caption{\label{fig:horizon} Left panel: We show the radial coordinate of the event horizon as a function of the angular coordinate $\theta$ for $M=1, a=1/2$ and $\ell_{\rm NP}=0$ (Kerr limit, continuous blue line),  $\ell_{\rm NP}= 4\cdot10^{-1}$ (continuous magenta line),  $\ell_{\rm NP}=\cdot 3\cdot 10^{-1}$ (dashed magenta line) and $\ell_{\rm NP}=2\cdot 10^{-1}$ (dotted magenta line). Right panel: We show the Killing (continuous) and event horizon (dot-dashed) for $M=1$, $a=9/10$ and $\ell_{\rm NP}= 10^{-1}$.}
\end{figure}

Going beyond vacuum solutions of GR, the Hawking rigidity theorem has been proven to hold within particular settings, e.g., \cite{Friedrich:1998wq,Alexakis:2009gi}, but it need not hold in general.  
In our case, the event and the Killing horizon only coincide in the equatorial plane and at the poles, with small deviations at other $\theta$.
In fact, a gap between the event and the Killing horizon opens up for $\ell_\text{NP}\neq 0$, cf.~Fig.~\ref{fig:horizon} which also highlights that at the quantitative level, the difference between those two is actually quite small.
\\

The impact of the new physics parameterized by $\ell_{\rm NP}>0$ results in characteristic features shared by the event and Killing horizon:
\\
First, the outer event horizon, if it exists\footnote{For sufficiently large $\ell_{\rm NP}$, no outer event horizon exists. In this paper, we do not explore the causal structure and features of the resulting spacetime.}, lies at a lower radial coordinate than for a Kerr black hole of the same spin and ADM mass. In other words, the black hole is more compact than its GR-counterpart.
This follows, as the new physics weakens the attractive gravitational force, i.e., as $M(r, \chi)\leqslant M$ for all $r<\infty$ and $\chi$. Moreover, the deviation grows with absolute values of the curvature invariants in the Kerr spacetime. Therefore, the horizon, i.e., the ``point of no return'' for infalling observers, must be located at lower values of $r(\chi)$ in comparison to a Kerr black hole. The same intuition was discussed in a quantum gravitational context in \cite{Held:2019xde}.
\\
Second, the location of the event horizon depends on $\theta$, i.e., it exhibits only axisymmetry, not spherical symmetry. Instead, it features a dent, i.e., $r_H(\theta=\pi/2)$ corresponds to a global minimum of the function $r_H(\theta)$. 
This is again connected to the physical intuition underlying the construction of our model:
The difference between $M$ and $M(r = \rm const, \chi)$, i.e., the weakening of gravity, is largest in the equatorial plane, where the curvature invariants in the Kerr spacetime are largest. Therefore, the increase in compactness of the event horizon is $\theta$ dependent, resulting in the characteristic dent.
 
The dent increases as a function of $\ell_{\rm NP}$, up to a critical value $\ell_\text{NP,\,crit}(a)$, with $\ell_{\rm NP, \, crit}(a=0.9) \approx 0.2556$, at which the outer event horizon is dissolved. At $a =0$, this occurs due to a collision between the outer and a new-physics induced inner event horizon.
In the present work, we do not further analyze the fate of the inner horizon as a function of $\ell_{\rm NP}$ at finite $a$, but only make a short comment here. 
We find that the inner solution to Eq.~\eqref{eq:horizonconditionH}, which is the inner horizon of the Kerr spacetime for $a \rightarrow 0$, develops a discontinuous derivative at both poles at finite $\ell_{\rm NP}$. 
A similar behavior can be observed for the outer horizon for other choices of mass functions, for instance, for $\beta=1$.
Let us remark in passing that the inner horizon is generically expected to be unstable to perturbations in GR \cite{Poisson:1989zz,Ori:1991zz}. Thus, even the causal structure inside a perturbed Kerr black hole is not yet properly understood. Furthermore, if an inner event horizon exists for our model, it might also be unstable to perturbations. In fact, within proposals for regular non-spinning black-hole spacetimes, e.g.,
\cite{Bardeen,Dymnikova:1992ux,Dymnikova:1996ux,AyonBeato:1998ub,Hayward:2005gi,Platania:2019kyx,Simpson:2019mud}, a potential instability of the inner horizon has been discussed in \cite{Carballo-Rubio:2018pmi,Carballo-Rubio:2019nel,Bonanno:2020fgp,Rubio:2021obb}.
The presence of such an instability depends on the beyond-GR dynamics, the exploration of which is beyond the scope of this paper. 

\subsubsection{Ergosphere of singularity-free black-hole spacetimes}\label{sec:ergosphere}
The ergosphere is the region outside the outer event horizon, in which frame-dragging occurs, i.e., timelike geodesics must show a change in $\phi$. This gives rise to the possibility of a Penrose \cite{Penrose:1971uk} and Blandford-Znajek \cite{Blandford:1977ds,McKinney:2005zw} process, in which energy is extracted from a black hole's rotation. It has been proposed \cite{2005ApJ...630L...5M} that this process can power the jets of supermassive black holes.
Therefore, the presence of an ergosphere may be an important phenomenological constraint on regular, spinning black-hole spacetimes. This motivates us to explore, whether an ergoregion exists for our black hole.

\begin{figure}[!t]
\centering
\includegraphics[width=0.6\linewidth]{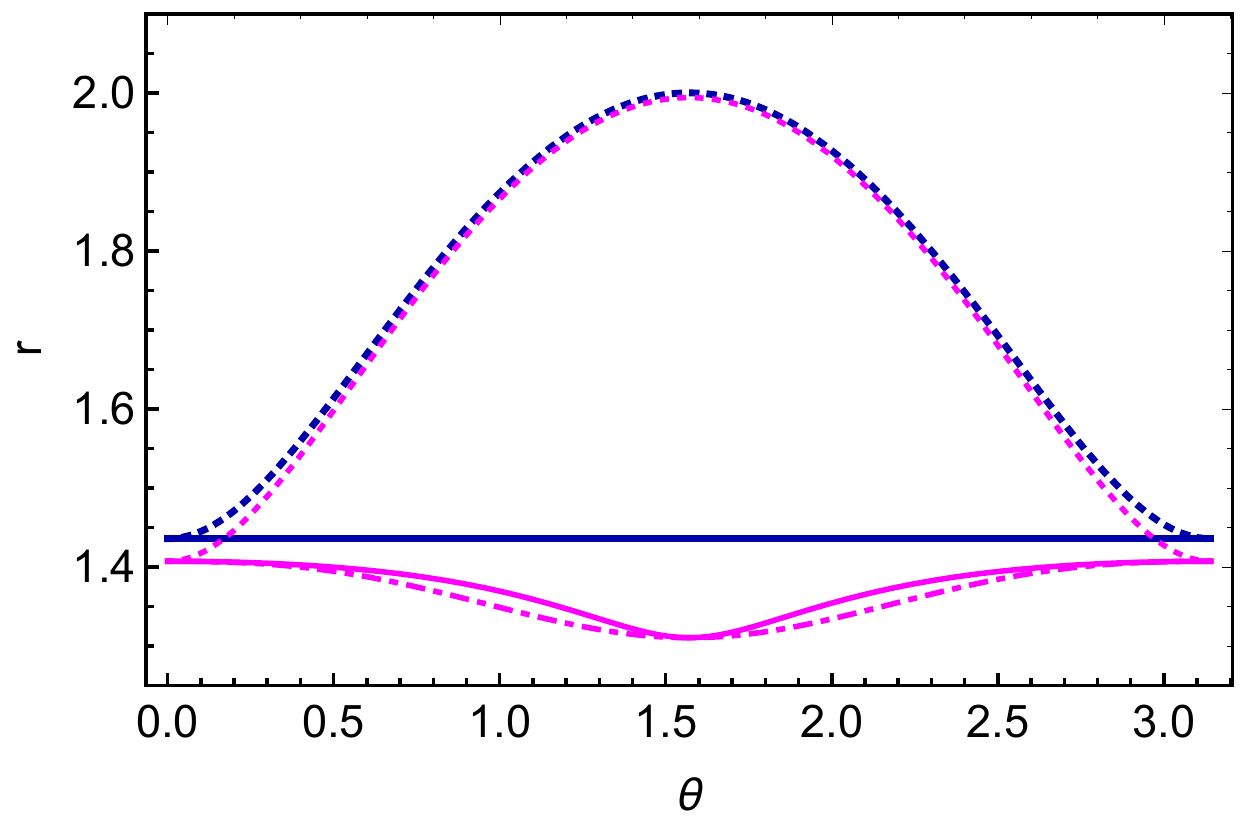}
\caption{\label{fig:ergohor} The radial coordinate of the ergoregion (dotted lines) $\tilde{r}_+$ and the event horizon (dot-dashed lines) and Killing horizon (continuous lines), as a function of $\theta$, for $M=1, a=9/10$ and $\ell_{\rm NP}=0$ (dark blue lines; Kerr case) and $\ell_{\rm NP}=2.5\cdot 10^{-2}$ (magenta lines).}
\end{figure}

In the ergoregion, orbits of the Killing vector $\partial_u$, associated to stationarity, are no longer timelike. In that region, a testparticle cannot remain stationary. Instead, a curve parameterized by the proper time $\tau$ can only be timelike if $d\phi/d\tau \neq 0$, since this enables the tangent vector $v^{\mu} = dx^{\mu}/d\tau$ to be timelike at fixed $r$. Thus, frame-dragging sets in at $\tilde{r}_+$, where $g_{uu}$ changes sign and forces the norm of the Killing vector field $\partial_u$ to vanish.
This condition cannot be solved analytically for our spacetime, but the corresponding location $\tilde{r}_+$ can be determined numerically and we find that $\tilde{r}_+ > r_H$. In Fig.~\ref{fig:ergohor}, we show the radial coordinate of the horizon as well as the ergoregion. We conclude that the candidate jet-launching mechanism that is likely associated with the jet of M87* \cite{2012Sci...338..355D,paper5} as well as further so-called FR1-sources (e.g., blazars) appears to also be available in the regular spinning black-hole spacetime we explore here. To confirm this idea, general relativistic
magnetohydrodynamic (GRMHD) simulations of an accreting disk and the associated magnetic field in the background of the new-physics spacetime are necessary. These can provide evidence for or against a jet-launching mechanism under the assumption that unlike the background spacetime, the disk physics is not impacted by $\ell_{\rm NP}$.

\subsubsection{Weak-gravity limit and Newtonian potential}
\label{sec:Newtonian-limit}
In order for the above metric to play a relevant role in phenomenology, its weak-field limit has to be in agreement with the Newtonian potential, see also \cite{Faraoni:2020stz, Faraoni:2020mdf}.
For spinning black holes, the Newtonian limit is typically extracted from $g_{tt}$ at large $r$ in Boyer-Lindquist coordinates. In the case of the mass function Eq.~\eqref{eq:massfunction} that depends on both $r$ and $\chi$, we do not find an appropriate coordinate transformation, see App.~\ref{app:BL-form}. Thus, we first consider the limit $r \rightarrow \infty$ and perform a Taylor expansion of the metric around this value. To $\mathcal{O}(r^{-6})$, the result agrees exactly with the classical Kerr metric in ingoing Kerr coordinates. Within the patch in which $\mathcal{O}(r^{-7})$ terms are negligible, a coordinate transformation to Boyer-Lindquist coordinates is therefore possible. Then, the Newtonian potential can be recovered from the $g_{tt}$ component of the metric in the standard way and agrees with the expected form. 

\subsection{Summary of the regular, spinning black-hole geometry with locality principle}

In summary, the regular black-hole spacetime is characterized by the following properties:
\begin{enumerate}
\item[i)] All special surfaces of the regular black hole lie at smaller values of $r$ than their corresponding classical counterparts. This increase in compactness of the object is a consequence of its effective weakened gravity.
\item[ii)] The black hole features an ergoregion outside the event horizon, which can give rise to the Blandford-Znajek mechanism and could power the observed jets of supermassive black holes, if these were indeed given by a regular black hole as we describe here.
\item[iii)] The black hole is characterized by an event horizon which no longer coincides with the Killing horizon, implying that no analogue of Hawking's rigidity theorem holds in a gravitational theory which features these regular black holes as solutions. Instead, at values of $\chi$ where the two locations do not coincide (i.e., away from the poles and the equator), the Killing horizon lies further outwards, cf.~Fig.~\ref{fig:horizon}.
\item[iv)] The location of the event horizon depends on $\chi$. It  features a ``dent'' in the equatorial plane, i.e., we are referring to a minimum of $r_H(\chi)$ at  $\chi=0$, cf.~Fig.~\ref{fig:horizon}.
Below, we contrast this with a large class of regular black holes in the literature and explain the physical principle at the heart of the dent.
\end{enumerate}

 \begin{figure}[!t]
\centering
\includegraphics[width=0.6\linewidth]{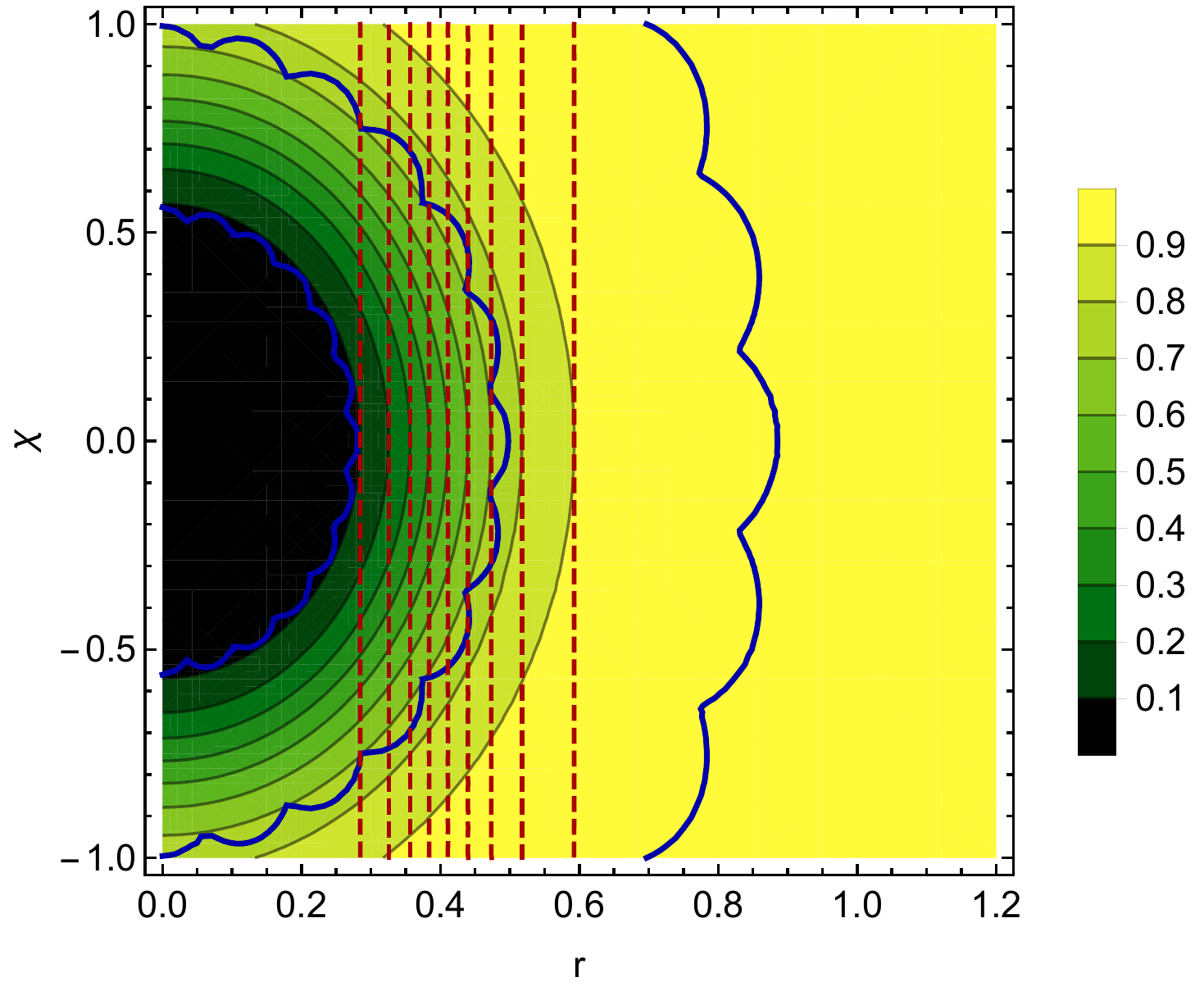}
\caption{\label{fig:massfunction} We show the massfunction Eq.~\eqref{eq:massfunction} as a function of $r$ and $\chi$ for $M=1, a=1/2, \ell_{\rm NP}=10^{-1}$. In addition, we show two contours for the maximum of the local curvature invariants, which constitute the ``wavy" dark blue continuous lines; in order to indicate that our mass-function follows the local curvature scale. This is contrasted with the situation in which a $\chi$-independent mass function is chosen, which can only coincide with the local curvature scale at a single value of the angle for a given radius. The contour lines of such a $\chi$-independent mass function are indicated in the dark red dashed lines.}
\end{figure}
The dent arises as a consequence of our locality principle, namely that the  local curvature scale determines when modifications from GR set in. This necessarily implies that, at any given radius, the modifications are largest in the equatorial plane. Thus, our mass function must be a function of $r$ and $\chi$, cf.~Fig.~\ref{fig:massfunction}.
Accordingly, the event horizon cannot be spherically symmetric and there is a minimum in $r_H(\chi)$ at $\chi=0$.
\\
This distinguishes our black-hole spacetime with a locality-based mass function $M(r, \chi)$ from those with $M =M(r)$. The latter can be motivated in various ways, e.g., by applying the Newman-Janis algorithm to a non-spinning regular black-hole-spacetime~\cite{Bambi:2013ufa,Azreg-Ainou:2014pra,Toshmatov:2014nya,Ghosh:2014hea,Abdujabbarov:2016hnw,Torres:2017gix,Lamy:2018zvj,Kumar:2019ohr,Shaikh:2019fpu,Contreras:2019cmf,Liu:2020ola,Junior:2021atr,Mazza:2021rgq}, or by specialized choices of Renormalization Group improvement~\cite{Reuter:2010xb,Litim:2013gga,Pawlowski:2018swz} of the Kerr metric.
Shadow images can be found, e.g., in ~\cite{Amir:2016cen,Abdujabbarov:2016hnw,Tsukamoto:2017fxq,Shaikh:2019fpu,Dymnikova:2019vuz}. A $\chi$-independent mass function cannot arise when a local curvature scale is used to determine the onset of new physics. Instead, spacetimes with $M=M(r)$ can only be obtained when a non-local notion of curvature, e.g., the curvature scale at a fixed angle $\chi$, or averaged over all angles, is used. In turn, the resulting $\chi$-independence implies a spherical horizon. Therefore, we conjecture that the existence of a dent is a  (not necessarily unambiguous) imprint of the locality principle. The fact that such a locality principle is a well-motivated physical assumption about the nature of new physics makes it of particular interest to explore whether a dent in the horizon results in visible features of near-horizon probes of the black-hole geometry.

\section{Image features of regular black-hole shadows}
\label{sec:images}
In order to pave the groud for making contact with EHT-observations, we proceed in three steps. First, we calculate the shape of the regular black-hole shadow and compare with that of a Kerr black hole, in order to identify its distinct features, cf.~Sec.~\ref{sec:shapes}. Second, we determine photon rings which probe an extended region of the black-hole spacetime -- and therefore the mass function $M(r,\chi)$ -- in the vicinity of the photon sphere, cf.~Sec.~\ref{sec:rings}.
Third, we take a first step towards more realistic images and account for emission from an analytical disk model to investigate the theoretical visibility of the previously identified image features and to produce intensity maps of the image, cf.~Sec.~\ref{sec:intensity}.

\subsection{Shape of the shadow}
\label{sec:shapes}
\begin{figure}[!t]
\centering
\includegraphics[width=0.5\linewidth]{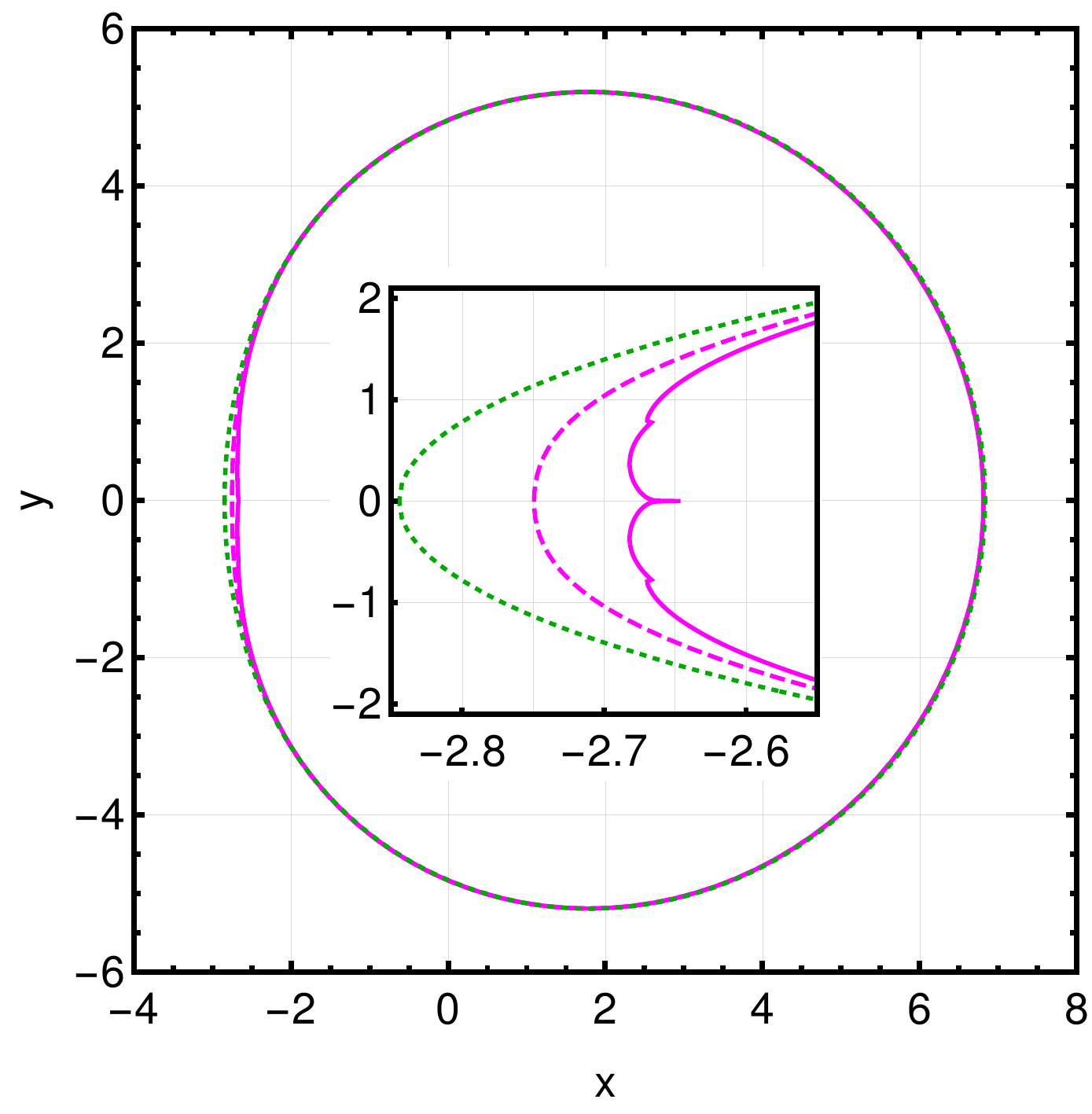}
\caption{\label{fig:thedent}
We show the deformation of the shadow boundary due to growing $\ell_\text{NP} = (0,\,0.2,\,0.2556 \approx \ell_\text{NP,\,crit})$ (green dotted, magenta dashed, magenta continuous) at $a=0.9$ and $\theta_\text{obs}=\pi/2$.
}
\end{figure}
A black hole's shadow, as seen by a distant observer, is an image of the black hole's photon sphere, i.e., the last marginally stable photon orbit \cite{Bardeen1973,Luminet1979}. The shadow is derived by solving the null geodesic equation
\begin{align}
	\label{eq:geodesicEq_null}
	\frac{d^2x^\mu}{d\lambda^2} = -\Gamma^\mu_{\alpha\beta}\frac{dx^\alpha}{d\lambda}\frac{dx^\beta}{d\lambda}\;,
\end{align}
for light rays which end on the observer's screen. Here, $x^\mu(\lambda)$ are the coordinates of the photon's world line as a function of the affine parameter $\lambda$. Since we do not find a Killing tensor and an associated `hidden' constant of motion (the Carter constant) in our spacetime,  cf.~App.~\ref{app:BL-form}, the geodesic equation cannot be solved analytically. Instead, we use numerical ray-tracing techniques detailed in App.~\ref{app:ray-tracing}. Our results are produced with a Mathematica-based \cite{Mathematica} numerical ray tracer.

The observer places her screen at a large distance to the black hole with the screen coordinates $(x,y)$ set up with their origin at $(r_\text{obs},\theta_\text{obs},\phi_\text{obs})$ in Kerr-coordinates.
Geodesics start perpendicularly to the screen and are integrated backwards from the observer towards the black hole, making use of the time-reversal symmetry of the system. 
Whenever the light ray falls into the photon sphere and thus inevitably into the event horizon, the image point lies within the imaged black-hole shadow. Whenever the light ray escapes to large radii, the image point lies outside of the imaged black-hole shadow. The shape of the shadow can thus be calculated by evaluating the above condition in nested intervals in the image plane.
\\

Due to its simplicity, the spherically symmetric spacetime with $a \rightarrow 0$ is a useful starting point for the analysis of the shadow. In astrophysics, this constitutes an idealized situation, as black holes are expected to have a non-vanishing spin \cite{Remillard2006,2003PhR...377..389R,2014SSRv..183..277R,Abbott:2016nmj,Abbott:2020gyp}. At $a>0$, a second parameter, namely the inclination, i.e., the angle between the spin vector and the viewing axis\footnote{We define the inclination $\theta_{\rm obs}$ to be zero when spin vector and viewing axis from the black hole to the observer are aligned. Note that conventions for the inclination vary in the literature.}, impacts the shape of the shadow. In the following, we include these phenomenologically relevant parameters successively.

For spherically symmetric spacetime, the shape of the shadow must be spherical. Its size is  smaller compared to a Schwarzschild black hole with the same ADM mass, see also \cite{Held:2019xde, EHJ2021}.
 This increased compactness of the shadow reflects the more compact horizon, cf.~Sec.~\ref{sec:horizon}. Both arise due to an effective weakening of gravity by the singularity-resolving new-physics effect encoded in $M(r, \chi)<M$.
\\
On its own, this increase in compactness is not detectable, as the shadow boundary can be matched exactly by a Schwarzschild black hole with $M'<M$. In \cite{Held:2019xde}, it was therefore proposed to combine a mass-measurement extracted from the black-hole shadow with a mass-measurement extracted from (post)-Newtonian orbits, such as in \cite{Ghez:2008ms}. In essence, these measurements access $M(r, \chi)$ at different $r$. Their combination\footnote{For an alternative suggestion how $M(r)$ can be constrained from X-ray binaries, see \cite{Zhou:2020eth}.} is thus sensitive to the fact that $M(r, \chi) \neq \rm const$. 
As we will now demonstrate with a specific example, the degeneracy of the shape of the shadow between our family of regular black-hole spacetimes and GR black holes is lifted in the spinning case, allowing to constrain $M(r, \chi)$ directly from the shadow.

\subsubsection{Effects of spin}\label{sec:effectsofspin}
We include a non-zero spin parameter, first fixing the inclination to $\theta_{\rm obs} =\pi/2$, such that the equatorial plane is orthogonal to the screen. Just like for a Kerr black hole, the shadow boundary is flattened on the prograde side, i.e., on the side where the black hole rotates towards the observer. There, backwards-traced null geodesics are pulled closer to the black hole due to framedragging, generating an asymmetry of the shadow boundary. Due to this asymmetry, the size of the new-physics effect is not constant along the shadow boundary: The increase in compactness from $\ell_{\rm NP}=0$ to $\ell_{\rm NP} >0$ is appreciably larger on the prograde than on the retrograde side, cf.~Fig.~\ref{fig:thedent}. This effect is strongest within the equatorial plane of the black hole, i.e., on the $y=0$ axis in the image. Therefore, a dent-like feature, corresponding to that in the event horizon (cf.~Fig.~\ref{fig:horizon}), can also appear in the black-hole shadow, cf.~Fig.~\ref{fig:thedent}. More specifically, at finite $a$ and for large enough $\ell_{\rm NP}$, $x(y)$ on the prograde side has a local minimum in the vicinity of $y=0$. This distinct feature of our model distinguishes its shadow boundary from that of a Kerr black hole.

\begin{figure}[!t]
\centering
\includegraphics[width=0.32\linewidth]{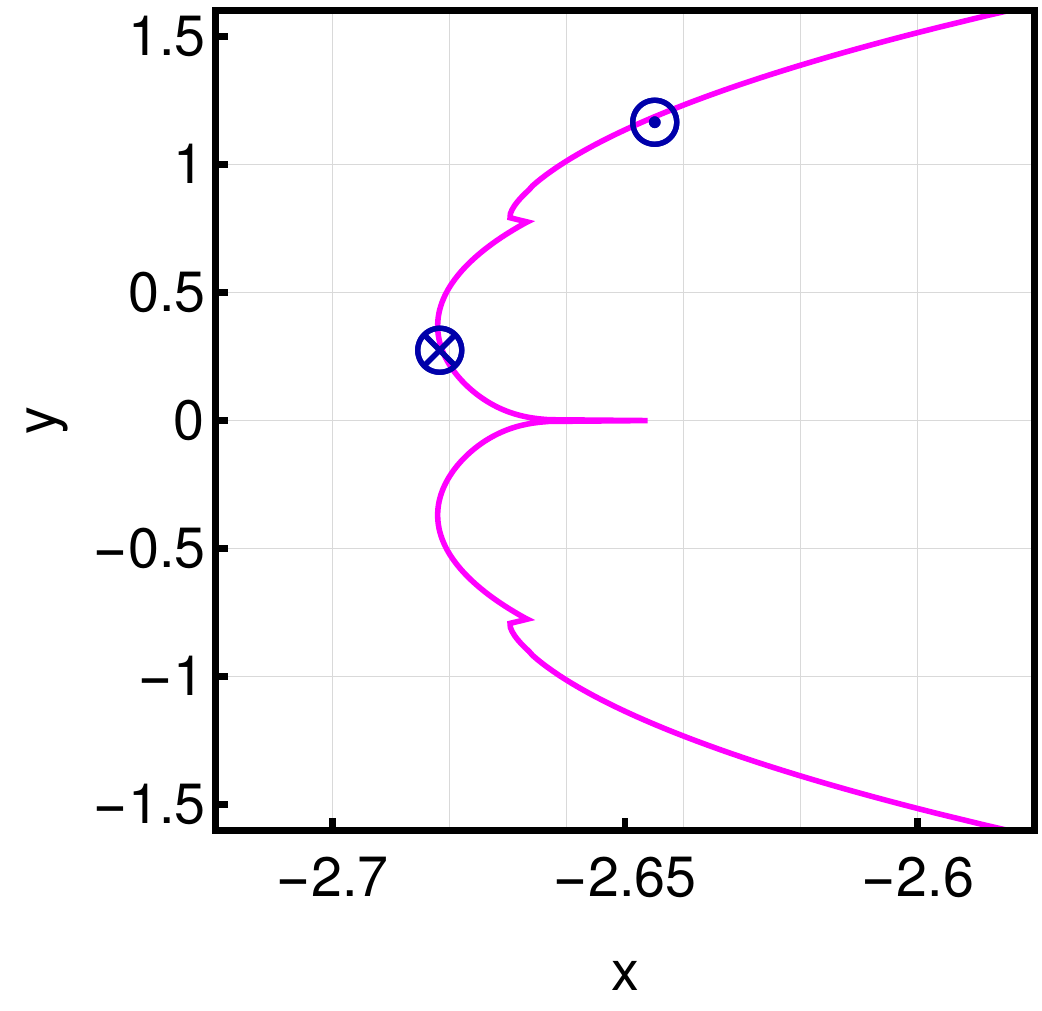}
\hfill
\includegraphics[width=0.33\linewidth]{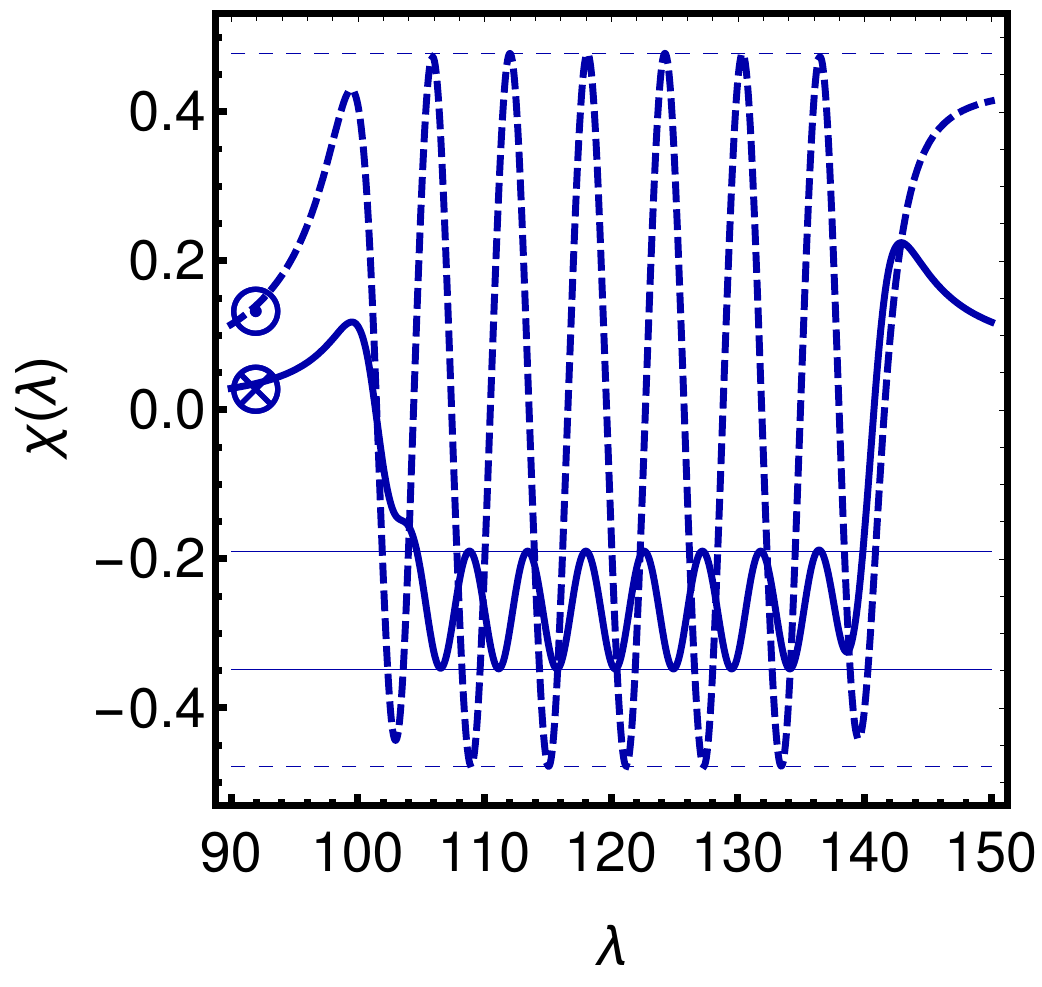}
\hfill
\includegraphics[width=0.33\linewidth]{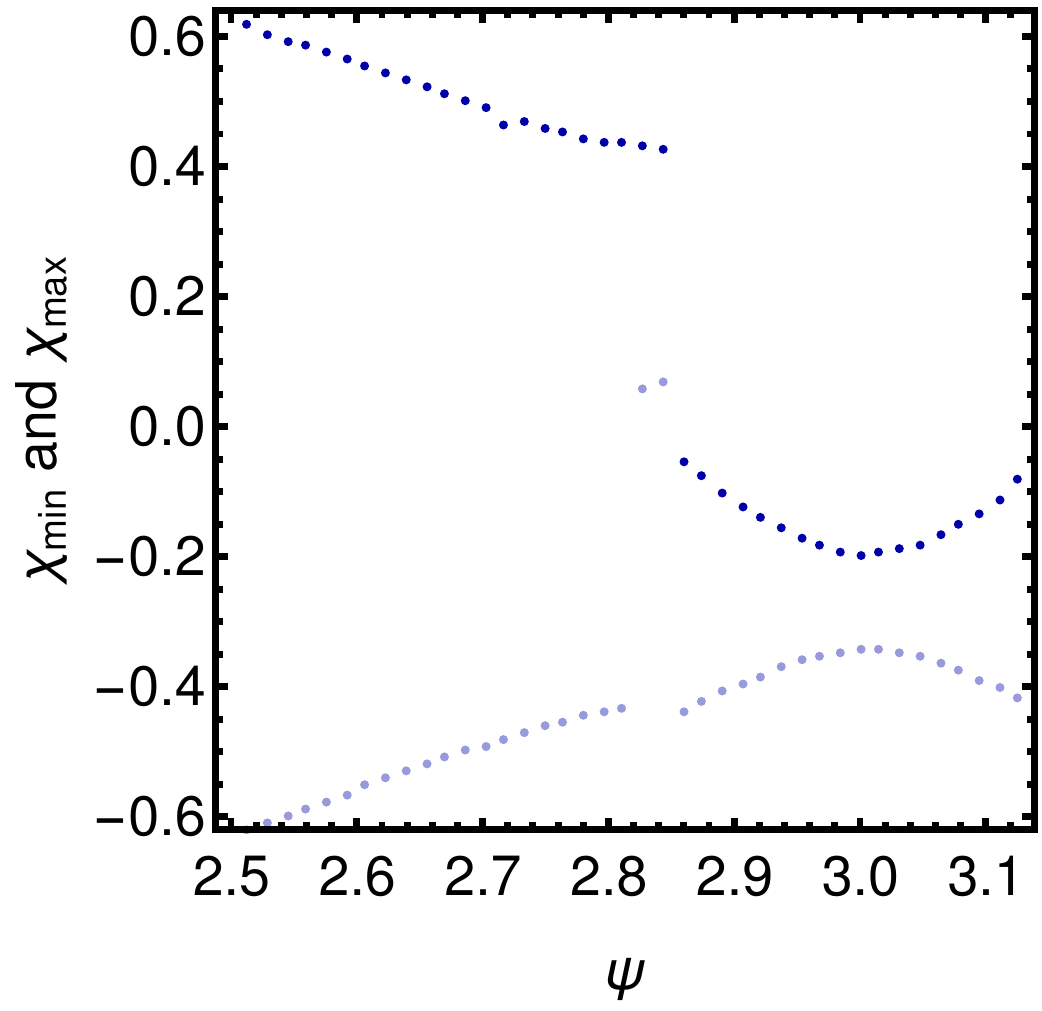}
\caption{\label{fig:cuspy-triptych}Left-hand panel: Close-up of the prograde side of the shadow boundary, with two exemplary points for distinct shadow-boundary sections below (circle-times) and above (circle-dot) a cusp. Central panel: Dependence of the polar coordinate $\chi(\lambda)$ of two null geodesics as a function of the affine parameter $\lambda$ that arrive at the two exemplary screen points identified in the left-hand panel.
Their respective $\chi_\text{min}$ and $\chi_\text{max}$ is indicated.
Right-hand panel: $\chi_\text{min}$ (lower light-shaded points) and $\chi_\text{max}$ (upper dark-shaded points) as a function of the image angle $\psi$, i.e., following the shadow boundary in the left-hand panel from the uppermost point to when it crosses the y-axis, see App.~\ref{app:chiMinMax} and for a detailed algorithm identifying $\chi_\text{min/max}$. All three panels are generated at $a=0.9$ and $\theta_\text{obs}=\pi/2$.
}
\end{figure}
Moreover, cusp-like features appear in the shadow boundary, cf.~inset in~Fig.~\ref{fig:thedent}. They reflect discontinuous changes that occur between geodesics arriving at neighboring screen points. These discontinuities arise, as geodesics are typically bounded in $\chi$ while orbiting the black hole. 
Therefore, the photon sphere covers different sections $\chi\in[\chi_\text{min},\chi_\text{max}]$.
In particular, in our model, the equatorial plane can serve as a boundary for selected orbits to the north and south of it, see also \cite{Held:2019xde}. Geodesics which probe only one of the two hemispheres probe a near-horizon geometry that has a smaller effective horizon-radius than the corresponding Kerr counterpart, ultimately as a direct consequence of the dented event horizon. Therefore, geodesics that are bound to one hemisphere arrive at an image location further inwards (i.e., at smaller Euclidean distance to $(x=0, y=0)$), than geodesics that probe both hemispheres, cf.~Fig.~\ref{fig:cuspy-triptych}. 
Cusps in the shadow boundary are the result of a discontinuous changes in $[\chi_\text{min},\chi_\text{max}]$. In effect, the shadow boundary is a combination of several shadow boundaries, see also \cite{Cunha:2017eoe,Wang:2017hjl,Qian:2021qow}.
For Kerr spacetime, geodesics are also bounded in $\chi$, \cite{Vazquez:2003zm,Gralla:2019ceu,Himwich:2020msm}, but the horizon is spherically symmetric, resulting in a continuous shadow boundary.

\begin{figure}[!t]
\centering
\includegraphics[width=0.49\linewidth]{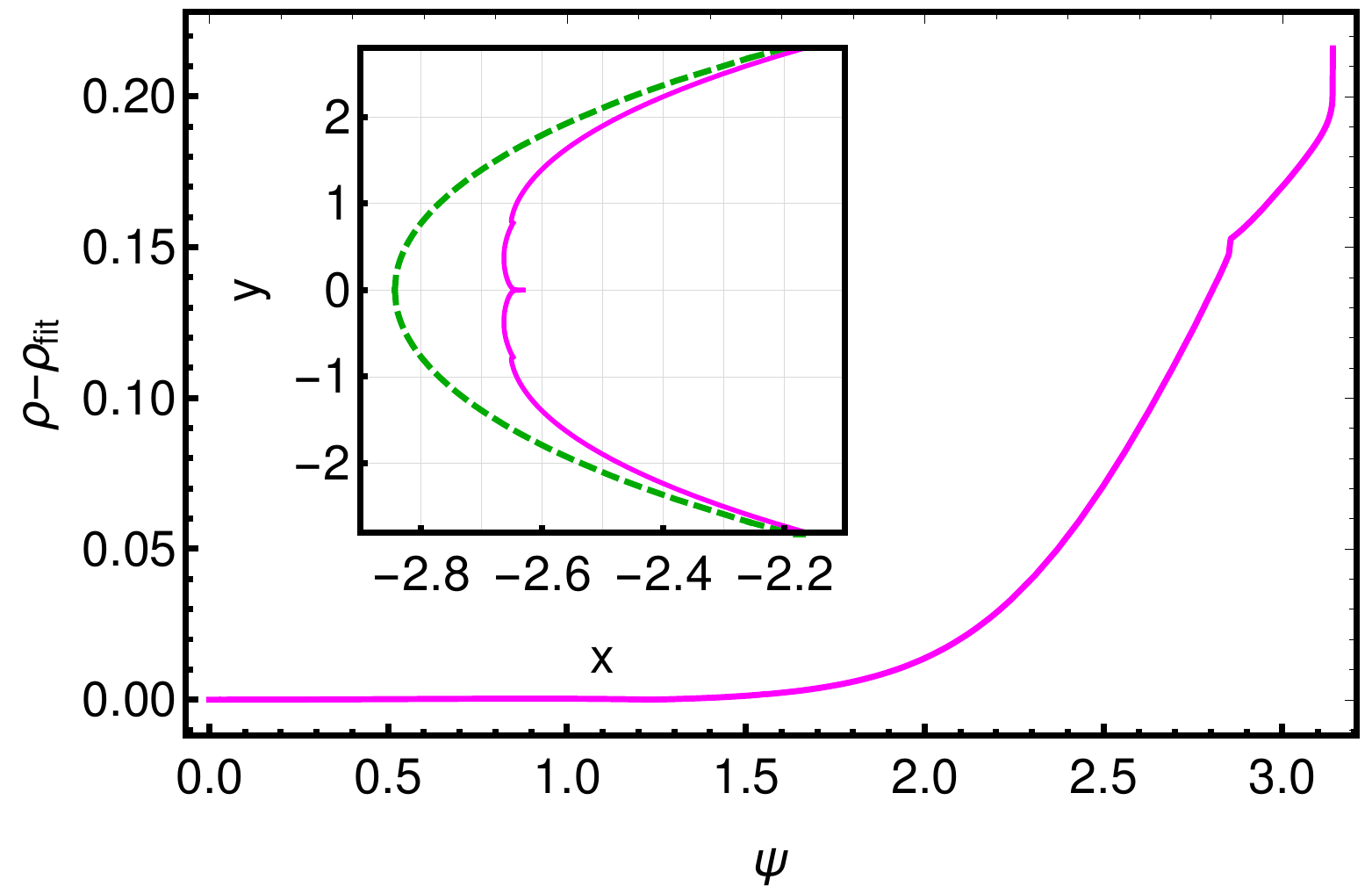}
\hfill
\includegraphics[width=0.49\linewidth]{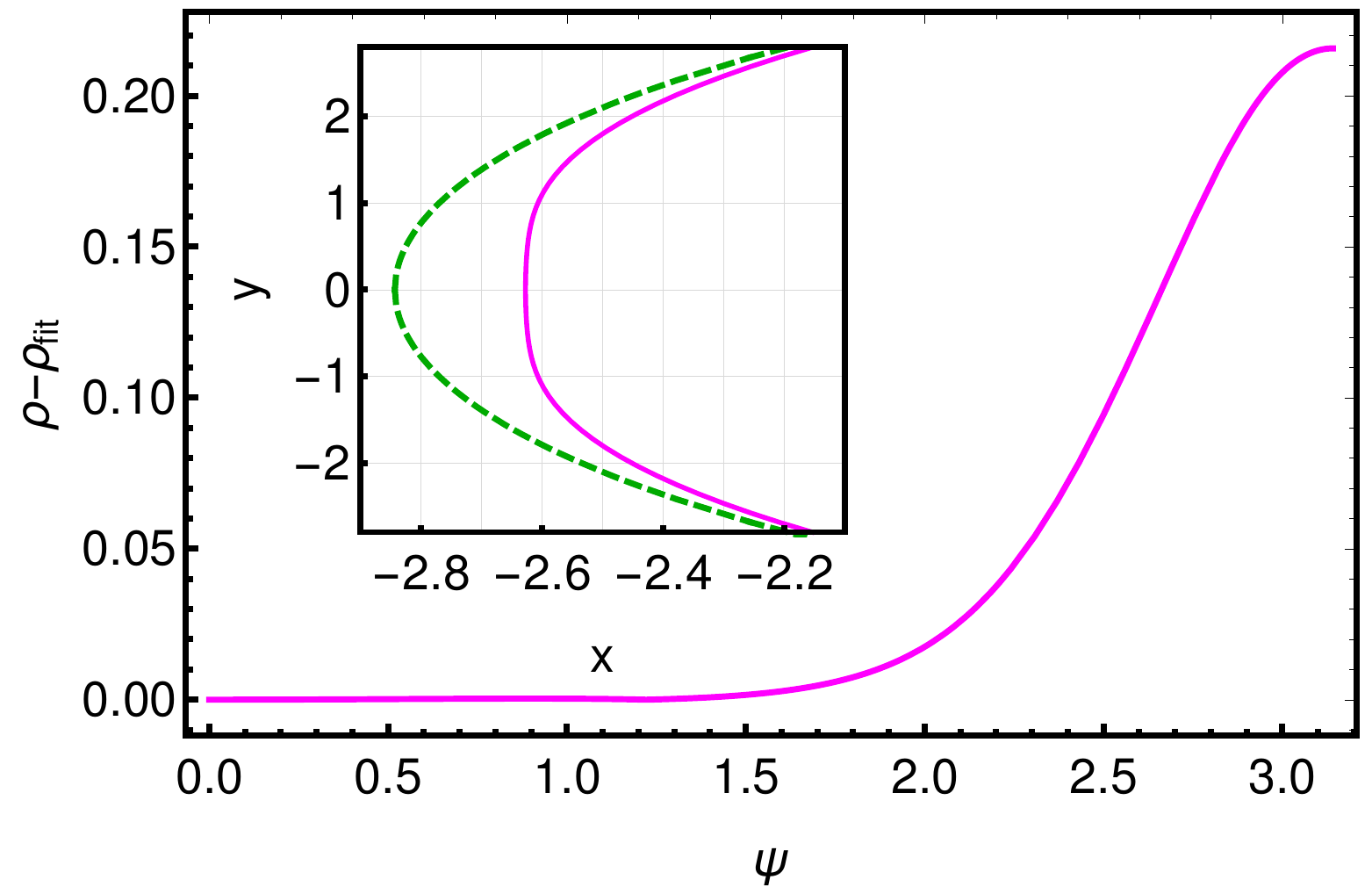}
\caption{\label{fig:NJ-comparison}
	Presence and absence of dents and cusps in the shadow boundary for new physics satisfying the locality principle with mass function $M(r,\chi)$, cf.~Eq.~\eqref{eq:massfunctionrep}, (left-hand panel) and not satisfying the locality principle with mass function $M(r) = M(r,0)$ (right-hand panel), respectively.
	We show the radial deviation between the regular shadow boundary $\rho(\psi)$ and a fitted classical boundary $\rho_\text{fit}(\psi)$ as a function of image angle $\psi$ for spin $a=0.9$, near-critical $\ell_\text{NP}=0.2556$, and inclination $\theta_\text{obs}=\pi/2$. The insets show the regular (continuous magenta) and fitted classical (dashed green) shadow boundaries on the prograde side of the $x$--$y$ image plane. The classical fits are obtained by shifting the classical shadow boundary to match the point with maximal $x$ and rescaling the classical mass ($M'=0.99963$ and $M'=0.99961$ for the left-hand and right-hand panel, respectively) to match the point with maximal $y$, cf.~App.~\ref{app:classicalEnvelope}.
}
\end{figure}

Due to the presence of these features (cusps and dent), the shape of the shadow boundary is distinct from the shape of the shadow of a Kerr black hole.
More specifically, one can choose a Kerr black hole with $M_\text{classical}$ and $a_\text{classical}$ such that the resulting Kerr shadow boundary is a minimal envelope of the regular one. More specifically, the fitted Kerr shadow boundary matches the regular one at three points, i.e., at the retrograde point $(x_\text{max},0)$ as well as at $(x(y_\text{max}),y_\text{max})$ and $(x(y_\text{min}),y_\text{min})$, i.e., the points with maximal and minimal screen coordinate $y$, cf.~App.~\ref{app:classicalEnvelope} for details. At all other points, the regular shadow boundary deviates from its Kerr envelope, cf.~left panel of Fig.~\ref{fig:NJ-comparison}.

Starting at $a =0$, the effect initially becomes more pronounced, the larger the spin becomes. Beyond a critical value $a \approx 0.91$, the distinctness of the features decreases, as the near-extremal case $a \rightarrow 1$ is actually a case where a very small $\ell_{\rm NP,crit}\rightarrow 0$ already leads to the formation of a horizon-less object.
Therefore, restricting to cases with an event horizon, features in the shadow boundary cannot be very pronounced for $a \approx 1$. 

At these values of $a$, the resulting opportunity to observationally access very small $\ell_{\rm NP}$ -- even close to the Planck scale $\ell_\text{Planck}$ -- will be spelled out in more detail in a forthcoming paper \cite{EH2021}. In brief, for a near-critical regular black hole, 
a tiny $\ell_\text{NP}$ is sufficient to cause a qualitative change in the spacetime structure: whereas the Kerr black-hole features an event horizon at $\ell_\text{NP}=0$, it becomes a horizonless object at $\ell_\text{NP}\sim \ell_\text{Planck}$.
There exists a spin $a_\text{crit}(\ell_\text{Planck} = \ell_\text{NP,crit})\lesssim 1$, leading to horizon dissolution. In turn, for a slightly larger $a$, $\ell_\text{Planck}$ is greater than $\ell_\text{NP,crit}(a)$. This suggests, at least in principle, a physical mechanism, by which Planck-sized effects in black holes could be probed: for a black hole with $a\lesssim a_\text{crit}$,
the additional angular momentum that an infalling flux of matter can add, could be sufficient to lead to a dissolution of the horizon.
This process is expected to result in a sudden change in the image features of the corresponding compact object.
\\

Both the dent and cusps arise due to the dependence of the mass function on the angular coordinate $\chi$ which impacts the images at non-zero spin, $a >0$. As Fig.~\ref{fig:NJ-comparison} highlights, the non-local case $M = M(r)$ features a smooth image without cusps or dent. We therefore conjecture that these shadow-features are an imprint of the locality principle that is encoded in $M(r, \chi)$. Accordingly, the absence or presence of such features might provide a hint on the nature of the fundamental physics underlying regular black holes. 

Let us comment on more general metrics in the family that we propose here, cf.~\cite{Eichhorn:2021etc}. For instance, considering general $\beta$ in Eq.~\eqref{eq:massfunctiongenbeta} provides a one-parameter family of regular black holes, as long as $\beta>1$.
Qualitatively similar image features can be expected for all members of the $\beta$-family.
All of the images in this paper have been generated for $\beta=2$, cf.~Sec.~\ref{sec:absenceofdiv}; for smaller $\beta$, distinct features would become more pronounced and thus in principle more easily accessible.

\subsubsection{Effects of inclination}
\begin{figure}[!t]
\centering
\includegraphics[width=0.49\linewidth]{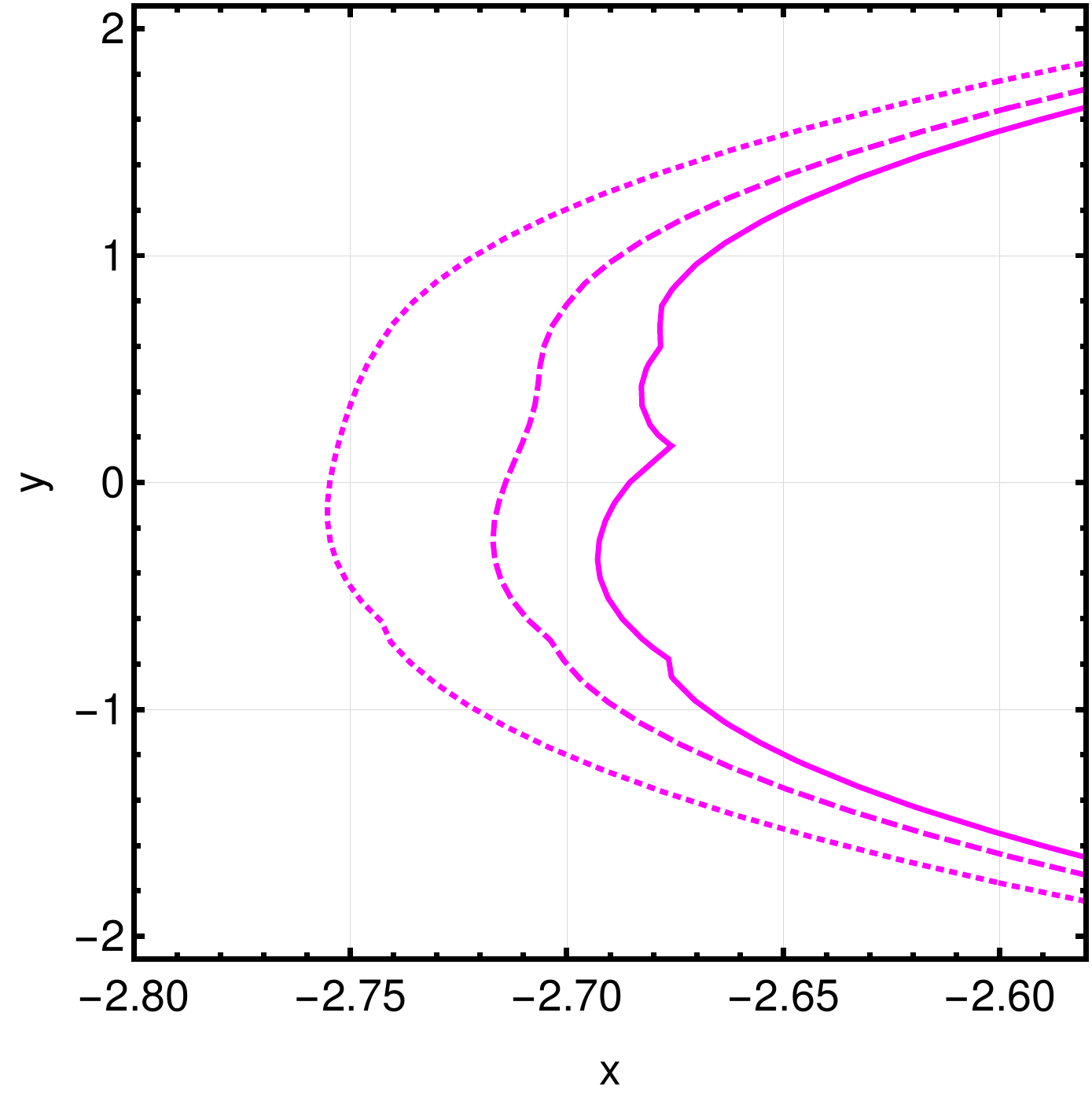}
\caption{\label{fig:inclination}
Detail plot of the prograde shadow boundary for different inclinations $\theta_\text{obs} =\frac{\pi}{2}\frac{95}{100}$ (continuous), $\theta_\text{obs} =\frac{\pi}{2}\frac{90}{100}$ (dashed), and $\theta_\text{obs} =\frac{\pi}{2}\frac{85}{100}$ (dotted). In all cases, $a=0.9$ and $\ell_\text{NP} = 0.2556$.
}
\end{figure}
Next, we explore the impact of the inclination on the image. Thereby we study potential degeneracies with Kerr black holes and account for the impact of an observationally relevant parameter on the image-features found above, see, e.g.,~\cite{Psaltis:2014mca} for a corresponding discussion in the Kerr case. We find that at $\theta_{\rm obs} \neq\{0, \pi/2, \pi\}$, the image is no longer symmetric under a reflection about $y=0$, i.e., it breaks the symmetry between northern and southern hemisphere. Heuristically, this follows because the imaged object, the black-hole horizon (or more precisely the associated photon sphere), is not spherically symmetric, but has a distinct upper and lower hemisphere. Tilting the axis of observation with respect to the symmetry-axis of the horizon results in an asymmetry of the image, as the upper half of the black-hole horizon is tilted towards/away from the observer. This effect is small in the images, cf.~Fig.~\ref{fig:inclination}.

In principle, the presence of the asymmetry allows an unambiguous determination of the inclination. In contrast, for a Kerr black hole, the two cases of the spin vector tilting towards and away from the observer are degenerate.

Further, the deviation from a face-on inclination (more precisely, with increasing $((\theta_\text{obs}-\pi/2) \mod \pi)$), reduces the distinct features (cusps and dent) in the shadow boundary. In the limiting cases of (anti-) aligned spin and inclination, i.e., for $\theta_{\rm obs} = \lbrace 0, \pi\rbrace $, the image approaches spherical symmetry. 
Thus, a case like M87*, with an estimated inclination of 
$\theta_{\rm obs} =  17 \pi/180$ is far from the ideal inclination to maximize the visibility of the identified features.
Additionally, for black holes accessible to current EHT observations, namely M87* and Sgr A*, the spin is not very well known \cite{Broderick:2016ewk}.
\\

The two limits $\theta_{\rm obs} = \lbrace 0, \pi\rbrace $ and $a \rightarrow 0$ motivate us to investigate how further information on $M(r, \chi)$ can be extracted from observations, at least in principle. In these limits, the shadow boundary is spherically symmetric and the corresponding null geodesics probe the spacetime at a fixed radius $r$. Thus, they are insensitive to the fact that $M(r, \chi) \neq \rm const$. Information on $M(r, \chi)$ can be extracted by additionally accounting for (post) Newtonian orbits, see \cite{Held:2019xde}. Yet, black-hole images actually contain more information than just on the shadow boundary itself. In particular, the shadow boundary is the limit of an exponentially stacked family of photon rings \cite{Beckwith:2004ae,Johnson:2019ljv}. These probe the spacetime at slightly different radii and can therefore, in principle, serve to reconstruct information on $M(r, \chi)$. This becomes of particular interest for the cases of small spin and very large or small inclination.

\subsection{Towards reconstructing $M(r,\chi)$ from photon rings} 
\label{sec:rings}
\begin{figure}[!t]
\centering
\includegraphics[height=0.4\linewidth]{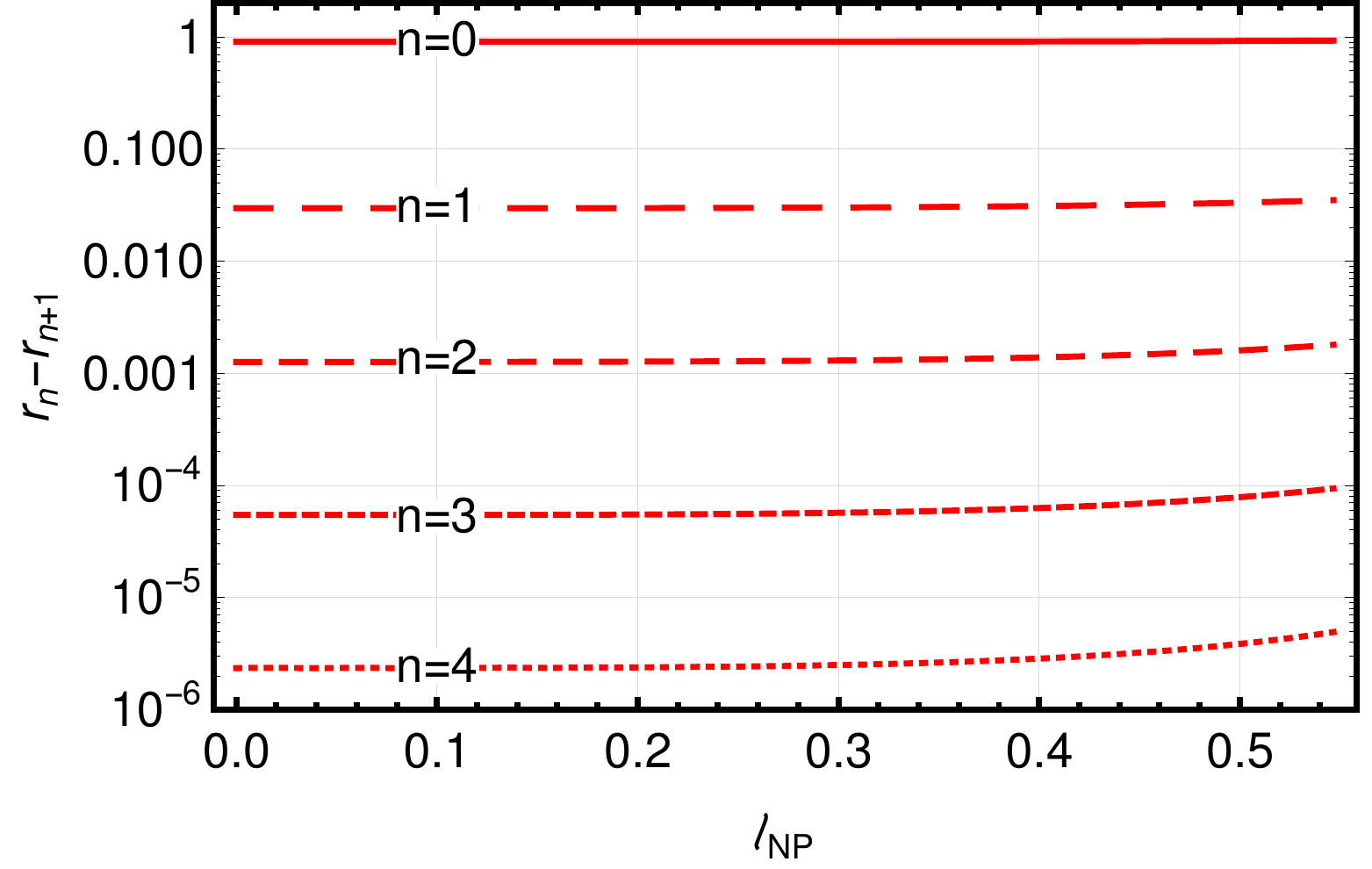}
\caption{\label{fig:lightRingsSpherical}
Difference between neighboring photon-ring radii $r_n$ (in units of $M$) as a function of $\ell_{\rm NP}$ in spherical symmetry.
}
\end{figure}
Strong gravitational lensing around the black hole implies that null geodesics can wind around the black hole several times before making their way to the observers' screen at asymptotic infinity. In a realistic astrophysical setting featuring an accretion disk, they can accordingly pierce through the disk multiple times. Thus, a given point in the accretion disk is imaged multiple times and an infinite number of nested images of the accretion disk appear on the observer's screen. The $(n=0)$-image-feature results from direct emission from the disk (and potentially a jet) and is therefore the only part of the image that is affected -- depending on the inclination -- by material in front of the black hole. This image feature is typically characterized by being rather diffuse.
The $(n=1)$ feature is likely to also contribute significantly to the photon flux in realistic settings \cite{Johnson:2019ljv}.
For increasing $n$, the image profiles become increasingly sharper, forming distinct, exponentially stacked photon rings on the image plane, see also \cite{Beckwith:2004ae}. These converge towards the $n \rightarrow \infty$ photon ring, a.k.a.~the shadow boundary. The existence of these features has been pointed out early on, see \cite{Bardeen1973,Luminet1979}. Their observational relevance in the context of VLBI has been discussed for example in \cite{Johnson:2019ljv,Himwich:2020msm}.
It has been argued in \cite{Johnson:2019ljv}, that the $n=1$ and $n=2$ photon rings could be extracted from future VLBI observations.
This motivates us to investigate whether, and if so how, the properties of the photon rings carry imprints of the new physics.
\\

In the following, we explore the geometric properties of these photon rings in an idealized setting which does not account for emission from a disk. In Sec.~\ref{sec:images} below, we account for such emission using a static disk model.

We define the $n^{\rm th}$ photon ring in a spherically-symmetric setting as follows: 
Within the equatorial plane, the angular change along a geodesic is purely azimuthal, and $n$ corresponds to the winding number of a geodesic. Outside of the equatorial plane, each geodesic oscillates within a bounded range of polar angles, $\chi_{\rm min} \leq \chi \leq \chi_{\rm max}$, as discussed in Sec.~\ref{sec:effectsofspin}. We follow \cite{Johnson:2019ljv} in defining $n$ by counting these $\chi$-oscillations, i.e., increasing $n$ by one, whenever $\chi'(\lambda)=0$, reflecting the number of times a geodesic pierces through the equatorial plane.

The properties of these photon rings depend on the spacetime metric. More specifically, each distinct photon ring effectively probes the near-horizon geometry at slightly different locations \cite{BroderickLoeb2006,Broderick:2016ewk,2020ApJ...892..132T}. Thus, within the idealized setting of spherical symmetry and ignoring the impact of an accretion disk, each $n$ is uniquely associated to a radius $R$ in the observational plane, which is an image of the radius $r$ of the corresponding photon orbit. For a Schwarzschild black hole, there is a unique relation between $n$ and $R$ that only depends on the mass of the black hole. Within the two-parameter family of $M(r, \chi)=M(r)$ given by Eq.~\eqref{eq:massfunction} at $a=0$, the radius associated to a given $n$ depends on $M$ and $\ell_{\rm NP}$.
Combining information across different $n$ therefore allows us to access the mass function $M(r) \neq \rm const$ that parameterizes the new physics in the spherically symmetric case. Specifically, the ratio of the radii of the photon rings depends on $\ell_{\rm NP}$, cf.~right panel of Fig.~\ref{fig:lightRingsSpherical}. A similar reasoning also applies in the axisymmetric case, i.e., for $M=M(r,\chi)$.
\\

This implies that the intensity images, as obtained by the EHT, could, in principle, be sufficient to detect the effect of our form of new physics. In \cite{Held:2019xde}, it was suggested that on its own, the EHT might not be able to constrain the new physics, at least in the regime where $a$ is far from extremal. This conclusion was based on the shadow boundary alone. 
Instead, it was proposed to combine EHT-measurements with mass-measurements from (post) Newtonian orbits. Here, we point out that alternatively, the added information from the distinct photon rings could be used to make a stronger observational case by providing information on $M(r, \chi)$ at more than two characteristic radii.

In particular, the combination of several distinct photon rings of spinning black holes also probes the effects of $\ell_{\rm NP}$ in the respective limits of near-aligned spin and observation axis, $\theta_{\rm obs} = 0,\pi$, as well as for vanishing spin. In both cases, the shadow boundary becomes spherically symmetric and the distinct new-physics features, i.e., dent and cusps, are not present. This is relevant for actual observations e.g., for M87*, where the line of sight forms a small angle to the spin axis, $\theta_{\rm obs} \approx 17$\textdegree, thus resulting in a nearly spherical shadow boundary. In such cases, just as in the spherically symmetric case, the ratio of the radii of different photon rings depends on $\ell_{\rm NP}$.

More specifically, the impact of $\ell_{\rm NP}$ is to increase the separation between neighboring photon rings, $r_n-r_{n-1}$, cf.~Fig.~\ref{fig:lightRingsSpherical}. Heuristically, one can explain this by accounting for the increase in compactness of the horizon, shadow boundary and photon rings, which is actually $r$ dependent. Thus, the increase in compactness grows with $n$. Accordingly, as $\ell_{\rm NP}$ increases, neighboring photon rings are pulled further apart. Coincidentally, this might even imply that at fixed EHT resolution, photon rings in the regular case are easier to resolve than in the Kerr case (when all other parameters are held fixed).

Our paper paves the way for an extensive study to assess whether EHT observations might allow to constrain $\ell_{\rm NP}$ by probing photon rings. Below, we will take a first step in a more realistic direction by accounting for the impact of a disk, described by a simple analytical model.

\subsection{Impact of disk-like structures}
\label{sec:intensity}

The shape of the shadow boundary, cf.~Sec.~\ref{sec:shapes}, and the shape of photon rings, cf. Sec.~\ref{sec:rings}, are idealized theoretical observables. In contrast, realistic shadow images account for, both, the spacetime geometry as well as the astrophysical environment surrounding the black hole.  In principle, accounting for the latter requires GRMHD simulations of the dynamical accretion disk \cite{Porth:2019wxk}, as well as other potential sources of infalling matter \cite{BroderickLoeb2006,2020ApJ...892..132T}. A simpler first step is to model effects of a  (static) disk by accounting for emission along a geodesic within the finite-density region of a disk profile.
We follow \cite{2020ApJ...897..148G} and implement the radiative transfer (Boltzmann) equation,
\begin{align}
\label{eq:radiative-transfer-eq}
	\frac{d}{d\lambda}\left(\frac{I_\nu}{\nu^3}\right) = \left(\frac{j_\nu}{\nu^2}\right) - (\nu\alpha_\nu)\left(\frac{I_\nu}{\nu^3}\right)\;,
\end{align}
which describes how the intensity $I_\nu$ changes along a null geodesic in dependence on the emissivity $j_\nu$ and absorptivity $\alpha_\nu$ of the relativistic fluid through which the light ray propagates. Herein, intensity, emissivity and absorptivity depend on the frequency $\nu$. In the above form, the light ray is parameterized by the affine parameter $\lambda$ and the equation is arranged such that each expression in parenthesis denotes a scalar quantity which may be calculated in any (and even in different) coordinate frames. 

Solving Eq.~\eqref{eq:radiative-transfer-eq} requires knowledge about the density and velocity profiles of the accretion disk. Instead of dynamically determining these via GRMHD, we work with a specific static disk model. We simplify the setting by neglecting the absorptivity ($A=0$ in the notation of \cite{2020ApJ...897..148G}). There are compelling indications that the accretion disks of supermassive black holes, e.g., M87* and Sgr A*, could indeed be optically thin~\cite{Johnson:2015iwg}. Further, we assume a frequency-independent emissivity ($\alpha=-2$ in the notation of \cite{2020ApJ...897..148G}). Such a frequency-independent disk model is not sensitive to the velocity profile of the disk and thereby radiative transfer only depends on $r$ and $\chi$. Given that EHT observations are essentially monochromatic \cite{paper1}, neglecting frequency-dependence is a reasonable first assumption. Frequency-independence guarantees that the investigated disk models can be properly implemented even in the vicinity of the horizon, where no Boyer-Lindquist form of the regular spacetime is available, cf.~App.~\ref{app:BL-form}. With these simplifications, the second term on the right-hand-side of Eq.~\eqref{eq:radiative-transfer-eq} vanishes, the first term is no longer frequency-dependent, and the radiative transfer equation reduces to
\begin{align}
\label{eq:radiative-transfer-eq-simplified}
	\frac{d}{d\lambda}\left(\frac{I_\nu}{\nu^3}\right) = C\,n\left(x^\mu(\lambda)\right)\;.
\end{align}
Here, $C$ denotes a dimensionful constant and $n(r,\chi)$ models the number density of the disk. In Eq.~\eqref{eq:radiative-transfer-eq-simplified}, $n(r,\chi)$ is evaluated on the photon world line $x^\mu(\lambda)$, obtained by numerical ray tracing of the geodesic equation as in Sec.~\ref{sec:shapes}. In principle, this setup allows us to investigate arbitrary number densities and accordingly disk profiles. Following \cite{2020ApJ...897..148G}, we model the number density by
\begin{align}
\label{eq:number-density}
	n(r\,,\chi) = n_0
	\times\exp\left[-\frac{1}{2}\left(\frac{r^2}{100}+h^2\chi^2\right)\right],
\end{align}
where the disk model parameter $h$ controls the $\chi$-dependence of the disk, i.e., $h=0$ denotes a spherical profile while for $h\rightarrow\infty$ the disk becomes infinitesimally thin. We remind the reader that we have set $M=1$ and $G=1$.
The dimensionful, constant combination $C\cdot n_0$ drops out once all resulting intensity profiles are normalized. In the following, we normalize each (set of) images to the brightest image point.

This simple disk model allows us to obtain not just the shadow shape but the relative intensity at every image point. In addition to spin $a$, inclination $\theta_\text{obs}$, and the new-physics parameter $\ell_\text{NP}$, the observed relative intensity now depends on the disk-parameter $h$.

\begin{figure}[!t]
\centering
\begin{minipage}{0.52\linewidth}
\includegraphics[width=\linewidth]{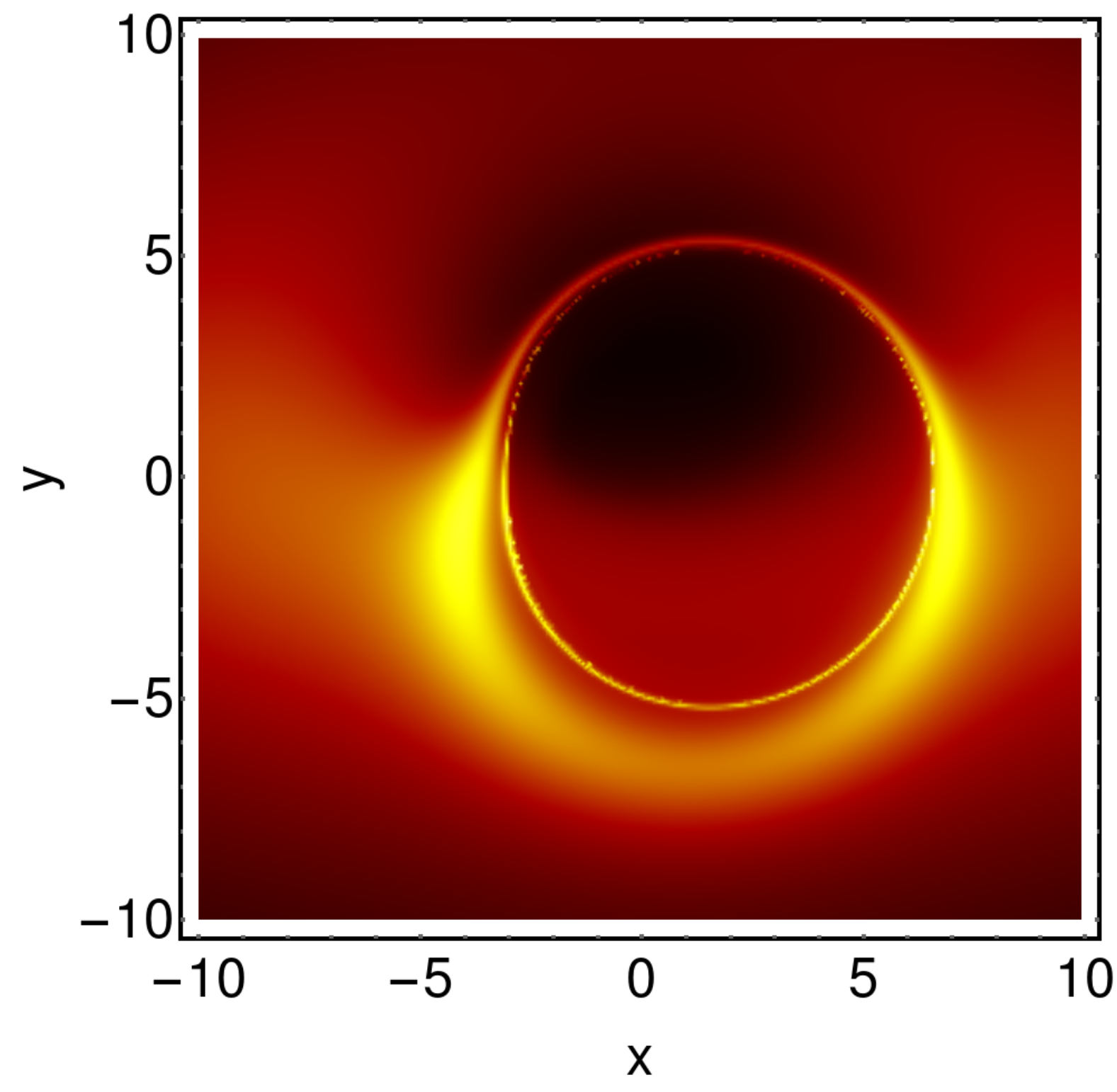}
\end{minipage}
\begin{minipage}{0.46\linewidth}
\includegraphics[width=\linewidth]{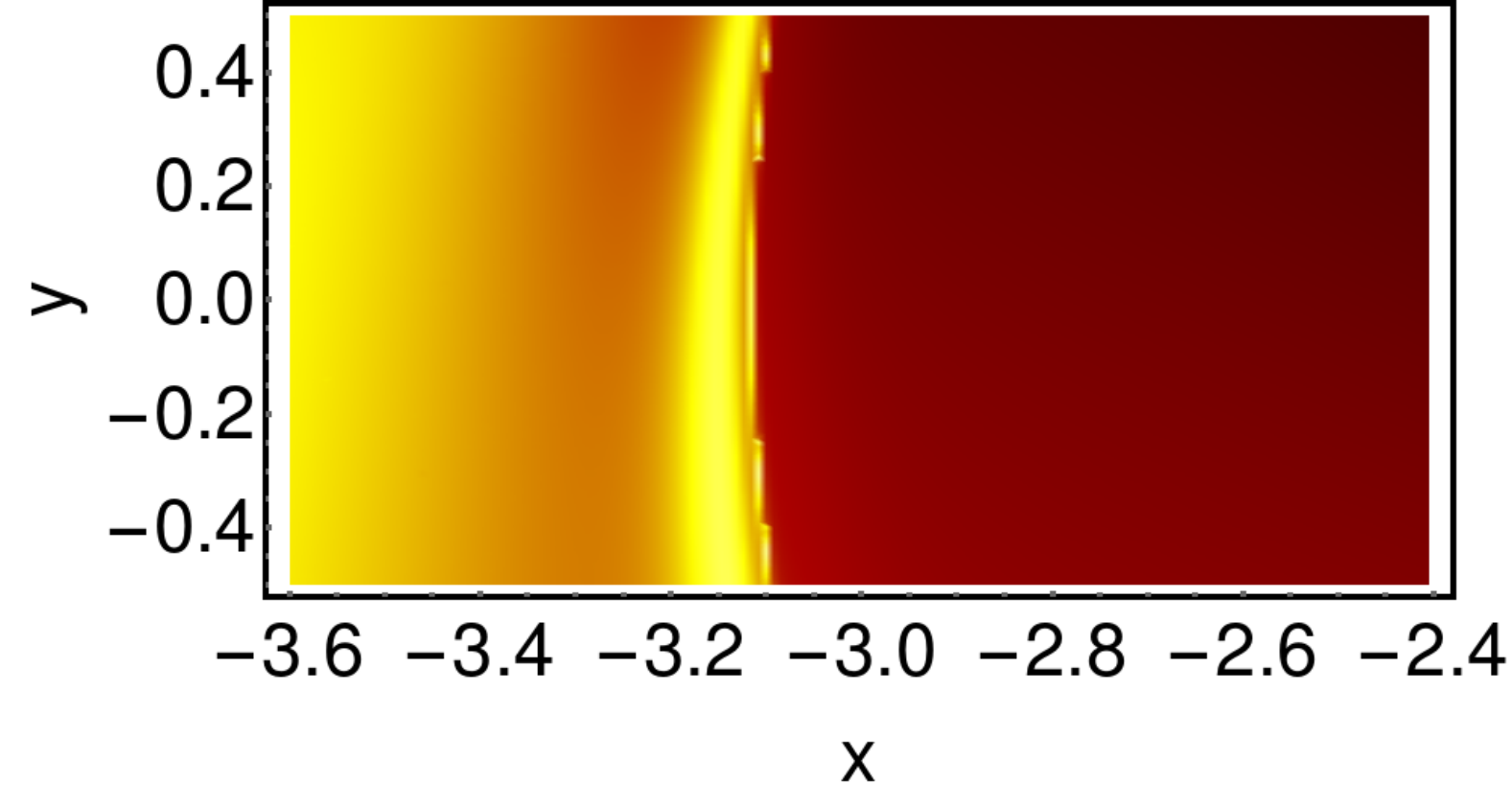}\\
\includegraphics[width=\linewidth]{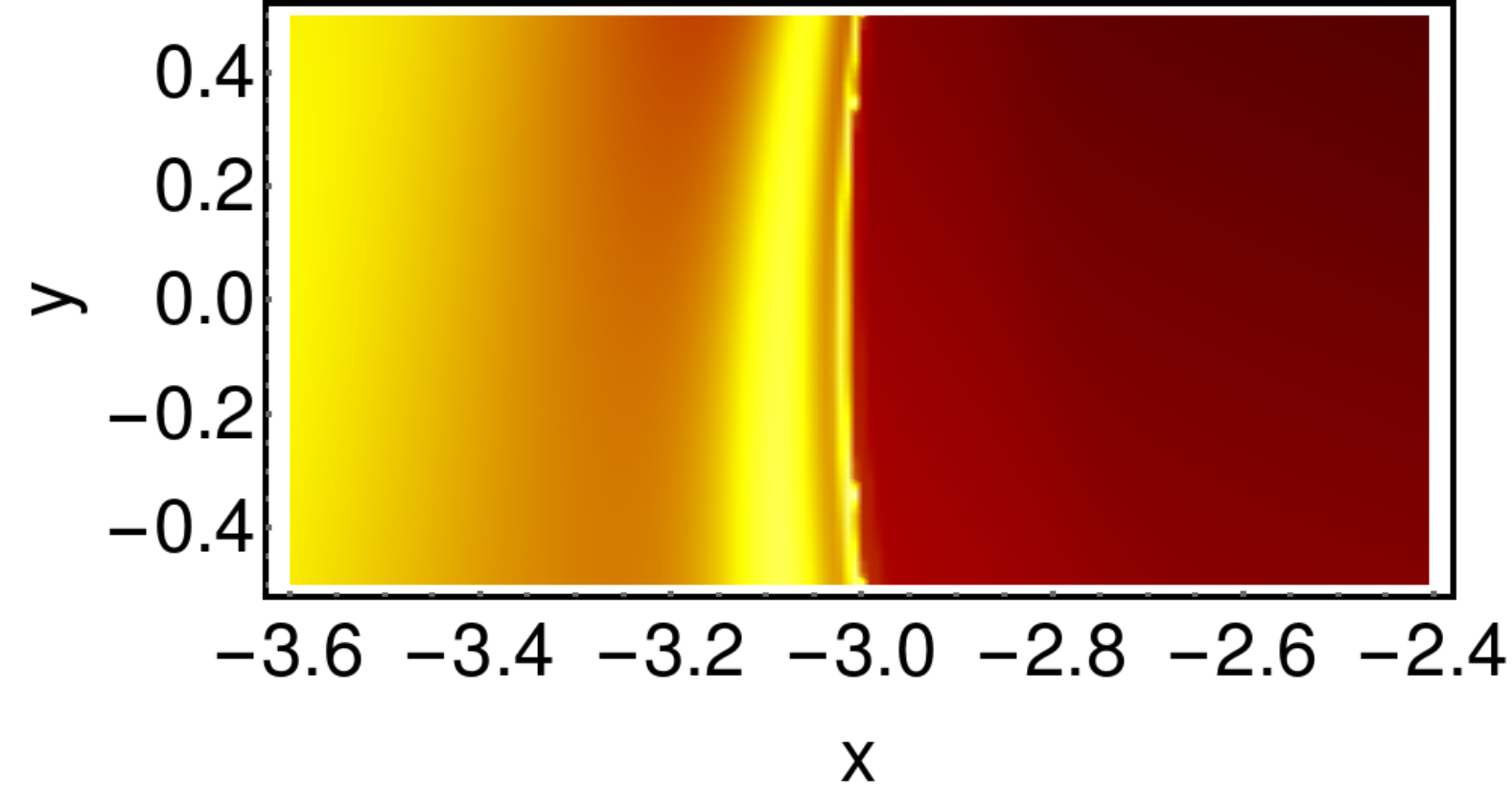}
\end{minipage}
\caption{\label{fig:images}
	Relative intensity on the image screen for disk models with $h=10/3$ for $\theta_\text{obs}=\pi/3$ and $a=9/10$. The left-hand panel is generated for $\ell_\text{NP} = 0.25$. The right-hand panels zoom in on the prograde region where the impact of singularity-resolving new physics is most clearly visible, with the upper panel showing the Kerr case and the lower panel the regular case. 
}
\end{figure}
\begin{figure}[!t]
\centering
\begin{minipage}{0.52\linewidth}
\includegraphics[width=\linewidth]{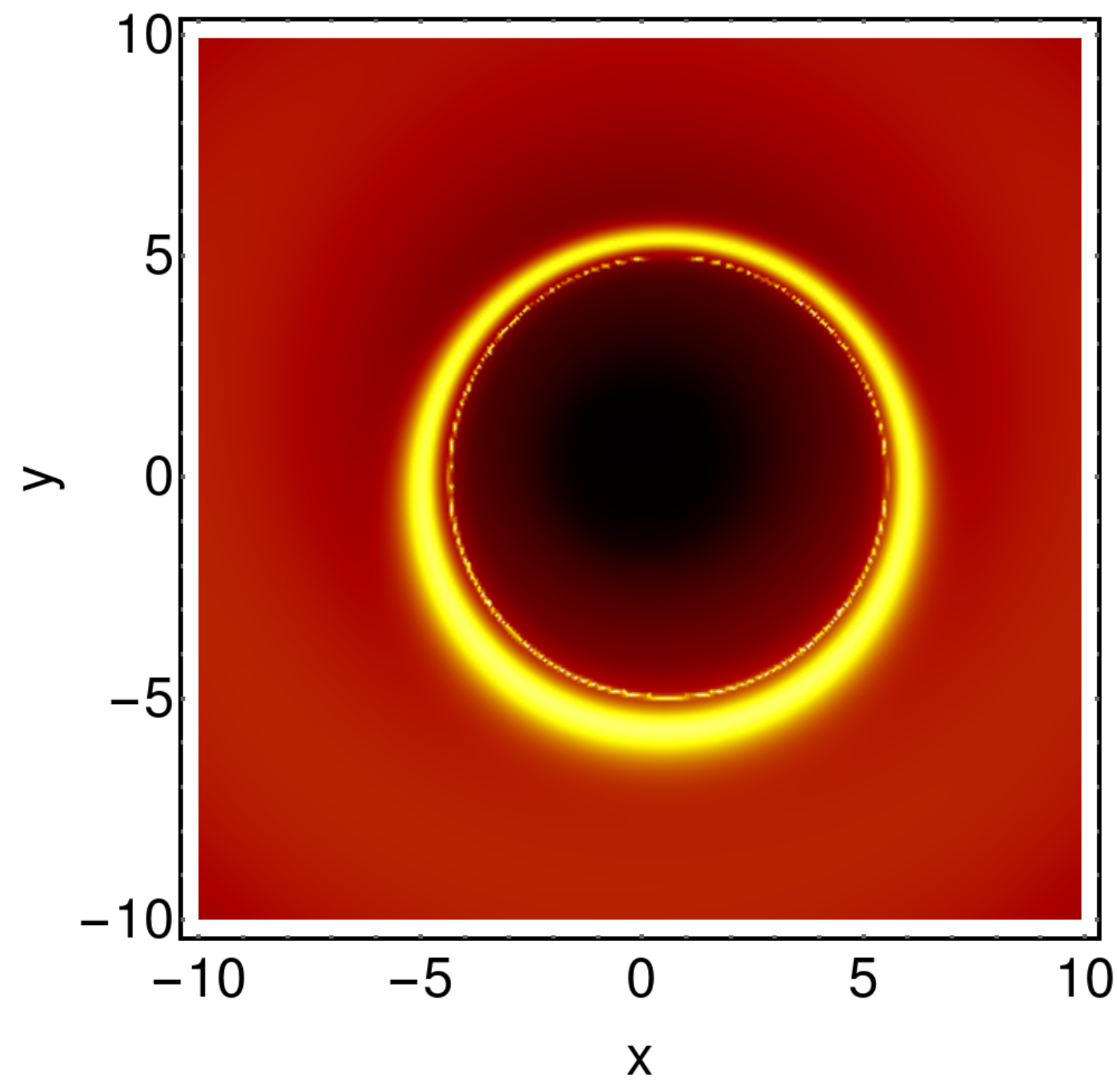}
\end{minipage}
\begin{minipage}{0.46\linewidth}
\includegraphics[width=\linewidth]{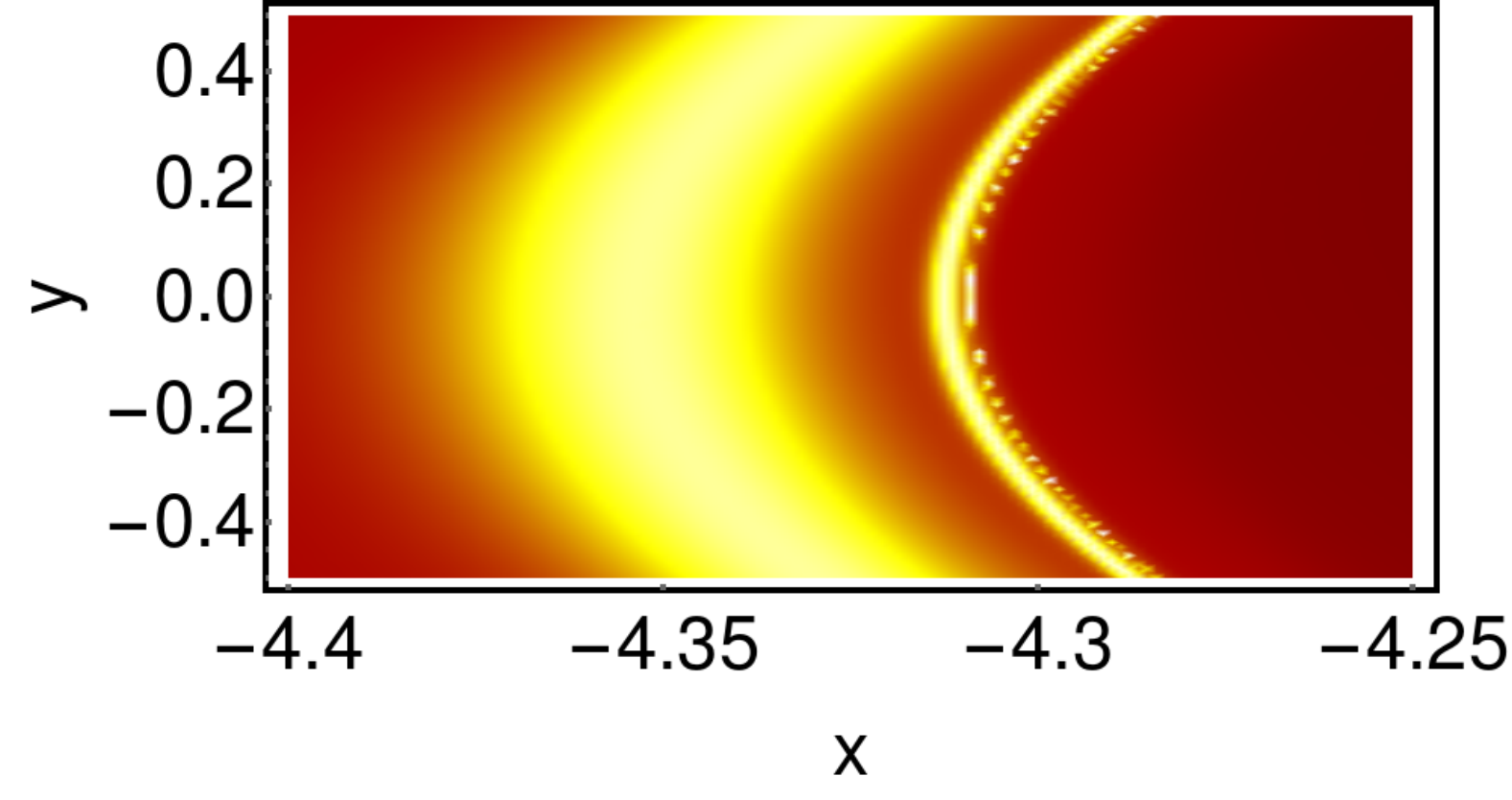}\\
\includegraphics[width=\linewidth]{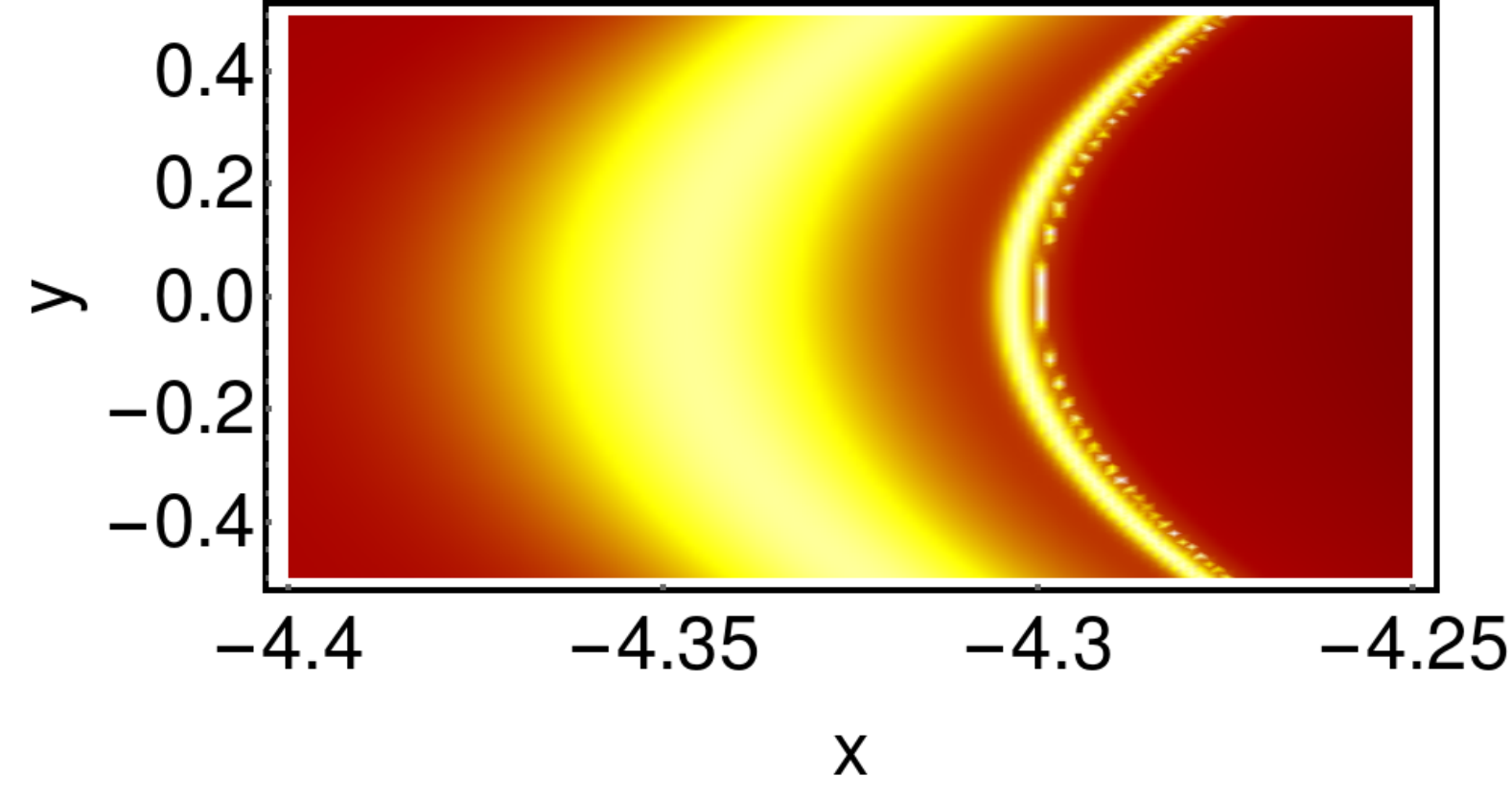}
\end{minipage}
\caption{\label{fig:imagesM87}
	We show the same as in Fig.~\ref{fig:images}, but for inclination $\theta_\text{obs} = 17\pi/180$ and $\ell_\text{NP} = 0.2556$.
	}
\end{figure}
\begin{figure}[!t]
\centering
\includegraphics[width=0.49\linewidth]{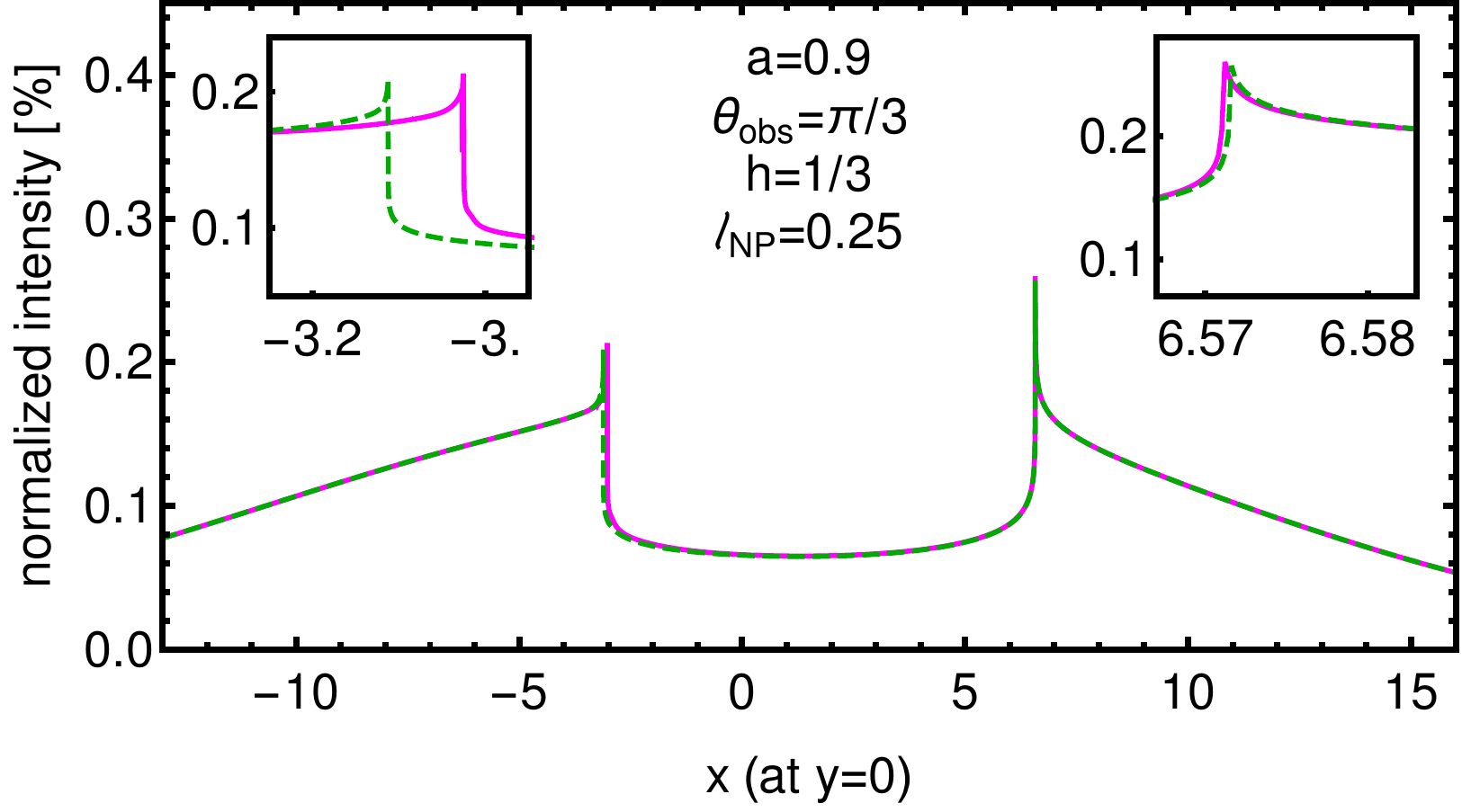}
\hfill
\includegraphics[width=0.49\linewidth]{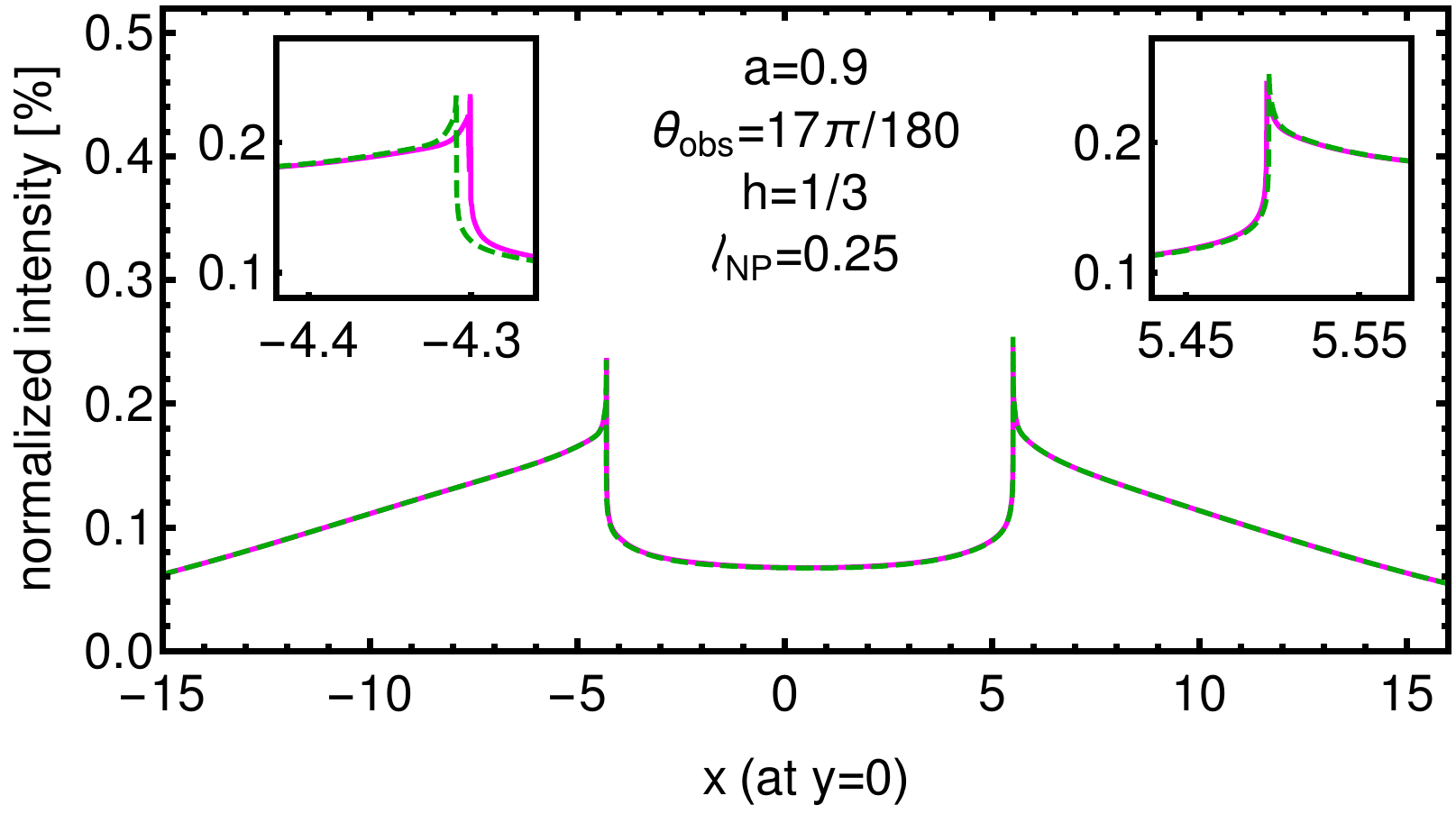}
\\
\includegraphics[width=0.49\linewidth]{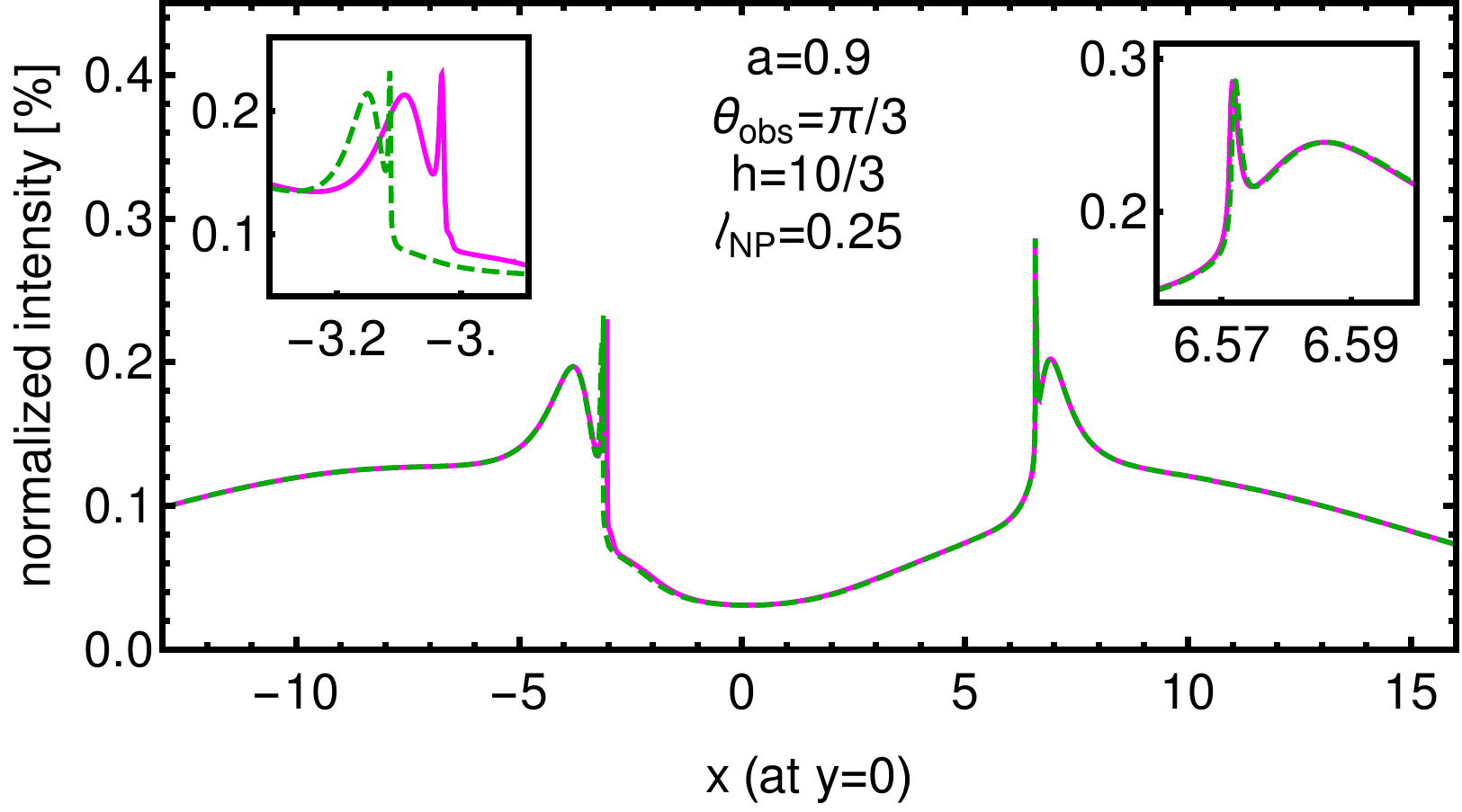}
\hfill
\includegraphics[width=0.49\linewidth]{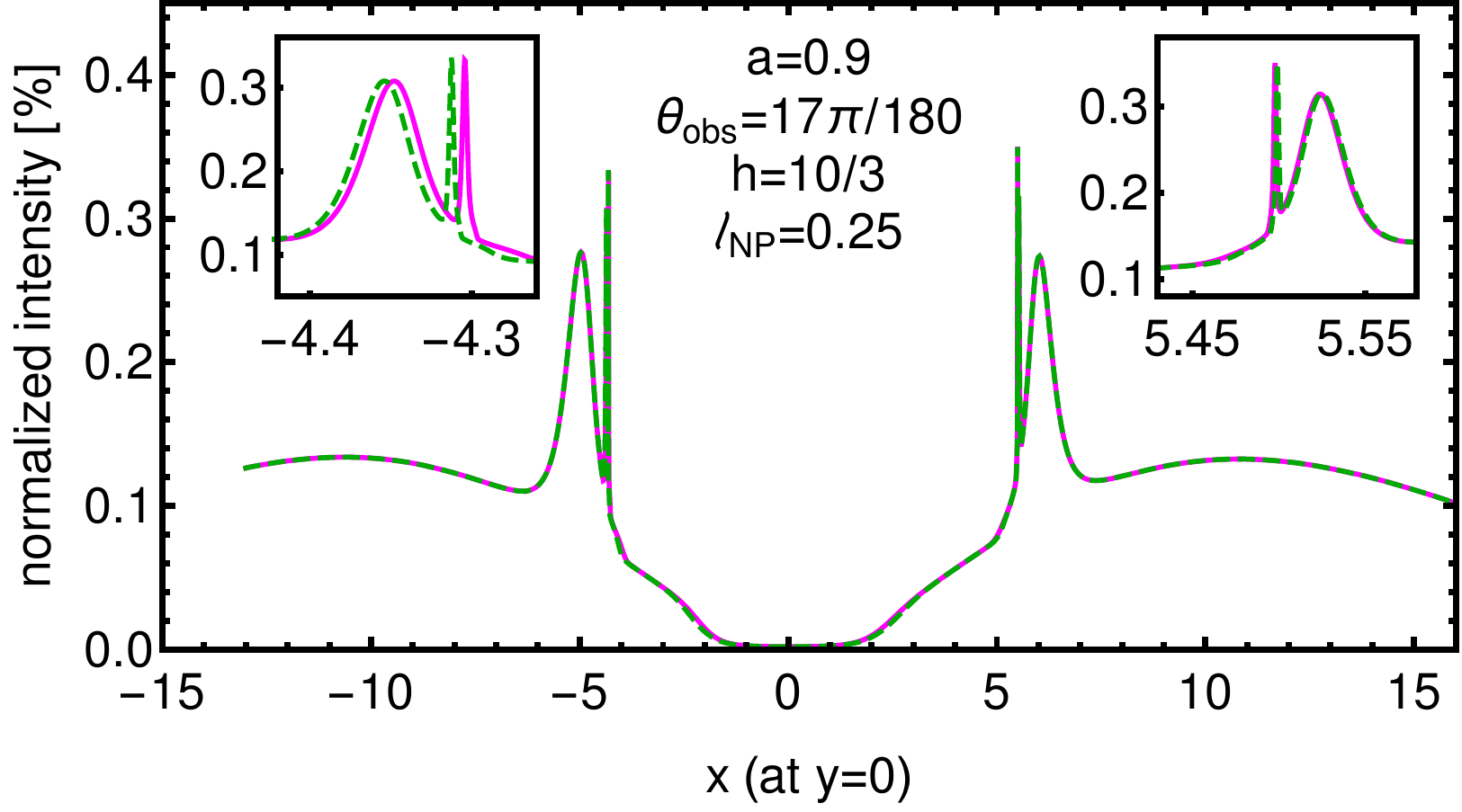}
\\
\includegraphics[width=0.49\linewidth]{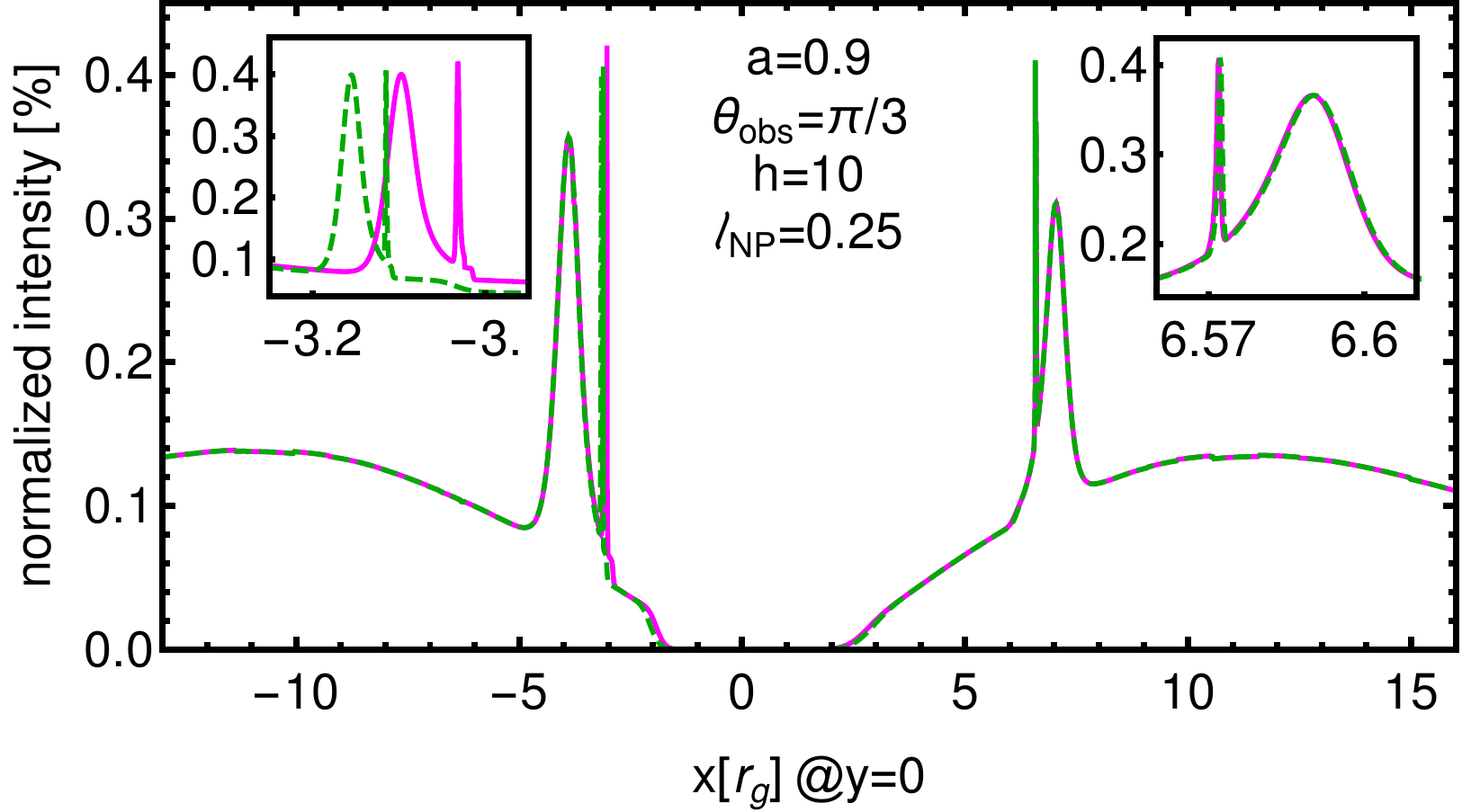}
\hfill
\includegraphics[width=0.49\linewidth]{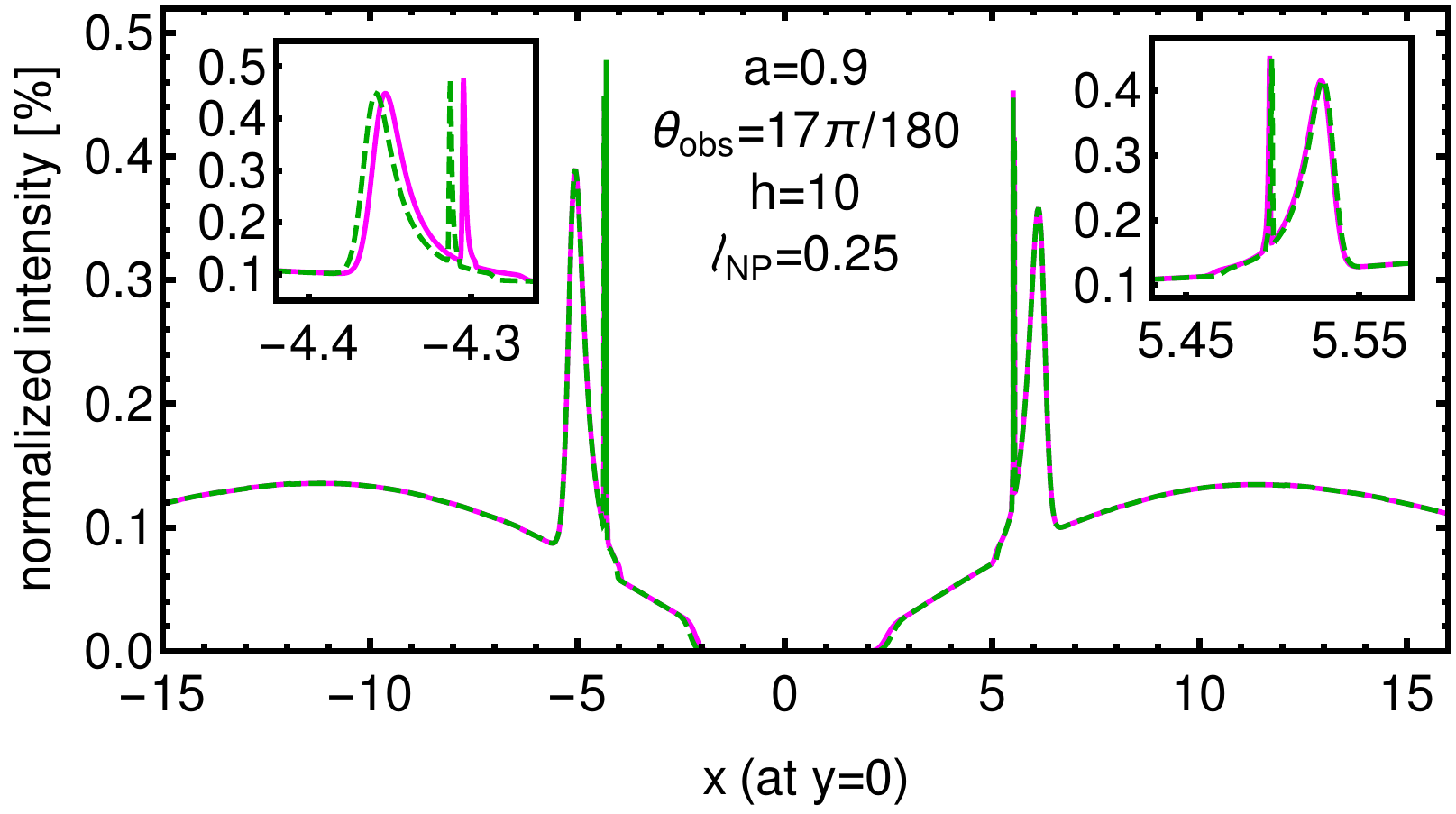}
\caption{\label{fig:sweeps}
	Image cross-sections at $y=0$ for different inclinations $\theta_\text{obs}$ and disk parameter $h$, all at spin parameter $a=0.9$. In all panels, we compare the singular Kerr (green dashed) to the regular new-physics spacetime with $\ell_\text{NP}=0.25$ (magenta continuous).
	In the left-hand column (right-hand column), $\theta_\text{obs}=\pi/3$ ($\theta_\text{obs}=17\pi/180$). In the upper, middle, and lower row the disk height is decreased successively, i.e., $h=1/3$, $h=10/3$, $h=10$, respectively.
	In each panel, the intensity has been normalized such that the depicted overall intensity integrates to one.
	We terminate the integration of the radiative transfer equations after $n=3$ minima in $\chi$ such that only the emission from the first three photon rings contributes to the intensity.
}
\end{figure}

We generate simulated black-hole images at inclination $\theta_\text{obs}=\pi/3$ and $\theta_\text{obs}=17\pi/180$, cf.~Fig.~\ref{fig:images} and Fig.~\ref{fig:imagesM87}, respectively. As one might have expected on the basis of the previous subsections, these images bear a lot of similarity to images of Kerr black holes with similar spin, inclination, mass and disk. Thus, the dramatic changes to the spacetime that occur within the horizon and even at the event horizon (cf.~Fig.~\ref{fig:horizon}, where the deviation between regular and Kerr horizon is $\sim \mathcal{O}(10\%)$), result in less significant deviations in the observed intensity images (cf.~Figs.~\ref{fig:images}~and~\ref{fig:imagesM87}, where the deviations between the regular and Kerr intensity image are of order $\sim \mathcal{O}(1\%)$), at least in our non-dynamical and frequency-independent model. 
In particular, theoretically distinct features, such as the cusps and dent in the shadow boundary, are challenging to resolve in an image with a realistic resolution, again, within the context of the present disk model. Therefore, exploring other observables, such as, e.g., properties of photon rings, might be a more promising route.

By zooming in on the prograde side, cf.~right-hand panels of Fig.~\ref{fig:images}, we observe that in the presence of the static disk image features distinguishing the regular and the Kerr black hole are present: In particular, the increased separation between the $n=1$ and $n=2$ ring, cf.~Fig.~\ref{fig:lightRingsSpherical}, in principle remains visible in the presence of the static disk. This also holds in the case of near-aligned spin and observation axis, cf.~Fig.~\ref{fig:imagesM87}.

This observation motivates us to explore image profiles, i.e., cross-sections through the image plane at $y=0$, for various inclinations $\theta_{\rm obs}$ and inverse disk heights $h$, cf.~Fig.~\ref{fig:sweeps}. In the case of the present disk model, the photon rings  become more pronounced, as the disk height decreases (i.e., as $h$ increases). Additionally, as the inclination approaches $\theta_{\rm obs}=\lbrace 0,\pi\rbrace$, the resemblance between regular image and the Kerr image increases. Specifically, the distance between neighboring photon rings  on the prograde side appears to differ most from Kerr for regular black holes viewed at $\theta_{\rm obs}=\pi/2$. 
Accordingly, observing black holes in situations of near-aligned spin and observation axis might not be the ideal case to place constraints on $\ell_{\rm NP}$ from EHT observations.

We caution that the above discussion applies only within the context of the present disk model, cf.~Eq.~\eqref{eq:number-density}, which does neither account for frequency-dependence, nor for the dynamics of an accreting disk. It is not excluded that the behavior of timelike geodesics close to the horizon deviates significantly from that in Kerr spacetime. If this were the case, accreting matter or hotspot-like features might carry additional imprints of the regular spacetime that our simplified treatment does not account for, cf.~\cite{BroderickLoeb2006,2020ApJ...892..132T}. We leave such intriguing questions for future work.

\section{Conclusions and outlook}\label{sec:conclusions}
Current observations show that compact objects with an uncanny resemblance to GR black holes exist. Despite passing all corresponding experimental tests, GR cannot provide a complete understanding of these objects since it predicts an infinite, hence unphysical, spacetime curvature at the center of black-hole solutions. Similar breakdowns of theoretical predictions need not be limited to the center of the black hole or even the inside of the horizon. 
This motivates the question: What is the true nature of these objects? This riddle can -- and should -- be tackled both theoretically and observationally, starting with the construction of well-motivated theoretical models and ending with a comparison to observable physics such as intensity images of compact objects.
\\

In the present paper and in \cite{Eichhorn:2021etc}, we take several steps along such a line of research:
(i) we develop a locality principle, giving rise to a well-motivated family of regular, spinning spacetimes; (ii) we investigate one member of the above family, based on a complete set of (scalar polynomial) curvature invariants and a study of Killing and outer event horizon; (iii) we explore the idealized shape of the resulting shadow boundary at finite spin and inclination; iv) going beyond the shadow boundary, we investigate the photon rings; and v) taking a step towards realistic images, we include an analytical model of a static disk and generate intensity images.
\\
In these steps, we parameterize deviations from Kerr by a spacetime-dependent mass function $M(r, \chi)$ with appropriate asymptotic behaviors to guarantee a Newtonian limit and regularity. Probing the new physics (not necessarily restricted to our principles of regularity and locality) thus requires probing $M(r, \chi)$. We point out three ways in which this can, in principle, be done:
\begin{itemize}
\item At finite spin and at inclination close to the equatorial plane, the shadow boundary itself deviates from Kerr.
\item At all spins and inclinations, the separation of neighboring photon rings carries imprints of the new physics.
\item Combining mass-measurements by the EHT from shadow boundary and/or photon rings with mass-measurements from (post) Newtonian stellar orbits or the dynamics of infalling gaseous matter measures $M(r,\chi)$ at significantly different radii, as, in part, already pointed out in \cite{Held:2019xde}.
\end{itemize}

In a shorter companion paper \cite{Eichhorn:2021etc}, we study an additional example for $M(r, \chi)$. This demonstrates that there exist other representatives in our family of regular spacetimes for which distinguishing image features remain qualitatively similar to those discussed here but can become quantitatively more pronounced.
\\

We highlight two points which distinguish the present developments from other related works.
\\
The distinctive construction principle of the family of regular, spinning black-holes is locality. More specifically, the locality principle ties the onset of new-physics modifications to the local curvature scale of the respective Kerr geometry. This results in specific image features, namely (i) asymmetry at non-face-on inclinations, (ii) cusps, and (iii) a dent in the shadow boundary. All of these features reflect a corresponding dent in the event horizon. This distinguishes the present model from numerous developments in the literature where spinning regular black holes are constructed by implementing the NJ-algorithm, all of which feature a spherical event horizon and violate our locality principle.
\\
Moreover, to implement the locality principle, we work in ingoing Kerr coordinates and the resulting metric cannot be brought into a Boyer-Lindquist form by a transformation of the passive coordinates, cf.~App.~\ref{app:BL-form}, except asymptotically, where it reproduces the Newtonian limit correctly. 
Therefore, a number of common parameterizations such as, \cite{Johannsen:2011dh,Cardoso:2014rha,Johannsen:2015pca,Lin:2015oan,Konoplya:2016jvv,Papadopoulos:2018nvd}
appear to not straightforwardly capture the new family of regular metrics and the underlying locality principle.
\\

A number of exciting theoretical and phenomenological questions are opened up by our investigations.
On the theoretical side, these include the following: Most importantly, the kinematical nature of the present study calls for the development of a potential underlying dynamics for the metric. This would enable a study of dynamical stability of the new family of spinning black holes. Further, studying the internal structure of the present class of metrics in more depth can answer, whether our construction principle can give rise to a spacetime for which Hawking's rigidity theorem holds, i.e., for which event and Killing horizon coincide. Finally, an investigation of the fate of the inner horizon under the impact of the new physics is called for.

On the phenomenological side, more realistic disk physics can be accounted for by adding frequency dependence and absorptivity to the present disk model. As a second step, GRMHD simulations can be performed in our modified spacetime, under the assumption that the new physics only impacts the spacetime structure, but not the dynamical equations for the accreting matter.
\\

\noindent \emph{Acknowledgements:}\\
 A.~E.~is supported by a research grant (29405) from VILLUM fonden. 
A.~H.~is supported by a Royal Society Newton fellowship [NIF/R1/191008].

\appendix
\section{Characterization of the regular spinning black hole} \label{app:curvatureinvnewmetric}

In this appendix, we characterize the constructed regular spacetime in terms of non-derivative curvature invariants, i.e., those built solely from the Riemann tensor and the metric following \cite{1997GReGr..29..539Z}.

\subsection{A algebraically complete basis of non-derivative curvature invariants}
\label{app:ZM-basis}
The construction of an algebraically complete basis of non-derivative curvature invariants\footnote{
One distinguishes between curvature invariants (built solely from the Riemann tensor) and derivative invariants (built from the Riemann tensor and its covariant derivatives). The number of functionally independent derivative invariants is limited by the spacetime dimension $d$. The number of algebraically independent invariants, i.e., those that do not satisfy a polynomial relation (called syzygy), has been determined in \cite{Thomas:255071}, see also [\cite{Stephani:2003tm},Sec.~9.1] for a review.
While the number of algebraically independent curvature invariants of a specific manifold is finite, the number of algebraically independent derivative invariants keeps growing with increasing number of covariant derivatives.}  
has been achieved for (electro-)vacuum and perfect fluid spacetimes in GR in \cite{1991JMP....32.3135C} and subsequently generalized to arbitrary spacetimes in \cite{1997GReGr..29..539Z,2002nmgm.meet..831C}. The latter set of Zakhary-McIntosh (ZM) invariants is constructed from the Weyl tensor $C_{\mu\nu\rho\sigma}$ and the Ricci-tensor $R_{\mu\nu}$ and consists of four real Weyl-invariants $I_{1/2}$ and $I_{3/4}$ (which can be combined into two complex invariants), four real Ricci-invariants $I_{5-8}$ as well as nine real mixed Ricci-Weyl invariants $I_{9/10}$, $I_{11/12}$, $I_{13/14}$, $I_{16/17}$ (combining into four complex invariants) and $I_{15}$, i.e., 
\begin{align}
	I_{1} &=C_{\mu\nu\rho\sigma}C^{\mu\nu\rho\sigma},
	\\
	I_{2} &=C_{\mu\nu\rho\sigma}\overline{C}^{\mu\nu\rho\sigma},
	\\
	I_{3} &=C_{\mu\nu}^{\phantom{\mu\nu}\rho\sigma}C_{\rho\sigma}^{\phantom{\rho\sigma}\alpha\beta }C_{\alpha\beta }^{\phantom{\alpha\beta }\mu\nu},
	\\
	I_{4} &=\overline{C}_{\mu\nu}^{\phantom{\mu\nu}\rho\sigma}C_{\rho\sigma}^{\phantom{\rho\sigma}\alpha\beta }C_{\alpha\beta }^{\phantom{\alpha\beta }\mu\nu},
	\\
	I_{5} &=R,
	\\
	I_{6} &=R_{\mu}^{\phantom{\mu}\nu}R_{\nu}^{\phantom{\nu}\mu},
	\\
	I_{7} &=R_{\mu}^{\phantom{\mu}\nu}R_{\nu}^{\phantom{\nu}\rho}R_{\rho}^{\phantom{\rho}\mu},
	\\
	I_{8} &=R_{\mu}^{\phantom{\mu}\nu}R_{\nu}^{\phantom{\nu}\rho}R_{\rho}^{\phantom{\rho}\sigma}R_{\sigma}^{\phantom{\sigma}\mu},
	\\
	I_{9} &=R^{\mu\nu}R^{\rho\sigma}C_{\mu\nu\rho\sigma},
	\\
	\label{eq:I10}
	I_{10} &=R^{\mu\nu}R^{\rho\sigma}\overline{C}_{\mu\nu\rho\sigma},
	\\
	I_{11} &=R^{\nu\rho}R_{\gamma\delta }\left(
		C_{\mu\nu\rho\sigma}C^{\mu \gamma\delta \sigma} 
		- \overline{C}_{\mu\nu\rho\sigma}\overline{C}^{\mu\gamma\delta \sigma}
	\right),
	\\
	\label{eq:I12}
	I_{12} &= 2 R^{\nu\rho}R_{\gamma\delta }C_{\mu\nu\rho\sigma}\overline{C}^{\mu\gamma\delta\sigma},
	\\
	I_{13} &=R_{\mu}^{\phantom{\mu}\gamma}R_{\gamma}^{\phantom{\gamma}\rho}R_{\nu}^{\phantom{\nu}\delta}R_{\delta}^{\phantom{\delta}\sigma}C^{\mu\nu}_{\phantom{\mu\nu}\rho\sigma},
	\\
	I_{14} &=R_{\mu}^{\phantom{\mu}\gamma}R_{\gamma}^{\phantom{\gamma}\rho}R_{\nu}^{\phantom{\nu}\delta}R_{\delta}^{\phantom{\delta}\sigma}\overline{C}^{\mu\nu}_{\phantom{\mu\nu}\rho\sigma},
	\\
	I_{15} &=\frac{1}{16}R^{\nu\rho}R_{\gamma\delta}\left(
		C_{\mu\nu\rho\sigma}C^{\mu\gamma\delta\sigma} 
		+ \overline{C}_{\mu\nu\rho\sigma}\overline{C}^{\mu\gamma\delta\sigma}
	\right),
	\\
	\label{eq:ZM-I16}
	I_{16} &=\frac{1}{32}R^{\rho\sigma}R^{\gamma\delta }C^{\mu\kappa\lambda\nu}\left(
		C_{\mu\rho\sigma\nu}C_{\kappa\gamma\delta\lambda} 
		+ \overline{C}_{\mu\rho\sigma\nu}\overline{C}_{\kappa\gamma\delta\lambda}
	\right),
	\\
	I_{17} &=\frac{1}{32}R^{\rho\sigma}R^{\gamma\delta }\overline{C}^{\mu \kappa\lambda \nu}\left(
		C_{\mu\rho\sigma\nu}C_{\kappa\gamma\delta\lambda} 
		+ \overline{C}_{\mu\rho\sigma\nu}\overline{C}_{\kappa\gamma\delta\lambda}
	\right)\;,
\end{align}
where $\overline{C}_{\mu\nu\rho\sigma} = 1/2\epsilon_{\mu\nu \kappa\lambda }C^{\kappa\lambda }_{\phantom{\kappa\lambda }\rho\sigma}$ is the (left-)dual Weyl tensor and $\epsilon$ is the totally anti-symmetric Levi-Civita tensor.
Since the Weyl-tensor is completely traceless, i.e., the metric-contraction of any pair of indices vanishes, the Ricci-tensor $R_{\mu\nu}$ may be substituted for its traceless counterpart $S_{\mu\nu} = R_{\mu\nu} - \frac{1}{4}g_{\mu\nu}R$ in all of the above \emph{mixed} expressions. 
Replacement of $I_{13}$ and $I_{14}$ with the real invariant $M_4 = 1/16\,R^{\mu\kappa}R^{\gamma\delta}R^{\rho}_{\phantom{\rho}\sigma}\left(C_{\mu\rho}^{\phantom{\mu\rho}\sigma\nu}C_{\nu\gamma\delta\kappa} + \overline{C}_{\mu\rho}^{\phantom{\mu\rho}\sigma\nu}\overline{C}_{\nu\gamma\delta\kappa}\right)$ recovers the set of 16 Carminati-McLenaghan (CM) invariants \cite{1991JMP....32.3135C}. 

Since this type of expressions is prone to typos, we have verified that the above expressions and their implementation recover the results for Kerr-Newman spacetime explicitly listed in \cite{Overduin:2020aiq}\footnote{Comparing to \cite{Overduin:2020aiq}, we find that it is necessary to correct the following typos in order to guarantee that the syzygies of Kerr-Newman spacetime, cf.~Eqs.~\eqref{siggi1}-\eqref{siggiLast}, hold: the Ricci-Weyl expression for invariant $I_{10}$ (see Eq.~\eqref{eq:I10}) has to come with opposite sign in order to combine into $\mathbb{K}$ as in Eq.~\eqref{eq:def-complex-invariants}; the Ricci-Weyl expression for invariant $I_{12}$ (see Eq.~\eqref{eq:I12}) contains an index-typo but the explicit expression on Kerr-Newman spacetime agrees with ours. (In adition, the respective definitions of $I_{16}$ and $I_{17}$ come with an opposite sign but this does not alter any syzygies.)
}. In particular, as a non-trivial cross check of our implementation we calculate the complex invariants, i.e.,
\begin{align}
\label{eq:def-complex-invariants}
	\mathbb{I} = I_1 + i I_2\;,
	\quad
	\mathbb{J} = I_3 + i I_4 \;,
	\quad
	\mathbb{K} = I_9 + i I_{10}\;,
	\quad
	\mathbb{L} = I_{11} + i I_{12}\;,
	\quad
	\mathbb{M}_2 = I_{16} + i I_{17}\;,
\end{align}
and verify the non-trivial syzygies of the Kerr-Newman spacetime, i.e.,
\begin{align}
	\mathbb{I}^3 &= 12\,\mathbb{J}^2\;,
	\label{siggi1}
	\\
	I_6^2 &= 4\,I_8\;,
	\label{siggi2}
	\\
	3\,\mathbb{L}^2 &= \mathbb{I}\,\mathbb{K}^2\;,
	\label{siggi3}
	\\
	16\,I_6\,I_{15} &= \mathbb{K}\,\overline{\mathbb{K}}\;,
	\\
	3072\,I_6^2\,\mathbb{M}_2^2 &=\mathbb{I}\,\mathbb{K}^2\overline{\mathbb{K}}^2\;.
	\label{siggiLast}
\end{align}
Syzygies, i.e., polynomial relations, among the ZM (or CM) invariants, are used to classify spacetimes according to their 6 Petrov (Weyl tensor) types \cite{1954UZKGU.114...55P} and 15 Segre (Ricci tensor) types, see, e.g.,~\cite{Stephani:2003tm}. Such a mathematical classification is of physical importance because spacetimes of the same type share important physical properties. For instance, regarding Kerr-Newman, the above syzygies imply that the spacetime is of Petrov type D and of Segre type $[11,(1,1)]$ (electro-vacuum solutions to GR).
For the Segre type corresponding to vacuum GR, i.e., for $R_{ab}=0$, a collection of theorems is known, see, e.g.,~\cite[Ch.~35.3.3]{Stephani:2003tm} for a review. In particular, these establish that every \emph{vacuum} Petrov type D metric (with one specific exception) implies the presence of a Killing tensor \cite{Walker:1970un, Hughston:1972qf}.
In contrast to Killing vectors, which encode an explicit symmetry of the spacetime, Killing tensors do not generate a spacetime isometry. The respective, so called `hidden symmetry' manifests itself only as a constant of motion (separability structure in mathematical terms), i.e., only in the dynamics of test-particles. 

Moreover, the special algebraic type has been tied to an explanation of the otherwise miraculous success of the Newman-Janis algorithm \cite{Newman:1965tw} which generates axisymmetric vacuum solutions from spherically symmetric vacuum solutions of the Einstein equations \cite{Gurses:1975vu, Drake:1998gf}.

\subsection{Invariant characterization of the spacetime with general mass function}
\label{app:gen-mass-fct}
Replacing $M\rightarrow M(r,\chi)$ in ingoing Kerr coordinates, cf.~Eq.~\eqref{eq:regular-spinning-BH-metric-ingoing-Kerr} for the explicit line element, introduces derivatives of the mass function $M(r,\chi)$ and therefore additional complexity in the set of ZM invariants, cf.~App.~\ref{app:ZM-basis}. 

Nevertheless, employing the Mathematica package xAct \cite{Martin-Garcia:2007bqa, MartinGarcia:2008qz} to calculate the set of invariants explicitly, we are able to identify several syzygies for the regular spacetime with general $M(r,\chi)$.
Two syzygies between the complex invariants which are also fulfilled in Kerr (as well as Kerr-Newman) spacetime, i.e., Eqs.~\eqref{siggi1} and \eqref{siggi3}, are still preserved.
Additionally, we identify two syzygies among the Ricci invariants
\begin{align}
	0 &= \frac{1}{8}\left(I_5^2 - 2\,I_6\right)^2 - \left(I_6^2 - 2\,I_8\right)\;,
	\\
	0 &= \frac{1}{8}I_5\left(I_5^2 - 6\,I_6\right) + I_7\;.
\end{align}
These syzygies are trivially fulfilled for Kerr (and even for Kerr-Newman) spacetime where vacuum (electro-vacuum) implies $I_5=I_6=I_7=I_8=0$ ($I_5=I_7=0$) and the above syzygies either vanish identically or reduce to Eq.~\eqref{siggi2}. In the regular spacetime, all of the Ricci invariants are non-vanishing, i.e., the spacetime is no longer a vacuum solution to GR, but the above syzygies still hold.

Making use of all the identified syzygies and writing complex combinations where available, cf.~Eq.~\eqref{eq:def-complex-invariants}, offers a way of writing the ZM invariants in a condensed form. In the following explicit expressions, we introduce the shorthand $M^{(n_r,n_\chi)}$ for the $n_r$-th and $n_\chi$-th partial derivatives of $M(r,\chi)$ with respect to $r$ and $\chi$.  
Introducing two fiducial invariants $\mathfrak{I}$ and $\mathfrak{K}$ to avoid having to rewrite large expressions twice, we find that the complex invariants can be written as
\begin{align}
	\mathfrak{I} &= \frac{2 (6 (r+i a \chi ) M(r,\chi )+(r-i a \chi ) (r(r-i a \chi)M^{(2,0)}(r,\chi )-2(2 r+i a \chi )M^{(1,0)}(r,\chi)))}{\sqrt{3} (r-i a \chi
   )^3 (r+i a \chi )}\;,
	\notag\\
	\mathfrak{K} &= \frac{\left(a^2 \chi ^2 (r M^{(2,0)}(r,\chi)+2 M^{(1,0)}(r,\chi))+r^2 (r M^{(2,0)}(r,\chi)-2 M^{(1,0)}(r,\chi))\right)^2}{3 \left(a^2 \chi ^2+r^2\right)^4}\;,
	\notag\\
	\mathbb{I} &= \mathfrak{I}^2\;,
	\\
	\mathbb{J} &= \frac{1}{\sqrt{12}}\mathfrak{I}^3\;,
	\\
	\mathbb{K} &= -\sqrt{3}\mathfrak{I}\mathfrak{K}\;,
	\\
	\mathbb{L} &= \mathfrak{I}^2\mathfrak{K}\;,
	\\
	\mathbb{M} &= 0\;,
	\\
	\mathbb{M}_2 &= -\frac{\left(r M^{(2,0)}(r,\chi)  \left(a^2 \chi ^2+r^2\right)-2 M^{(1,0)}(r,\chi)  (r-a \chi ) (a \chi +r)\right)^2}{108 (r-i a \chi )^{11} (r+i a \chi )^9}\cdot
\notag\\&
		\cdot\left(r M^{(2,0)}(r,\chi)  (r-i a \chi )^2-2 M^{(1,0)}(r,\chi)  (2 r+i a \chi ) (r-i a \chi )+6 M(r,\chi) (r+i a \chi )\right)^2\cdot
\notag\\&
		\cdot\left((r+i a \chi ) \left(r M^{(2,0)}(r,\chi) (r+i a \chi )+M^{(1,0)}(r,\chi)  (-4 r+2 i a \chi )\right)+6 M(r,\chi) (r-i a \chi )\right)
	\;.
\end{align}
These are supplemented by the real invariants
\begin{align}
	I_5 &= \frac{2 (r M^{(2,0)}(r,\chi) +2 M^{(1,0)}(r,\chi) )}{a^2 \chi ^2+r^2}\;,
	\\
	I_6 &= 
	\frac{2 r^2 (M^{(2,0)}(r,\chi))^2}{\left(a^2 \chi ^2+r^2\right)^2}
	+\frac{8 a^2 r \chi ^2 M^{(1,0)}(r,\chi)  M^{(2,0)}(r,\chi) }{\left(a^2 \chi ^2+r^2\right)^3}
	+\frac{8 (M^{(1,0)}(r,\chi))^2 \left(a^4 \chi ^4+r^4\right)}{\left(a^2 \chi^2+r^2\right)^4}
	\\
  	I_7 &= 
  	-\frac{1}{8}I_5\left(I_5^2 - 6\,I_6\right)
  	\;,
	\\
	I_8 &= 
	\frac{1}{16}\left(-I_5^4 + 4\,I_5^2\,I_6 + 4\,I_6^2\right)
	\;,
	\\
	I_{15} &= \frac{\left(r M^{(2,0)}(r,\chi)  \left(a^2 \chi ^2+r^2\right)-2 M^{(1,0)}(r,\chi)  (r-a \chi ) (a \chi +r)\right)^2}{36 \left(a^2 \chi ^2+r^2\right)^8}\Bigg[
		36 M(r,\chi)^2 \left(a^2 \chi ^2+r^2\right)
\notag\\&\hspace{2em}
		+4 (M^{(1,0)}(r,\chi))^2 \left(a^4 \chi ^4+5 a^2 r^2 \chi ^2+4 r^4\right)
\notag\\&\hspace{2em}
		+4 r M^{(1,0)}(r,\chi) \left(M^{(2,0)}(r,\chi)  \left(a^4 \chi ^4-a^2 r^2 \chi ^2-2 r^4\right)-12 M(r,\chi) r^2\right)
\notag\\&\hspace{2em}
		+r^2 M^{(2,0)}(r,\chi)  \left(M^{(2,0)}(r,\chi)  \left(a^2 \chi ^2+r^2\right)^2+12 M(r,\chi) \left(r^2-3 a^2 \chi ^2\right)\right)
	\Bigg]\;.
\end{align}
Note that all derivatives with respect to $\chi$ cancel out. We provide the equivalent but significantly longer expressions for all real ZM invariants in a supplementary notebook.

Having identified two complex syzygies, two real syzygies, and one vanishing complex invariant, the regular spacetime is described by (at most) nine (algebraically) independent non-vanishing real ZM invariants which can all be extracted from the above. The syzygies imply that our class of spacetimes is of Petrov type D, but no longer a vacuum solution to GR. We cannot exclude the existence of additional syzygies.

\subsection{Mass function from classical curvature invariants}
\label{app:KGRfromClassicalKerrInvariants}
\begin{figure}[!t]
\centering
\includegraphics[width=0.49\linewidth]{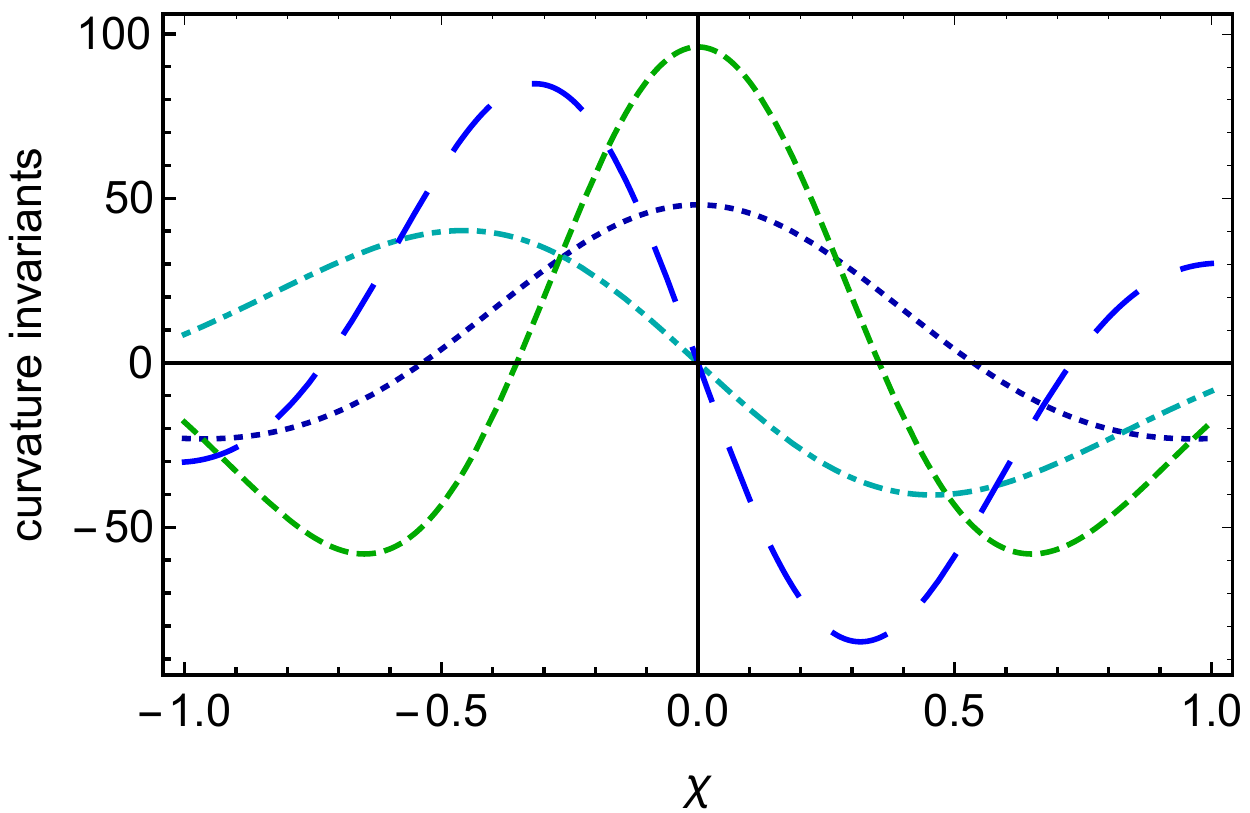}\hfill \includegraphics[width=0.49\linewidth]{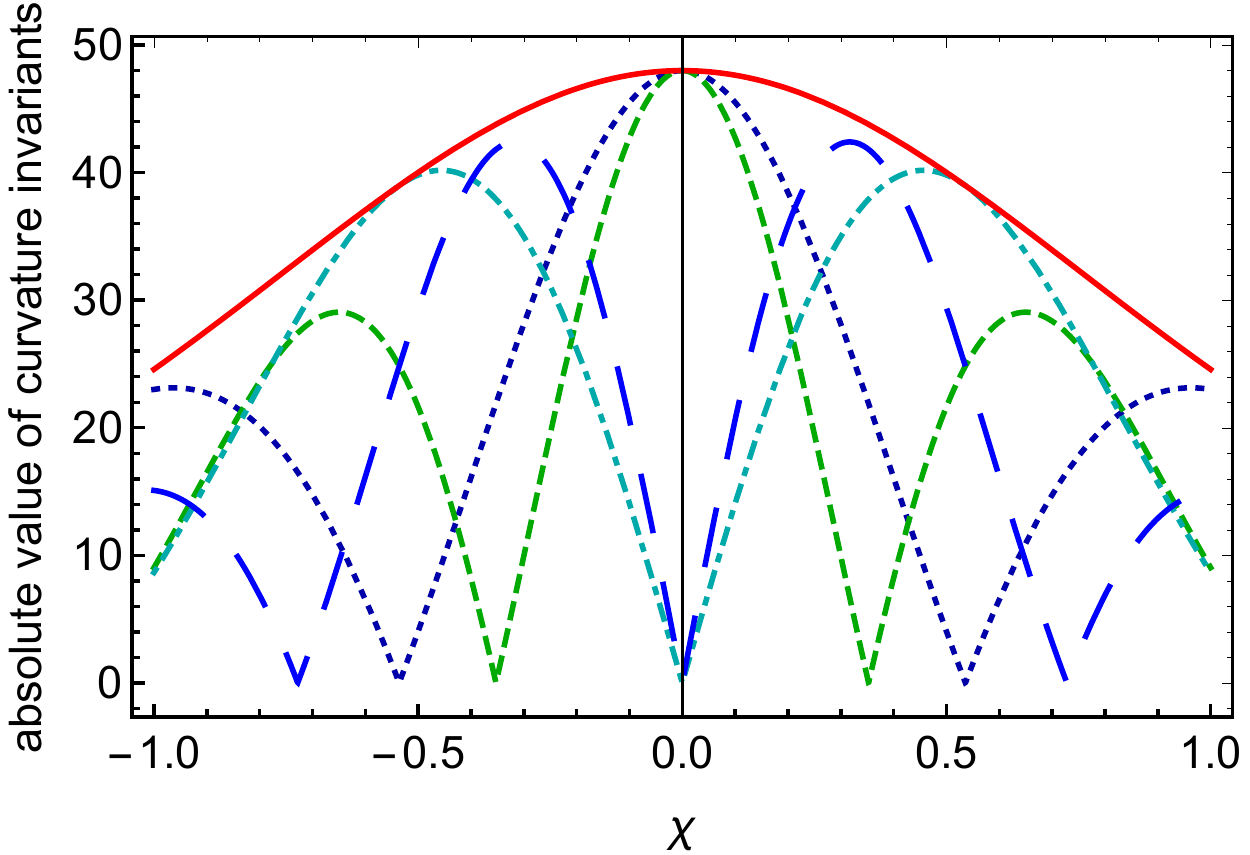}
\caption{\label{fig:classicalinvariants} Left panel: We show $I_1$ (dark-blue, dotted), $I_2$ (cyan, dot-dashed), $I_3$ (green, dashed) and $I_4$ (long dashes). Right panel: The absolute values of the curvature invariants (with an additional factor of 1/2 for $I_3$ and $I_4$) are shown in comparison to the interpolation $K_{\rm GR} =\sqrt{I_1^2+I_2^2}$ (red, continuous).}
\end{figure}

For the Kerr solution, obtained by setting $M(r,\chi) = M$ in
App.~\ref{app:gen-mass-fct}, 
only the pure Weyl invariants $I_1$, $I_2$, $I_3$, and $I_4$ are non-vanishing, i.e.,
\bea
I_1&=&\frac{48 M^2}{(r^2+a^2 \chi^2)^6} \left(r^6-15r^4 a^2 \chi^2 +15 r^2 a^4 \chi^4- a^6 \chi^6\right),\label{eq:I1Kerr}\\
I_2&=&\frac{-96 M^2}{(r^2+a^2\chi^2)^6} \left(3r^4-10r^2a^2\chi^2+3 a^4 \chi^4 \right),\label{eq:I2Kerr}\\
I_3&=&\frac{96 M^3}{(r^2+a^2\chi^2)^9} \left(r^9-36r^7a^2\chi^2 +126 r^5 a^4 \chi^4 \right), \label{eq:I3Kerr}\\
I_4&=&\frac{-96 M^3}{(r^2+a^2\chi^2)^9} \left( 9r^8 a \chi- 84 r^6a^3 \chi^3 +126 r^4 a^5\chi^5 - 36 r^2 a^7 \chi^7 +a^9 \chi^9\right).
\eea
In fact, due to the syzygy in Eq.~\eqref{siggi1}, the above invariants are non-linearly related amongst each other by
\begin{align}
	I_1(I_1^2 - 3\,I_2^2) &= 12(I_3^2 - I_4^2)\;, \label{eq:relnI1}
	\\
	I_2(3\,I_1^2 - I_2^2) &= 24\,I_3\,I_4 \label{eq:relnI2}
	\;,
\end{align}
such that only two of them are (algebraically) independent.
All four invariants take positive as well as negative values, inside and also outside of the horizon. 

We assume that the sign of the local curvature is not relevant to determine the onset of new physics, i.e., that the onset of new physics is determined by the absolute value of the maximum of the local curvature invariants. 
Therefore we construct an approximation of the maximum local curvature scale by setting
\be
K_{\rm GR} = \sqrt{I_1^2+I_2^2}.
\ee
 In fact, $K_{\rm GR}$ provides an envelope function for the function $\max[|I_1|,|I_2|, 1/2 |I_3|, 1/2|I_4|]$, cf.~Fig.~\ref{fig:classicalinvariants}. Herein, the factors of 1/2 are chosen such that $K_{\rm GR} \rightarrow I_1$ for the spherically symmetric limit $a \rightarrow 0$. Given the relations in Eq.~\eqref{eq:relnI1} and \eqref{eq:relnI2}, the choice of such coefficients does not change the full characterization of the spacetime, as all independent invariants are actually accounted for by including $I_1$ and $I_2$.

\subsection{Regularity of curvature invariants}\label{app:reginvs}
As in the main text, we specify to the case
\bea
M(r, \chi)& =& \frac{M}{1+\ell_{\rm NP}^4 \cdot K_{\rm GR}} = \frac{M}{1+\ell_{\rm NP}^4 \frac{48 M^2}{(r^2+a^2\chi^2)^3}}, \label{eq:massfunctionrep}
\eea
This mass function is shown in Fig.~\ref{fig:massfunction}. It interpolates between the classical mass $M$, normalized to one in the plot, at large radial geodesic distance to the center, and a vanishing mass in the center. In fact, it has to vanish sufficiently fast in order to ensure the regularity of curvature invariants in the center.
We now investigate the finiteness of all curvature invariants, focusing in particular on the limit $r \rightarrow 0, \chi \rightarrow 0$, where the classical Kerr spacetime features a ring singularity. 
It is known that the finite-spin counterpart of the Hayward black hole, which is obtained using the Newman-Janis algorithm, i.e., a $\chi$-independent mass function, features a multiply-valued Kretschmann scalar \cite{Bambi:2013ufa} at this point. The same property is true for a subset of higher curvature invariants \cite{Borissova:2020knn}. Therefore, we will investigate the regularity as well as multivaluedness of the full basis of non-derivative curvature invariants at $r=0,\chi=0$.\\
Further, as in \cite{Held:2019xde}, curvature invariants can become singular at the horizon, when deviations from the Kerr metric are introduced, see also \cite{Johannsen:2013rqa}. Thus, for each curvature invariant, we consider its radial dependence in and away from the equatorial plane.

First, we explore the limit $(r \rightarrow 0, \chi \rightarrow 0)$. This is the location of the ring singularity for the Kerr black hole and the location of a multi-valued Kretschmann scalar in, e.g, the spinning counterpart of the Hayward black hole.

Here, we provide a list of the leading-order behavior of the Weyl-invariants which are part of the basis of non-derivative interactions as $r \rightarrow 0$, from which the limit $\chi \rightarrow 0$ can be taken. These are the invariants that become singular as $\ell_{\rm NP} \rightarrow 0$. We have convinced ourselves that the order of the limits does not matter in our case (it would for the choice $\beta=1$), i.e., all curvature invariants are single-valued.
Specifically, we have
\bea
I_1&=& -\frac{48 M^2 a^6 \chi^6}{(48 \ell_{\rm NP}^4 M^2 + a^6 \chi^6)^2} + \mathcal{O}(r^2),\\
I_2&=& - \frac{288 M^2 a^{11}\chi^{11}}{(48 \ell_{\rm NP}^4 M^2 + a^6 \chi^6)^3}r+\mathcal{O}(r^3),\\
I_3&=& \frac{864 M^3 a^{14}\chi^{14}}{(48 \ell_{\rm NP}^4 M^2 + a^6 \chi^6)^4}r +\mathcal{O}(r^3),\\
I_4&=& - \frac{96 M^3 a^9 \chi^9}{(48 \ell_{\rm NP}^4 M^2 + a^6 \chi^6)^3} +\mathcal{O}(r^2).
\eea
From these expressions, it is clear how the introduction of a finite $\ell_{\rm NP}$ ensures a regular behavior at the origin.

At the horizon, all curvature invariants stay finite. This follows from the fact that the mass-function and its derivatives are finite everywhere, and thus the only divergences can come from the denominators. These take the form $(r^2 +a^2\chi^2)^n$, $n>0$ and accordingly only vanish for $r \rightarrow 0,\, \chi \rightarrow 0$, but stay finite at finite $r$, including at the horizon.

\subsection{Boyer-Linquist form}\label{app:BL-form}

Due to its special algebraic nature (Petrov type D vacuum solution of GR), Kerr spacetime can be described by a special form of coordinates found by Boyer and Lindquist \cite{1967JMP.....8..265B}. 
We analyze whether the regular black-hole metric which was constructed in horizon-penetrating coordinates, cf.~Eq.~\ref{eq:regular-spinning-BH-metric-ingoing-Kerr}, can be brought to a Boyer-Lindquist-like form away from an expansion to order $\mathcal{O}(r^{-6})$ around asymptotic infinity, where a BL-form is possible, see Sec.~\ref{sec:Newtonian-limit}.
\\
In the literature, two distinct and successively more restrictive conditions are both being referred to as Boyer-Lindquist form.
First, one may demand a form in which the Killing vectors of axisymmetry and stationarity are manifest (we denote the corresponding coordinates by $t$ and $\phi_\text{BL}$, respectively) and the only off diagonal component of the metric is $g_{t\phi_\text{BL}}$. 
Second, one may ask for a more specialized form which implies the existence of a Killing tensor and thereby a separability structure and an associated hidden constant of motion. For Kerr spacetime this constant of motion is called the Carter constant \cite{Carter:1968rr}.
\\
We focus on the first requirement first and then proceed to the more specialized question of a hidden constant of motion below.

In the following, we explicitly show that there does not exist a coordinate transformation of the passive coordinates, i.e., those that do not occur in any metric component, which leads to even the less-restrictive Boyer-Lindquist form, i.e., to coordinates in which $g^{t\phi}$ is the only off-diagonal component of the metric.\\
Given the regular black-hole metric in Eq.~\eqref{eq:regular-spinning-BH-metric-ingoing-Kerr}, where metric components only depend on coordinates $r$ and $\chi$, we ask whether a general coordinate transformation of the other two passive coordinates $u$ and $\phi$, i.e.,
\begin{align}
	du &=
	\mathcal{F}_t(r,\, \chi) dt 
	+ \mathcal{F}_r(r,\,\chi) dr 
	+ \mathcal{F}_\chi(r,\,\chi) d\chi 
	+\mathcal{F}_{\phi}(r,\, \chi) d\phi_{\rm BL}\;,
	\label{eq:BLtrafo1}
	\\
	d\phi &= 
	\mathcal{G}_{\phi}(r, \, \chi)d\phi_\text{BL} 
	+ \mathcal{G}_r(r,\,\chi) dr 
	+ \mathcal{G}_\chi(r,\,\chi) d\chi
	+\mathcal{G}_{t}(r,\, \chi)dt\;,
	\label{eq:BLtrafo2}
\end{align}
can lead to a form of the metric as in Eq.~\eqref{eq:Carter-form}.

We perform the general coordinate transformation and invert the resulting metric. Demanding that all off-diagonal components except $g^{t\phi_\text{BL}}$ vanish results in four constraints (a fifth one is trivially fulfilled), 
\begin{align}
	0&=\frac{(\chi -1) \left(\mathcal{G}_{\chi } \mathcal{F}_{\phi }-\mathcal{G}_{\phi } \mathcal{F}_{\chi }\right)}{\left(a^2 \chi^2+r^2\right) \left(\mathcal{G}_t \mathcal{F}_{\phi }-\mathcal{G}_{\phi } \mathcal{F}_t\right)}\;,
	\\
	0&=\frac{(\chi -1) \left(\mathcal{G}_t \mathcal{F}_{\chi }-\mathcal{G}_{\chi } \mathcal{F}_t\right)}{\left(a^2 \chi ^2+r^2\right)\left(\mathcal{G}_t \mathcal{F}_{\phi }-\mathcal{G}_{\phi } \mathcal{F}_t\right)}\;,
	\\
	0&=\frac{\mathcal{F}_{\phi } \left(\mathcal{G}_r \left(a^2-2 M(r,\chi) r+r^2\right)-a\right)+\mathcal{G}_{\phi } \left(-\mathcal{F}_r\left(a^2-2 M(r,\chi) r+r^2\right)+a^2+r^2\right)}{\left(a^2 \chi ^2+r^2\right) \left(\mathcal{G}_t \mathcal{F}_{\phi}-\mathcal{G}_{\phi } \mathcal{F}_t\right)}\;,
	\\
	0&=
	\frac{2 \left(\mathcal{F}_t \left(\mathcal{G}_r \left(a^2-2 M(r,\chi) r+r^2\right)-a\right)+\mathcal{G}_t \left(-\mathcal{F}_r
   \left(a^2-2 M(r,\chi) r+r^2\right)+a^2+r^2\right)\right)}{\left(a^2 \chi ^2+r^2\right) \left(\mathcal{G}_t \mathcal{F}_{\phi
   }-\mathcal{G}_{\phi } \mathcal{F}_t\right)}\;,
\end{align}
where we have dropped all arguments of $\mathcal{F}_i$ and $\mathcal{G}_i$.
These constraints only have a single special solution
\begin{align}
	\mathcal{F}_r &= \frac{a^2 + r^2}{r^2 - 2M(r,\chi) + a^2}\;,\nonumber
	\\
	\mathcal{G}_r &= \frac{a}{r^2 - 2M(r,\chi) + a^2}\;,\nonumber
	\\
	\mathcal{F}_\chi &= \mathcal{G}_\chi = 0\;,
	\label{eq:specialSols}
\end{align}
leaving $\mathcal{F}_t$, $\mathcal{F}_{\phi_\text{BL}}$, $\mathcal{G}_t$, and $\mathcal{G}_{\phi_\text{BL}}$ unconstrained.
However, unless $M(r,\chi) = M(r)$ is $\chi$-independent, this algebraic solution does not result in an exact differential form $du$ and $d\phi$ because $\partial_r \mathcal{F}_\chi \neq \partial_\chi \mathcal{F}_r$ and $\partial_r \mathcal{G}_\chi \neq \partial_\chi \mathcal{G}_r$.\\

More generally, it is known that whenever the spacetime admits an inverse metric of the form \cite{1979GReGr..10...79B}
\begin{align}
\label{eq:Carter-form}
	g^{ab}\partial_{a}\partial_{b}  =\frac{1}{S_{r}(r)+S_{\theta}(\theta)}
	\Big[
		\left(G_{r}^{ij}(r)+G_{\theta}^{ij}(\theta)\right)\partial_{\sigma_i}\partial_{\sigma_j}
		+\Delta_{r}(r)\partial_{r}^{2}
		+\Delta_{\theta}(\theta)\partial_{\theta}^{2}
	\Big]\;,
\end{align}
where the $\sigma_i = (t,\,\phi)$ are the coordinates associated with the Killing vectors, this implies the existence of a Killing tensor
\begin{align}
	K^{ab}\partial_{a}\partial_{b} = \frac{1}{S_{r}+S_{\theta}}
	\Big[
		\left(S_{r}G_{\theta}^{ij} - S_{\theta}G_{r}^{ij}\right)\partial_{\sigma_i}\partial_{\sigma_j}
		+S_{r}\Delta_{\theta}\partial_{\theta}^{2}
		-S_{\theta}\Delta_{r}\partial_{r}^{2}
	\Big]
\end{align}
and an associated constant of motion. It can be checked explicitly that Kerr spacetime (in Boyer-Lindquist coordinates) fulfills even the more restrictive of the above demands.

Since we have already established that there does not exist a coordinate transformation of the passive coordinates that leads to even the less-restrictive Boyer-Lindquist form, it is also clear that we cannot identify a respective Killing tensor for our family of regular spinning black holes.

\section{Numerical ray tracing and shadow analysis}
\label{app:ray-tracing}

For both, the numerical ray tracing as well as the integration of the radiative transfer equation, we make use of the internal numerical integration techniques available in Mathematica \cite{Mathematica}.

\subsection{Initial conditions on the screen}
\label{app:initialConditions}
The observer's screen is placed at sufficiently large radial distance $r_\text{obs}\gg M$ to the black hole (we work with $r_\text{obs}=100M$ throughout the paper), where, as Sec.~\ref{sec:Newtonian-limit} shows, the modified geometry is well-approximated by Kerr spacetime and exhibits a Newtonian limit. In particular, this justifies that the initial conditions may conveniently be set in Boyer-Lindquist (BL) coordinates and can subsequently be transformed to ingoing Kerr coordinates by the respective \emph{classical} coordinate transformation.

The origin of the screen is placed at $(r_\text{obs},\,\theta_\text{obs},\,\phi_{BL,\rm obs})$ in BL coordinates and the image plane is set up perpendicular to the vector pointing from $(r_\text{obs},\,\theta_\text{obs},\,\phi_{BL,\rm obs})$ to the center of the black hole. The image-plane coordinates $(x,y)$ can be expressed in terms of Cartesian coordinates $(X,Y,Z)$ centered around the black hole via
\begin{align}
    X &= \mathcal{D}\,\cos\phi_\text{BL,obs} - x\,\sin\phi_\text{BL,obs}\;,
    \\
    Y &= \mathcal{D}\,\sin\phi_\text{BL,obs} + x\,\cos\phi_\text{BL,obs}\;,
    \\
    Z &= r_\text{obs}\,\cos(\theta_\text{obs})  + y\,\sin(\theta_\text{obs}) \;,
\end{align}
where $\mathcal{D} = \sin(\theta_\text{obs})\sqrt{r^2_\text{obs} + a^2} - y\,\cos(\theta_\text{obs})$. These can be transformed to Boyer-Lindquist coordinates $(t,\,r,\,\theta,\,\phi_{\rm BL})$ by
\begin{align}
    \label{eq:screenCoordsInBL}
    r^2 = \sigma + \sqrt{\sigma^2 + a^2\,Z^2)^{1/2}}
    \;,\;
    \cos{\theta} = Z/r
    \;,\;
    \tan{\phi_\text{BL}} = Y/X\;,
\end{align}
where $\sigma = \left(X^2+Y^2+Z^2-a^2\right)/2$. The light rays are initialized perpendicular to the screen. Together with the condition that 
the trajectories are lightlike, this fixes the initial momentum vector, which can be determined by differentiating Eq.~\eqref{eq:screenCoordsInBL}.
These initial conditions are subsequently transformed into ingoing Kerr coordinates $(u,\,r,\,\theta,\,\phi)$ by the classical transformation, i.e.,
\begin{align}
	\label{eq:BLtoKerr}
	t &= u - \left[
		r + M\,\log\left(\frac{r^2 + a^2 - 2 M r}{4 M^2}\right) 
		+ \frac{M^2}{\sqrt{M^2 - a^2}}\log\left(\frac{r-M-\sqrt{M^2-a^2}}{r-M+\sqrt{M^2-a^2}}\right) 
	\right]\;,
	\\
	\phi_{BL} &= \phi - \frac{a}{2\,\sqrt{M^2 - a^2}}\log\left(\frac{r-M-\sqrt{M^2-a^2}}{r-M+\sqrt{M^2-a^2}}\right)\;.
\end{align}
The numerical ray tracing as well as the integration of the radiative transfer equation are performed in ingoing Kerr coordinates.

\subsection{Boundary bisections and angular boundary curves}
\label{app:boundaryCurves}
The shadow boundary is determined by iterative bisection of radial intervals. More specifically, we define the polar image coordinates
\begin{align}
	\psi = \tan(x/y),
	\\
	\rho = \sqrt{x^2 + y^2},
\end{align}
such that the shadow boundary in the image plane is described by a boundary curve $\rho(\psi)$. At each image angle $\psi$, we numerically approximate the shadow boundary with desired precision $\epsilon$ by the following algorithm:
\begin{enumerate}
	\item
	Select a suitable initial interval $\rho = [\rho^{(0)}_1,\,\rho^{(0)}_2]$ such that a light ray initialized at $(\psi,\,\rho^{(0)}_1(\psi))$ falls into the black hole, while one that is initialized at $(\psi,\,\rho^{(0)}_2(\psi))$ is scattered by the black hole and escapes to radial infinity.
	\item
	Check whether the interval $\epsilon<\rho^{(n)}_2-\rho^{(n)}_1$ has reached the desired precision. If so, the algorithm terminates and the shadow boundary is guaranteed to lie within the interval $[\rho^{(n)}_1,\,\rho^{(n)}_2]$. Otherwise proceed to the next step.
	\item
	Determine whether the light ray falls into the black hole or escapes when initialized at $(\psi,\,(\rho^{(n)}_2-\rho^{(n)}_1)/2)$.
	If the light ray with bisected initial conditions falls into the black hole, repeat the previous step with $[\rho^{(n+1)}_1,\rho^{(n+1)}_2] = [\rho^{(n)}_1,\,(\rho^{(n)}_2-\rho^{(n)}_1)/2]$.
	If the light ray with bisected initial conditions escapes, repeat the previous step with $[\rho^{(n+1)}_1,\,\rho^{(n+1)}_2] = [(\rho^{(n)}_2-\rho^{(n)}_1)/2,\,\rho^{(n)}_2]$.
\end{enumerate}
If the algorithm converges to either of the initially chosen values $\rho^{(0)}_1$ or $\rho^{(0)}_2$, no shadow boundary occurs within the initial interval. Throughout this work it is sufficient to use the initial interval $[\rho_1,\rho_2] = [0,10M]$.

Due to the cusp-like features in the shadow boundary, it can happen that one radial bisection contains multiple boundary points. Whenever this happens, the algorithm only converges to one of these points, typically the outermost one.

\subsection{Matching to an enveloping shadow boundary of Kerr spacetime}
\label{app:classicalEnvelope}
Due to the existence of a hidden constant of motion, cf.~App.~\ref{app:ZM-basis}, and the respective Carter constant, the classical shadow boundary of Kerr black holes is known analytically, cf.~\cite{Bardeen:1973tla}, and can be written as a parameterized curve in terms of the parameter $t$
\begin{align}
	x_\text{Kerr}(t) &=\frac{M}{a\,\sin(\theta_\text{obs})}\left[t^2 + a^2 - 3 - \frac{2(1-a^2)}{t}\right]\;,
	\\
	y_\text{Kerr}(t) &=\pm\sqrt{\frac{(1+t)^3}{a^2}\left(3 - t - \frac{4(1-a^2)}{t^2} + \left(a^2 - \frac{x^2}{M^2}\right)\cos(\theta_\text{obs})^2\right)}M\;.
\end{align}
The range of $t$ can be determined by demanding that the expression under the square-root is positive. 
We shift the classical shadow boundary in $x$-direction to match with the regular one at the retrograde boundary point at $(x>0,\,y=0)$. In addition, an arbitrary second point on the classical boundary curve can be matched by rescaling the classical mass $M$. In polar image coordinates this matching procedure results in a classical boundary curve $\rho_\text{fit}(\psi)$ which matches with the regular one $\rho(\psi)$ at two points $\psi\in[0,\pi]$. 

 The classical shadow boundary is an enveloping function to the regular one, i.e., $\rho(\psi) - \rho_\text{fit}(\psi)<0$ for all $\psi$,  if we pick the point with maximal $y_\text{max} = \max_{\psi\in[0,\pi]}y(\psi)$ as the second matching point. 

According to the above, we define an integrated measure of deviation to an enveloping Kerr shadow boundary, i.e.,
\begin{align}
	\text{dev} = \int_0^\pi d\psi\,|\rho(\psi) - \rho_\text{fit}(\psi)|\;.
\end{align}
We find that this integrated deviation is maximal at around $a\approx 0.91$.

\subsection{Numerical algorithm to identify $\chi_\text{min}$ and $\chi_\text{max}$}
\label{app:chiMinMax}

The shadow boundary corresponds to the image of the photon sphere of a black hole. The photon sphere is the region in spacetime in which closed photon orbits are possible.
While the photon sphere of Schwarzschild spacetime is a spherical 2-surface located at $r=3\sqrt{3}M$, the photon sphere of Kerr spacetime covers an extended spacetime volume. These null geodesics exhibit an oscillating behavior between $\chi_\text{min}$ and $\chi_\text{max} = -\chi_\text{min}$. 

For the regular spacetime presented in the main text, we cannot analytically determine the photon sphere, because we could not identify a Killing tensor which guarantees separability of geodesic motion, cf.~App.~\ref{app:BL-form}. Nevertheless, ray tracing allows us to numerically approximate the closed photon orbits. Image points outside the shadow, which approximate the shadow boundary with increasing precision $\epsilon$ (with $\epsilon$ a dimensionless measure of the distance between the given image point and the idealized shadow boundary) correspond to photon trajectories which wind around the black hole (both in $\chi$ and $\phi$) an increasing number of times $n_\chi(\epsilon)$. In the limit $\epsilon\rightarrow0$, these trajectories converge towards a specific closed photon trajectory within the photon sphere, before escaping to asymptotic infinity. In particular $n_\chi(\epsilon)\rightarrow\infty$ for $\epsilon\rightarrow 0$. 

Our aim is to identify $\chi_{\rm min}$ and $\chi_{\rm max}$ associated to the closed photon trajectory that a given light ray originating in our image screen is approximating. This is complicated by additional oscillations in $\chi$ that occur on such a light ray: Before and after (in affine-parameter time) such a near-critical trajectories approximate the closed photon orbit, it enters from and leaves towards asymptotic infinity. During these transitions between approach/departure and approximate closed photon orbit, they potentially pass additional local minima or maxima in $\chi$. These are not part of the approximated closed photon orbit. Their occurrence can be understood as follows. At precision $\epsilon$, a slight perturbation of the initial conditions at the observer's screen will only slightly perturb the first $n_\chi(\epsilon)$ oscillations. In contrast, the escape to infinity may vary drastically. The same argument applies if the endpoint of the resulting near-critical trajectory is slightly perturbed and subsequently integrated again -- now forward in time. While these different photon trajectories can exhibit vastly different behavior before and after orbiting the black hole, they share the same characteristic intermediate behavior, since they all approximate the same closed photon orbit. Here, we present an algorithm to identify the associated $\chi_\text{min}$ and $\chi_\text{max}$:
\begin{enumerate}
	\item[1.)]
	Approximate a point on the shadow boundary with sufficient precision $\epsilon$ by use of nested intervals as detailed in App.~\ref{app:boundaryCurves}. The resulting light ray is initialized on the observer's image plane at large values of
	$r_\text{obs}$ with affine parameter $\lambda_\text{min} = 0$, is then scattered by the black hole, and is integrated further until it escapes back to $r_\text{obs}$ at $\lambda_\text{max}$.
	\item[2.)]
	Identify \emph{all} local minima $\chi^{(n)}_\text{min}$ and maxima $\chi^{(n)}_\text{max}$ which occur along the trajectory in the interval $[\lambda_\text{min},\lambda_\text{max}]$.
	\item[3.)]
	Identify the $\chi^{(m)}_\text{min}$ for which at least one $\chi^{(n\neq m)}_\text{min}$ exists such that the deviation $|\chi^{(m)}_\text{min} - \chi^{(n\neq m)}_\text{min}|<\widetilde{\epsilon}$ is smaller than a given $\widetilde{\epsilon}$. Apply the corresponding procedure to identify $\chi^{(m)}_\text{max}$. If no such $\chi^{(m)}_\text{min}$ or $\chi^{(m)}_\text{max}$ are identified, decrease $\epsilon$ in comparison to $\widetilde{\epsilon}$ and re-start from step 1.). Otherwise proceed to the next step.
	\item[4.)]
	The mean values of the $\chi^{(m)}_\text{min}$ and the $\chi^{(m)}_\text{max}$ are identified as $\chi_\text{min}$ and $\chi_\text{max}$, respectively.
\end{enumerate}
We employ the above algorithm to generate the right-hand panel of Fig.~\ref{fig:cuspy-triptych} where we choose $\epsilon = 10^{-15}$ (in units of the classical black-hole mass) and $\widetilde{\epsilon}=10^{-3}$.

\bibliography{References}
	
\end{document}